%% file: article.tex
\pdfoutput=1
\documentclass{aa}  

\usepackage{txfonts}
\usepackage{graphicx}
\usepackage{empheq}
\usepackage{cases}
\usepackage{tabularx}

\usepackage{subfig} 
\usepackage{array}
\usepackage{color}
\usepackage[squaren,Gray]{SIunits}

\newcommand{\arc}{\arcsecond\per\mathrm{yr}}
\newcommand{\Gyr}{\mathrm{Gyr}}
\newcommand{\Myr}{\mathrm{Myr}}
\newcommand{\kyr}{\mathrm{kyr}}
\newcommand{\MYR}{\,\mathrm{Myr}}
\newcommand{\KYR}{\,\mathrm{kyr}}
\newcommand{\yr}{\mathrm{yr}}
\newcommand{\h}{\mathrm{h}}
\newcommand{\km}{\mathrm{km}}
\newcommand{\AU}{\mathrm{AU}}
\newcommand{\mas}{\mathrm{mas}}
\newcommand{\La}{La2011}

\newcommand{\Ce}{Ceres2017}

\newcommand{\Msol}{M_{\odot}}
\newcommand{\degday}{\mathrm{\degree/day}}
\newcommand{\C}{\overline{C}}
\newcommand{\I}{\overline{I}}
\newcommand{\kgm}{\mathrm{kg\,m^{-3}}}
\newcolumntype{L}[1]{>{\raggedright}p{#1}}
\newcolumntype{R}[1]{>{\raggedleft}p{#1}}
\newcolumntype{C}[1]{>{\centering}p{#1}}

\begin{document}

   \title{Long-term orbital and rotational motions of Ceres and Vesta}

   \subtitle{}

   \author{T.~Vaillant \and J.~Laskar \and N.~Rambaux \and M.~Gastineau}

   \institute{ASD/IMCCE, Observatoire de Paris, PSL Université, Sorbonne Université, 77 avenue Denfert-Rochereau,
75014 Paris, France \\e-mail: timothee.vaillant@obspm.fr}

   \date{Received ; accepted }
   
 
  \abstract
   {The dwarf planet Ceres and the asteroid Vesta have been studied by the Dawn space mission. They are the two heaviest bodies of the main asteroid belt and have different characteristics.
   Notably, Vesta appears to be dry and inactive with two large basins at its south pole. Ceres is an ice-rich body with signs of cryovolcanic activity.}
   {The aim of this paper is to determine the obliquity variations of Ceres and Vesta and to study their rotational stability.}
   {The orbital and rotational motions have been integrated by symplectic integration. The rotational stability has been studied by integrating secular equations and by computing the diffusion of the precession frequency.}
   {The obliquity variations of Ceres over $[-20:0]\MYR$ are between $2$ and $20\degree$ and the obliquity variations of Vesta are between $21$ and $45\degree$. The two giant impacts suffered by Vesta modified the precession constant and could have put Vesta closer to the resonance with the orbital frequency $2s_6-s_V$. Given the uncertainty on the polar moment of inertia, the present Vesta could be in this resonance where the obliquity variations can vary between $17$ and $48\degree$.}
   {Although Ceres and Vesta have precession frequencies close to the secular orbital frequencies of the inner planets, their long-term rotations are relatively stable. The perturbations of Jupiter and Saturn dominate the secular orbital dynamics of Ceres and Vesta and the perturbations of the inner planets are much weaker. The secular resonances with the inner planets also have smaller widths and do not overlap, contrary to the case of the inner planets.}

   \keywords{celestial mechanics - planets and satellites: dynamical evolution and stability - minor planets, asteroids: individual: Ceres - minor planets, asteroids: individual: Vesta}

   \maketitle
%

\section{Introduction}

Ceres and Vesta are the two heaviest bodies of the main asteroid belt. They have been studied by the Dawn space mission which has determined their shapes, gravity fields, surface compositions, spin rates, and orientations \citep{russell2012,russell2016}. However, the precession frequency  of their spin axes has not be determined and there are still uncertainties about their internal structures \citep[e.g.,][]{park2014,ermakov2014,park2016,ermakov2017b,konopliv2018}. No satellites have been detected from observations with the Hubble Space Telescope and the Dawn space mission around these bodies \citep{mcfadden2012,mcfadden2015,demario2016}.

The long-term rotation of the bodies in the solar system can be studied with the secular equations \citep{kinoshita1977,laskar1986,laskarrobutel1993} or with a symplectic integration of the orbital and rotational motions \citep{toumawisdom1994}. Secular equations are averaged over the mean longitude and over the proper rotation, which is generally fast for the bodies of the solar system, and their integration is much faster. They were used by \cite{laskarjoutelrobutel1993} and \cite{laskarrobutel1993} to study the stability of the planets in the solar system. 

The method of \cite{laskarrobutel1993} has been applied by \cite{skoglov1996} to study the stability of the rotation and the variations of the obliquity for Ceres and nine asteroids including Vesta. At this time, however, the initial conditions for the spin axes were not determined precisely and the knowledge of the internal structure was not sufficient to constrain the precession frequencies. \cite{skoglov1996} assumed that the bodies are homogeneous and concluded that their long-term rotations are relatively stable. By using secular equations and a secular model for the orbital motion, \cite{bills2017} determined the obliquity variations of Ceres. \cite{ermakov2017a} obtained the obliquity variations of Ceres for different polar moments of inertia by realizing the symplectic integration of the rotational and orbital motions.

Asteroid impacts and close encounters can influence the long-term rotation of bodies in the main asteroid belt. Vesta has suffered two giants impacts \citep{marchi2012,schenk2012} that have significantly modified its shape and its spin rate \citep{fu2014,ermakov2014}. \cite{laskar2011a} obtained an orbital solution of Ceres and Vesta, called La2010, which takes into account mutual interactions between bodies of the main asteroid belt, and \cite{laskar2011b} showed that close encounters in the solution La2010 are the cause of the chaotic nature of the orbits of Ceres and Vesta. These close encounters can affect their long-term rotation.

For Ceres, the obliquity drives the ice distribution on and under the surface. Ceres possesses cold trap regions that do not receive sunlight during a full orbit. This prevents the sublimation of the ice, which can accumulate \citep{platz2016}. The surface area of these cold traps depends on the value of the obliquity. \cite{ermakov2017a} determined that the obliquity of Ceres varies between $2$ and $20\degree$ and that the cold trap areas for an obliquity of $20\degree$ correspond to bright crater floor deposits that are likely water ice deposits. \cite{platz2016} determined that one bright deposit near a shadowed crater is water ice. In addition, the Dawn mission gave evidence of the presence of ice under the surface of Ceres from the nuclear spectroscopy instrument \citep{prettyman2017} and from the morphology of the terrains \citep{schmidt2017}. The ice distribution and the burial depth with respect to the latitude depend on the history of the obliquity \citep{schorghofer2008,schorghofer2016}. For Vesta, studies of the long-term evolution of the obliquity were not performed with the initial conditions of the spin axis and the physical characteristics determined by the Dawn space mission.

The main purpose of this article is to investigate the long-term evolution of the rotational motions of Ceres and Vesta.
First, we explore the obliquity variations of these bodies for a range of possible precession constants obtained from the data of the Dawn mission. Then, the stability of their spin axes is studied.

In this paper for the orbital motion we consider the solutions \La{} and La2010 \citep{laskar2011a}, which do not include the rotation of Ceres and Vesta. To compute the obliquity variations, we follow the symplectic method of \cite{farago2009} by averaging the fast proper rotation. This method avoids integrating the fast rotation and allows us to use a large step to reduce the computation time. We call the long-term rotational solution obtained \Ce{}. The orbital and rotational equations are integrated simultaneously in a symplectic way and the effects of the rotation on the orbital motions are considered. We consider the close encounters of Ceres and Vesta with the bodies of the main asteroid belt used in \cite{laskar2011b} and estimate with a statistical approach their effects on the long-term rotation of Ceres and Vesta. In order to determine the secular frequencies and identify the possible secular resonances on the orbital and rotational motions, the solutions are studied by the method of the frequency map analysis \citep{laskar1988,laskar1990,laskar1992,laskar1993,laskar2003}. Moreover, to study the effects of the close secular orbital resonances, we compute a secular Hamiltonian from the method of \cite{laskarrobutel1995}. We obtain a secular model, which reproduces the secular evolution of the solution \La{} and allows us to investigate the effects of the secular resonances.

The stability of the spin axes is studied by using secular equations with a secular orbital solution obtained from the frequency analysis of the solution \La{}. We verify beforehand that they allow us to reproduce the obliquity variations computed by the symplectic method and have the same stability properties. We study the stability of the rotation in the vicinity of the range of possible precession constants to identify the secular resonances between the orbital and rotational motions.
Vesta has suffered two giant impacts that have changed its shape and its spin rate \citep{fu2014,ermakov2014} and also its precession constant. We investigate whether this possible evolution of precession constant changed the stability properties. Following the method of \cite{laskarrobutel1993}, we finally realize a stability map of the spin axes of Ceres and Vesta.

In section \ref{SEC:methods} we present the methods used in this paper to obtain the long-term rotation. In section \ref{SEC:prec} we estimate the precession constants deduced from Dawn space mission and their possible variations during the history of Ceres and Vesta. In section \ref{SEC:orbobliquite} we analyze the long-term solutions obtained for the orbital and rotational motions. In section \ref{SEC:secularmodels} we study the effects of the orbital secular resonances with a secular Hamiltonian model. In section \ref{SEC:stab} we study the stability of the rotation axes from the secular equations of the rotation.

\section{Methods for the integration of the rotation\label{SEC:methods}}

The spin rates of Ceres and Vesta are relatively fast (see section \ref{SEC:prec}). We thus average the fast rotation using the method of \cite{farago2009} in order to integrate in a symplectic way the angular momentum of a rigid body.

When we need many integrations with different initial conditions or parameters, we use the secular equations from \cite{bouelaskar2006} in order to speed up the computation.

\subsection{Symplectic integration of the angular momentum\label{sec:oblisymp}}

We consider a planetary system of $n+1$ bodies with a central body $0$ and $n$ planetary bodies, where the body of index $1$ is a rigid body and the other planetary bodies point masses. We note the vectors in bold. The Hamiltonian $H$ of the system is \citep{bouelaskar2006}
\begin{equation}
H=H_{N}+H_{I,0}+\sum^{n}_{k=2}H_{I,k}+H_{E},
\end{equation}
with $H_{N}$ the Hamiltonian of $n+1$ point masses. The Hamiltonian $H_{E}$ of the free rigid body is
\begin{equation}
H_{E}=\frac{\left(\textbf{G}.\textbf{I}\right)^{2}}{2A}+\frac{\left(\textbf{G}.\textbf{J}\right)^{2}}{2B}+\frac{\left(\textbf{G}.\textbf{K}\right)^{2}}{2C},
\end{equation}
where ($\mathbf{I}$,$\mathbf{J}$,$\mathbf{K}$) is the basis associated with the principal axes of moments of inertia respectively $A$, $B$, $C$, where $A\leq B\leq C$, and $\mathbf{G}$ the angular momentum of the rigid body. The Hamiltonians $H_{I,0}$ and $H_{I,k}$ are respectively the interactions without the point mass interactions of the central body $0$ and of the planetary body $k$ with the rigid body $1$ and are obtained with a development in Legendre polynomials \citep{bouelaskar2006}
\begin{equation}
\begin{split}
H_{I,0} = & -\frac{\mathcal{G}m_{0}}{2r_{1}^{3}}\left[\left(B+C-2A\right)\left(\frac{\textbf{r}_{1}.\textbf{I}}{r_{1}}\right)^{2}+\left(A+C-2B\right)\left(\frac{\textbf{r}_{1}.\textbf{J}}{r_{1}}\right)^{2} \right. \\
& \left. +\left(A+B-2C\right)\left(\frac{\textbf{r}_{1}.\textbf{K}}{r_{1}}\right)^{2}\right], \label{eq:hamI0}
\end{split}
\end{equation}  
\begin{equation}
\begin{split}
H_{I,k} = & -\frac{\mathcal{G}m_{k}}{2r_{1,k}^{3}}\left[\left(B+C-2A\right)\left(\frac{\textbf{r}_{1,k}.\textbf{I}}{r_{1,k}}\right)^{2}+\left(A+C-2B\right)\left(\frac{\textbf{r}_{1,k}.\textbf{J}}{r_{1,k}}\right)^{2} \right. \\
& \left. +\left(A+B-2C\right)\left(\frac{\textbf{r}_{1,k}.\textbf{K}}{r_{1,k}}\right)^{2}\right],
\end{split}
\end{equation}
with ($\textbf{r}_k$,$\tilde{\textbf{r}}_k$) the heliocentric position and the conjugate momemtum of the body $k$, $m_{k}$ the mass of the body $k$, $\textbf{r}_{1,k}=\textbf{r}_{1}-\textbf{r}_{k}$, $r_{1}$ and $r_{1,k}$ the norms of $\textbf{r}_{1}$ and $\textbf{r}_{1,k}$, and $\mathcal{G}$ the gravitational constant. By averaging over the fast Andoyer angles $g$, the angle of proper rotation, and $l$, the angle of precession of the polar axis $\textbf{K}$ around the angular momentum $\textbf{G}$ \citep{bouelaskar2006}, $H_{E}$ becomes constant and the averaged total Hamiltonian $\mathcal{H}$ is 
\begin{equation}
\mathcal{H}=\left\langle H\right\rangle_{g,l}=H_{N}+\mathcal{H}_{I,0}+\sum^{n}_{k=2}\mathcal{H}_{I,k}, \label{eq:hamI0mean}
\end{equation} 
where
\begin{equation}
\mathcal{H}_{I,0}=\left\langle H_{I,0} \right\rangle_{g,l}=-\frac{\mathcal{C}_{1}m_{0}}{r_{1}^{3}}\left(1-3\left(\frac{\textbf{r}_{1}.\textbf{w}}{r_{1}}\right)^{2}\right),
\end{equation} 
\begin{equation}
\mathcal{H}_{I,k}=\left\langle H_{I,k} \right\rangle_{g,l}=-\frac{\mathcal{C}_{1}m_{k}}{r_{1,k}^{3}}\left(1-3\left(\frac{\textbf{r}_{1,k}.\textbf{w}}{r_{1,k}}\right)^{2}\right),
\end{equation}   
with 
\begin{equation}
\textbf{w}=\frac{\textbf{G}}{G},
\end{equation}
\begin{equation}
\mathcal{C}_{1}=\frac{\mathcal{G}}{2}\left(C-\frac{A+B}{2}\right)\left(1-\frac{3}{2}\sin^{2}J\right)\label{eq:C_1},
\end{equation}
$J$ the Andoyer angle between $\textbf{w}$ and $\mathbf{K}$, and $G$ the norm of the angular momentum $\mathbf{G}$.

The Hamiltonian $\mathcal{H}=H_{N}+\mathcal{H}_{I,0}+\sum^{n}_{k=2}\mathcal{H}_{I,k}$ can be split into several parts. The Hamiltonian $H_{N}$ of $n$ point masses can be integrated with the existing symplectic integrators \citep[e.g.,][]{wisdom1991,laskar2001,farres2013}.

For a planetary system where a planet is located much closer to the central star than the other planets,
\cite{farago2009} averaged its fast orbital motion to obtain a Hamiltonian of interaction between the orbital angular momentum of the closest planet and the other more distant planets. Because the Hamiltonians $\mathcal{H}_{I,0}$ and $\mathcal{H}_{I,k}$ are analogous to this Hamiltonian, we can use the symplectic method developed by \cite{farago2009} for this case. We detail explicitly how this method can be applied here. 

The Hamiltonian $\mathcal{H}_{I,0}$ gives the equations of the motion \citep{bouelaskar2006}
\begin{empheq}[left={\empheqlbrace}]{align}
\dot{\textbf{r}}_{1} &= \textbf{0}, \nonumber \\
\dot{\tilde{\textbf{r}}}_{1} &= -\nabla_{\textbf{r}_{1}} \mathcal{H}_{I,0} \nonumber \\
&= -\frac{3\mathcal{C}_{1}m_{0}}{r_{1}^{5}}\left(\left(1-5\left(\frac{\textbf{r}_{1}.\textbf{w}}{r_{1}}\right)^{2}\right)\textbf{r}_1+2\left(\textbf{r}_{1}.\textbf{w}\right)\textbf{w}\right), \label{eq:eqdiffh0} \\
\dot{\textbf{w}} &= \frac{1}{G}\nabla_{\textbf{w}} \mathcal{H}_{I,0} \times\textbf{w} = \frac{6\mathcal{C}_{1}m_{0}}{Gr_{1}^{5}}\left(\textbf{r}_{1}.\textbf{w}\right)\textbf{r}_{1}\times\textbf{w}. \nonumber
\end{empheq}
Here $\textbf{r}_{1}$ is conserved and because of $\textbf{r}_{1}.\dot{\textbf{w}}=0$, $\textbf{r}_{1}.\textbf{w}$ is also constant. With the angular frequency $\Omega_{0}=6\mathcal{C}_{1}m_{0}\left(\textbf{r}_{1}.\textbf{w}\right)/(Gr_{1}^{4})$ as in \cite{farago2009}, the solution for $\textbf{w}$ is
\begin{equation}
\textbf{w}\left(t\right)=R_{\textbf{r}_{1}}\left(\Omega_{0} t\right)\textbf{w}\left(0\right),
\end{equation}  
where $R_{\textbf{x}}\left(\theta\right)$ is the rotation matrix of angle $\theta$ around the vector $\textbf{x}$. The solution for $\tilde{\textbf{r}}_{1}$ is \citep{farago2009}
\begin{equation}
\begin{split}
\tilde{\textbf{r}}_{1}\left(t\right)= & \tilde{\textbf{r}}_{1}\left(0\right)-\frac{3\mathcal{C}_{1}m_{0}}{r_{1}^{5}}\left(\left(1-3\left(\frac{\textbf{r}_1.\textbf{w}}{r_1}\right)^{2}\right)t\textbf{r}_{1} \right.\\
 & \left.+\frac{2\textbf{r}_1.\textbf{w}}{\Omega_{0}r_1}\left(\textbf{w}\left(t\right)-\textbf{w}\left(0\right)\right)\times\textbf{r}_{1}\right).
\end{split}
\end{equation}
We have then an exact solution for the Hamiltonian $\mathcal{H}_{I,0}$.

The equations of motion for the Hamiltonian $\mathcal{H}_{I,k}$ are similar. However, this Hamiltonian modifies the variables of the body $k$. The equations are then
\begin{empheq}[left={\empheqlbrace}]{align}
\dot{\textbf{r}}_{1} &= \textbf{0}, \nonumber \\
\dot{\tilde{\textbf{r}}}_{1} &= -\frac{3\mathcal{C}_{1}m_{k}}{r_{1,k}^{5}}\left(\left(1-5\left(\frac{\textbf{r}_{1,k}.\textbf{w}}{r_{1,k}}\right)^{2}\right)\textbf{r}_{1,k}+2\left(\textbf{r}_{1,k}.\textbf{w}\right)\textbf{w}\right), \nonumber \\
\dot{\textbf{r}}_{k} &= \textbf{0}, \label{eq:eqdiffhI} \\
\dot{\tilde{\textbf{r}}}_{k} &= \frac{3\mathcal{C}_{1}m_{k}}{r_{1,k}^{5}}\left(\left(1-5\left(\frac{\textbf{r}_{1,k}.\textbf{w}}{r_{1,k}}\right)^{2}\right)\textbf{r}_{1,k}+2\left(\textbf{r}_{1,k}.\textbf{w}\right)\textbf{w}\right), \nonumber \\
\dot{\textbf{w}} &= \frac{6\mathcal{C}_{1}m_{k}}{Gr_{1,k}^{5}}\left(\textbf{r}_{1,k}.\textbf{w}\right)\textbf{r}_{1,k}\times \textbf{w}, \nonumber
\end{empheq}
which have the solution
\begin{empheq}[left={\empheqlbrace}]{align}
\tilde{\textbf{r}}_{1}\left(t\right) &= \tilde{\textbf{r}}_{1}\left(0\right)-\frac{3\mathcal{C}_{1}m_{k}}{r_{1,k}^{5}}\left(\left(1-3\left(\frac{\textbf{r}_{1,k}.\textbf{w}}{r_{1,k}}\right)^{2}\right)t\textbf{r}_{1,k} \right. \nonumber \\
& \left. +\frac{2\textbf{r}_{1,k}.\textbf{w}}{\Omega_{k} r_{1,k}}\left(\textbf{w}\left(t\right)-\textbf{w}\left(0\right)\right)\times\textbf{r}_{1,k}\right), \nonumber \\
\tilde{\textbf{r}}_{k}\left(t\right) &= \tilde{\textbf{r}}_{k}\left(0\right)+\frac{3\mathcal{C}_{1}m_{k}}{r_{1,k}^{5}}\left(\left(1-3\left(\frac{\textbf{r}_{1,k}.\textbf{w}}{r_{1,k}}\right)^{2}\right)t\textbf{r}_{1,k} \right. \\
& \left. +\frac{2\textbf{r}_{1,k}.\textbf{w}}{\Omega_{k} r_{1,k}}\left(\textbf{w}\left(t\right)-\textbf{w}\left(0\right)\right)\times\textbf{r}_{1,k}\right), \nonumber \\
\textbf{w}\left(t\right) &= R_{\textbf{r}_{1,k}}\left(\Omega_{k}t\right)\textbf{w}\left(0\right), \nonumber
\end{empheq}
with the angular frequency $\Omega_{k}=6\mathcal{C}_{1}m_{k}\left(\textbf{r}_{1,k}.\textbf{w}\right)/(Gr_{1,k}^{4})$.

The symplectic scheme for the total Hamiltonian is \citep{farago2009}
\begin{equation}
S\left(t\right)=e^{\frac{t}{2}L_{\mathcal{H}_{I,0}}}e^{\frac{t}{2} L_{\mathcal{H}_{I,2}}}\ldots e^{\frac{t}{2} L_{\mathcal{H}_{I,n}}} e^{t L_{H_{N}}}e^{\frac{t}{2} L_{\mathcal{H}_{I,n}}}\ldots e^{\frac{t}{2} L_{\mathcal{H}_{I,2}}}e^{\frac{t}{2}L_{\mathcal{H}_{I,0}}},
\end{equation}
where $L_{X}$ represents the Lie derivative of a Hamiltonian $X$. This scheme gives a symplectic solution for the long-term evolution of the angular momentum of the rigid body. 

It is possible to neglect the effects of the rotation on the orbital motion by keeping $\dot{\tilde{\textbf{r}}}_{1}=0$ and $\dot{\tilde{\textbf{r}}}_{k}=0$ in Eqs. (\ref{eq:eqdiffh0}) and (\ref{eq:eqdiffhI}). This allows us to obtain multiple solutions for the long-term rotation with different initial conditions for the angular momentum by computing only one orbital evolution. In this case, the total energy is still conserved, but it is not the case for the total angular momentum.

By averaging over the fast rotation of Ceres and Vesta, this method is used in section \ref{SEC:orbobliquite} to obtain the long-term evolution of the angular momenta of Ceres and Vesta where the torques are exerted by the Sun and the planets.

\subsection{Secular equations for the angular momentum\label{sec:oblisec}}

In order to speed up the computation, we average the Hamiltonian (Eq. (\ref{eq:hamI0mean})) over the mean longitude of the rigid body. By considering only the torque exerted by the Sun, we obtain the secular Hamiltonian for the rotation axis \citep{bouelaskar2006}
\begin{equation}
H=-\frac{G\alpha}{2\left(1-e^2\right)^{3/2}} \left(\bf{w}.\bf{n}\right)^2,
\end{equation}
with $G=C\omega$ for the spin rate $\omega$. The motion of the angular momentum is forced by a secular orbital solution, from which the normal to the orbit $\bf{n}$ and the eccentricity $e$ are computed. The precession constant $\alpha$ can be written
\begin{equation}
\alpha=\frac{3}{2}\frac{\mathcal{G}M_{\odot}}{C \omega a^{3}}\left(C-\frac{A+B}{2}\right)\left(1-\frac{3}{2}\sin^{2}J\right)
\end{equation}
with $a$ the semi-major axis and $\Msol$ the mass of the Sun.
The moments of inertia can be normalized by
\begin{equation}
\overline{A}=\frac{A}{mR^2},\ \overline{B}=\frac{B}{mR^2},\ \C=\frac{C}{mR^2},\ \I=\frac{I}{mR^2},\label{eq:Cnorm}
\end{equation}
with $I=(A+B+C)/3$ the mean moment of inertia, $m$ the mass of the solid body, and $R$ the reference radius used for the determination of the gravity field. The gravitational flattening $J_2$ depends on the normalized moments of inertia with
\begin{equation}
J_2=\C-\frac{\overline{A}+\overline{B}}{2}.
\end{equation}
The precession constant can then be written
\begin{equation}
\alpha=\frac{3}{2}\frac{\mathcal{G}M_{\odot}J_2}{\C \omega a^{3}}\left(1-\frac{3}{2}\sin^{2}J\right).\label{eq:alpha_const}
\end{equation}
The secular equation for the angular momentum $\bf{w}$ is then \citep[e.g.,][]{colombo1966,bouelaskar2006}
\begin{equation}
\dot{\bf{w}} =\frac{\alpha}{\left(1-e^2\right)^{3/2}} \left(\bf{w}.\bf{n}\right) \bf{w}\times\bf{n}.\label{eq:integsec}
\end{equation}
The angle between the normal to the orbit, $\mathbf{n}$, and the angular momentum, $\mathbf{w}$, is the obliquity $\epsilon$.
 
Equation (\ref{eq:integsec}) is used in section \ref{SEC:stab} to study the stability of the spin axes of Ceres and Vesta.

\section{Precession constants and initial conditions\label{SEC:prec}}

To determine the quantity $\mathcal{C}_1$ (Eq. (\ref{eq:C_1})) and the precession constant $\alpha$ (Eq. (\ref{eq:alpha_const})), the polar moment of inertia $C$, the spin rate $\omega$, the gravitational flattening $J_2$, and the Andoyer angle $J$ are necessary.

\subsection{Estimation of the Andoyer angle $J$}

The Andoyer angle $J$ is the angle between the angular momentum $\mathbf{G}$ and the polar axis $\mathbf{K}$.

The Dawn space mission determined the principal axes of Ceres and Vesta and measured the gravitational field in these frames. To obtain the precision of the determination of the principal axes, we estimate the angle $\gamma$ between the polar axis and its determination by Dawn with the expression 
\begin{equation}
\gamma\approx\sqrt{C_{21}^2+S_{21}^2}/\C. \label{eq:angleJ}
\end{equation}
The spherical harmonic gravity coefficients of second degree and first-order $C_{21}$ and $S_{21}$ were determined with their uncertainties by Dawn for Ceres \citep{park2016} and Vesta \citep{konopliv2014}.
Because $C_{21}$ and $S_{21}$ are smaller than their uncertainties and than the other coefficients of second degree for both bodies, \cite{park2016} and \cite{konopliv2014} deduce that this angle is negligible. By replacing $C_{21}$ and $S_{21}$ by their uncertainties in Eq. (\ref{eq:angleJ}) and $\C$ by the values of the sections \ref{sec:Cceres} and \ref{sec:Cvesta}, the angle $\gamma$ is about $7\times 10^{-5}\,\degree$ and $1\times 10^{-6}\,\degree$ respectively for Ceres and Vesta.

Using the basis ($\mathbf{I}$,$\mathbf{J}$,$\mathbf{K}$) associated with the principal axes of moments of inertia, the rotational vector $\mathbf{\Omega}$ can be expressed as
\begin{equation}
\mathbf{\Omega}= \omega\begin{pmatrix}
m_1 \\
m_2 \\
1+m_3 
\end{pmatrix},
\end{equation}
where $m_1$ and $m_2$ describe the polar motion and $m_3$ the length of day variations, which were estimated by \cite{rambaux2011} and \cite{rambaux2013} respectively for Ceres and Vesta. The amplitude of the polar motion is about $0.4\,\mas$ for Ceres and $0.8\,\mas$ for Vesta. \cite{rambaux2011} assumed that Ceres is axisymmetric and obtained an amplitude  of about $8\times 10^{-4}\,\mas$ for $m_3$. \cite{rambaux2013} considered a triaxial shape for Vesta and obtained an amplitude for $m_3$ of about $0.1\,\mas$. The angle between the rotational vector $\mathbf{\Omega}$ and the polar axis is about $1\times 10^{-7}\,\degree$ for Ceres and $2\times 10^{-7}\,\degree$ for Vesta and the polar motion is then negligible. 

The rotational vector can also be approximated by $\mathbf{\Omega}=\omega \mathbf{K}$ and $\mathbf{G}$ verifies $\mathbf{G}=C\omega\mathbf{K}$. Therefore, we can neglect $\sin^2J$ in Eq. (\ref{eq:alpha_const}) and the precession constant becomes
\begin{equation}
\alpha=\frac{3}{2}\frac{\mathcal{G}M_{\odot}J_2}{\C \omega a^{3}}\label{eq:prec_const}.
\end{equation}

\subsection{Precession constant of Ceres}

\subsubsection{Physical parameters}

From the Dawn data, \cite{park2016} determined $J_2$
\begin{equation}
J_2=2.6499\times10^{-2}\pm 8.4\times10^{-7} \label{eq:J2ceres}
\end{equation} for the reference radius  
\begin{equation}
R=470\,\km.
\end{equation}
\cite{park2016} also refined the spin rate to 
\begin{equation}
\omega=952.1532\pm0.0001\degday.
\end{equation}

\subsubsection{Polar moment of inertia\label{sec:Cceres}}

\begin{figure}
\centering
\includegraphics[width=9cm]{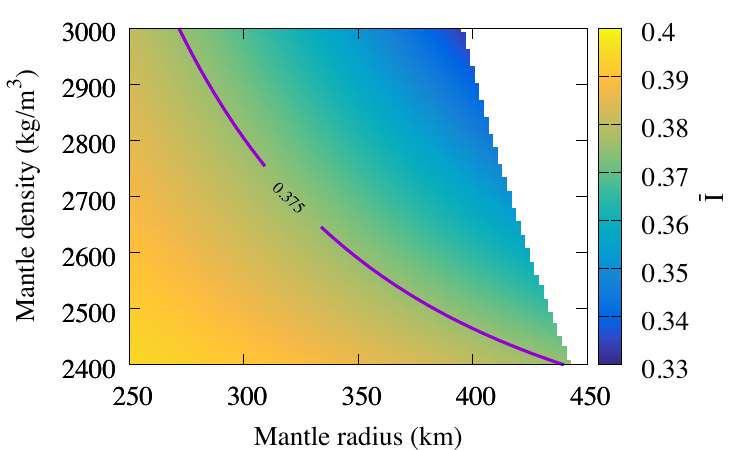}
\caption{\label{fig:interceres}Normalized mean moment of inertia $\I$ assuming a spherical shape with respect to the density and radius of the mantle. The purple line represents the numerical solutions of Clairaut's equations which reproduce the observed gravitational flattening $J_2$ \citep{park2016}.}
\end{figure}
The polar moment of inertia can be estimated from a model of internal structure. \cite{park2016} proposed a set of internal models with two layers by numerically integrating Clairaut's equations of hydrostatic equilibrium. The mantle of density $2460-2900\,\kgm$ has a composition similar to those of different types of chondrites and the outer shell of density $1680-1950\,\kgm$ is a blend of volatiles, silicates, and salts.
\cite{ermakov2017b} used the gravity field and the shape obtained by the Dawn space mission and took into account the effect of the isostasy to constrain the internal structure of Ceres. Their favored model has a crust density of $1287^{-87}_{+70}\,\kgm$, a crust thickness of $41.0^{-4.7}_{+3.2}\,\km$, a mantle density of $2434^{-8}_{+5}\,\kgm$, and a mantle radius of $428.7^{+4.7}_{-3.2}\,\km$.

In figure \ref{fig:interceres} the purple curve represents the numerical solutions of Clairaut's equations which reproduce the observed gravitational flattening $J_2$ \citep{park2016}. In figure \ref{fig:interceres} the normalized mean moment of inertia $\I$ is computed by assuming a spherical shape, as in Eq. (1) of \cite{rambaux2011}. For a mantle density of $2460-2900\,\kgm$, we have $\I=0.375$. The normalized polar moment of inertia $\C$ can be deduced by
\begin{equation}
\C=\frac{2J_ 2}{3}+\I.\label{eq:CIJ}
\end{equation}
With Eq. (\ref{eq:J2ceres}) and $\I=0.375$, we found $\C=0.393$\footnote{If we take into account the nonspherical shape of Ceres to compute the normalized mean moment of inertia, the normalized polar moment of inertia becomes $\C=0.395$.}.

The gravitational flattening possesses a nonhydrostatic component $J_{2}^{nh}$, which causes an uncertainty on $\C$. \cite{park2016} estimated $J_{2}^{nh}$ with
\begin{equation}
\frac{J_{2}^{nh}}{J_2}=\frac{\sqrt{\C_{22}^{2}+\overline{S}_{22}^{2}}}{\overline{J}_2}
\end{equation}
for the normalized spherical harmonic gravity coefficients of second degree and second-order $\C_{22}$ and $\overline{S}_{22}$ and $\overline{J}_2$ the normalized value of $J_2$. By deriving the Radau-Darwin relation \citep[e.g.,][]{rambaux2015,ermakov2017b}, we obtain the uncertainty $\Delta \I$ on $\I$
\begin{equation}
\Delta \I=\frac{2k}{3\sqrt{\left(4-k\right)\left(1+k\right)^3}}\frac{J_{2}^{nh}}{J_2}. \label{eq:DI}
\end{equation}
The fluid Love number $k$ verifies $k=3J_2/q$ \citep{ermakov2017b} with $q=\omega^2 R_{vol}^3/(\mathcal{G}m)$, $R_{vol}$ the volume-equivalent radius, and $m$ the mass.
For $R_{vol}=469.7\,\km$ \citep{ermakov2017b}, Eq. (\ref{eq:DI}) gives the uncertainty $\Delta \I=0.0047$.
With Eq. (\ref{eq:CIJ}), we have $\Delta \C=\Delta \I+2J^{nh}_2/3=0.0053$. We keep $\Delta \C=0.005$ as in \cite{ermakov2017a}.

For the integration of the obliquity, we choose $\C=0.393$ and $0.005$ for its uncertainty. The interval of uncertainty on $\C$ is then $[0.388:0.398]$.
\cite{ermakov2017a} obtained the value $\C=0.392$ for a radius of $R=469.7\,\km$, which corresponds to $\C\approx0.3915$ for $R=470\,\km$. The value $\C=0.393$ is then consistent with the value of \cite{ermakov2017a} given the uncertainties.

\subsubsection{Precession constant\label{sec:preconstceres}}

We take a constant semi-major axis to compute the precession constant. We use the average of the semi-major axis of the solution La2011 on $\left[-25:5\right]\MYR$, which is about $a\approx2.767\,\AU$. On this interval, the semi-major axis can move away to $\Delta a =0.005\,\AU$ from this value. From Eq. (\ref{eq:prec_const}) and previous values and uncertainties, we deduce the precession constant
\begin{equation}
\alpha=6.40\pm0.12\arc.
\end{equation}

\subsubsection{Early Ceres\label{sec:earlyceresalpha}}

\cite{mao2018} estimated that Ceres should spin about $7\pm4\%$ faster to be in hydrostatic equilibrium with the observed present shape. They supposed that Ceres had been in hydrostatic equilibrium in the past and had then slowed down due to some phenomenon like significant asteroid impacts.
They obtained for this higher spin rate an internal structure of normalized mean moment of inertia $\overline{I}=0.353\pm 0.009$ for a reference radius $R=470\,\km$. With Eq. (\ref{eq:CIJ}), it corresponds to a normalized polar moment of inertia $\C=0.371\pm0.009$.
The corresponding precession constant is
\begin{equation}
\alpha=6.34\pm0.43\arc
\end{equation}
with the value of the semi-major axis of the section \ref{sec:preconstceres}.
With the present spin rate and considering a normalized polar moment of inertia of $\C=0.371\pm0.009$, the precession constant of the present Ceres would be
\begin{equation}
\alpha=6.78\pm0.20\arc.
\end{equation}

\subsection{Precession constant of Vesta}

\subsubsection{Physical parameters}

From the Dawn data, \cite{konopliv2014} gave the normalized value for $J_2$
\begin{equation}
\overline{J}_2=3.1779397\times10^{-2}\pm 1.9\times10^{-8}
\end{equation}
for the reference radius 
\begin{equation}
R=265\,\km.
\end{equation}
It corresponds to about $J_2=\sqrt{5}\times\overline{J}_2=7.1060892\times10^{-2}$. The rotation rate has been refined by \cite{konopliv2014} with 
\begin{equation}
\omega=1617.3331235\pm0.0000005\,\degday.
\end{equation}

\subsubsection{Polar moment of inertia\label{sec:Cvesta}}

\begin{table}
\centering
\begin{tabular}{lcl}
\hline
		& $\rho$ $(\kgm)$	& semi-principal axes $\left(\km\right)$ \\
\hline
Crust (a)   & 2900	&	$a=b=280.9$ $c=226.2$ \\
Mantle   	& 3200	&	$a=b=253.3$ $c=198.8$ \\
Core 	    & 7800	& 	$a=b=114.1$ $c=102.3$ \\
\hline
Crust (b)	& 2970	&	$a=284.50$ $b=277.25$ $c=226.43$ \\
Mantle 		& 3160	&	$a=b=257$ $c=207$ \\
Core 		& 7400	& 	$a=b=117$ $c=105$ \\
\hline
Crust (c) 	& 2970	&	$a=284.50$ $b=277.25$ $c=226.43$ \\
Mantle 		& 3970	&	$a=b=213$ $c=192$ \\
\hline
\end{tabular}
\caption{\label{tab:paramstrucvesta}Densities and semi-principal axes for different models of internal structure of Vesta. (a) corresponds to the reference ellipsoids of table 3 in \cite{ermakov2014}, (b) and (c) respectively to the three-layer and two-layer models of \cite{park2014}. For (b) and (c), the dimensions of the crust are given by the best-fit ellipoid of \cite{konopliv2014}.}
\end{table}

The polar moment of inertia $C$ of Vesta could not be obtained from the observation of the precession and nutation of its pole \citep{konopliv2014}. Following \cite{rambaux2013}, we determine $C$ from an internal model composed of ellipsoidal layers of semi-axes $a_i$, $b_i$,  and $c_i$, and uniform densities $\rho_i$, where $a_i$, $b_i$, and $c_i$ are respectively the major, intermediate, and minor semi-axes. For a three-layer model constituted of a crust (1), a mantle (2), and a core (3), $C$ is
\begin{eqnarray}
C &=& \frac{4\pi}{15}\left(a_1b_1c_1\left(a_1^2+b_1^2\right)\rho_1+ a_2b_2c_2\left(a_2^2+b_2^2\right)\left(\rho_2-\rho_1\right)\right. \nonumber \\ 
& & \left.  + a_3b_3c_3\left(a_3^2+b_3^2\right)\left(\rho_3-\rho_2\right)\right). \label{eq:polarC}
\end{eqnarray}

\cite{ermakov2014} and \cite{park2014} proposed internal models from the gravity field and the shape model of \cite{gaskell2012}. \cite{ermakov2014} determined the interface between the crust and the mantle. The densities and the reference ellipsoids used by \cite{ermakov2014} to compare their model are in table \ref{tab:paramstrucvesta}. For these parameters, Eq. (\ref{eq:polarC}) gives $\C=0.4061$. If we use instead of the biaxial crust in table \ref{tab:paramstrucvesta}, the triaxial best-fit ellipsoid of \cite{ermakov2014} determined from the shape model of \cite{gaskell2012} with $a=284.895\,\km$, $b=277.431\,\km$, and $c=226.838\,\km$, we obtain $\C=0.4086$.

\cite{park2014} proposed three-layer and two-layer models (table \ref{tab:paramstrucvesta}). For the form of the crust, we use the best-fit ellipsoid of \cite{konopliv2014} instead of the shape of \cite{gaskell2012}. Equation (\ref{eq:polarC}) gives the approximate values $\C=0.4089$ for the three-layer model and $\C=0.4218$ for the two-layer model.

We keep $\C=0.409$ obtained for the three-layer model of \cite{park2014} with the uncertainty $0.013$, given from the uncertainty interval $[0.406:0.422]$.

\subsubsection{Precession constant\label{sec:preconstvesta}}

As we did for Ceres, we consider a mean value for the semi-major axis. With $a\approx2.361\,\AU$ and $\Delta a=0.002\,\AU$, Eq. (\ref{eq:prec_const}) gives
\begin{equation}
\alpha=15.6\pm0.6\arc.
\end{equation}

\subsubsection{Early Vesta\label{sec:earlyvestaalpha}}

The southern hemisphere of Vesta has a large depression with two basins, Veneneia and Rheasilvia, created by two giant impacts \citep{marchi2012,schenk2012}. \cite{fu2014} fitted the regions of the northern hemisphere not affected by the giant impacts with an ellipsoid of principal axes of dimensions $a=280.6\,\km$, $b=274.6\,\km$ and $c=236.8\,\km$. By extrapolating this shape to the two hemispheres of the early Vesta supposed hydrostatic, \cite{fu2014} obtained a paleorotation period of $5.02\,\h$ and \cite{ermakov2014} a paleorotation period between $4.83\,\h$ and $4.93\,\h$ for respectively the most and least differentiated internal structures. 

By replacing the shape of the previous models by the supposed shape of the early Vesta determined by \cite{fu2014}, Eq. (\ref{eq:polarC}) gives $\C=0.4055$ for the three-layer model of \cite{ermakov2014}, and $\C=0.4081$ and $\C=0.4210$ respectively for the three-layer and two-layer models of \cite{park2014}. We choose $\C=0.408$ with an uncertainty of $0.013$. The corresponding gravitational flattening is $J_2=0.0559\pm0.0003$, where the uncertainty is given from the gravitational flattenings of the three different models of internal structure. We choose the paleorotation period of $5.02\,\h$ of \cite{fu2014}. We use the value of the semi-major axis of section \ref{sec:preconstvesta}, and Eq. (\ref{eq:prec_const}) gives the precession constant for the early Vesta
\begin{equation}
\alpha=11.6\pm0.9\arc.
\end{equation}

\subsection{Initial conditions}

The Dawn space mission refined the orientation of the rotation axes of Ceres and Vesta. We use the coordinates given under the form right ascension/declination in the ICRF frame for the epoch J2000 by \cite{park2016} for Ceres and by \cite{konopliv2014} for Vesta listed in table \ref{tab:CI}. From these coordinates and their uncertainties, we obtain the obliquities $\epsilon_C$ and $\epsilon_V$ of respectively Ceres and Vesta at the epoch J2000
\begin{equation}
\epsilon_C=3.997\pm0.003\degree,
\end{equation}
\begin{equation}
\epsilon_V=27.46784\pm0.00003\degree.
\end{equation}
\begin{table}
\centering
\begin{tabular}{ccc}
\hline
& Ceres 	& Vesta \\
\hline
R.A. $(\degree)$ & $291.421\pm0.007$ & $309.03300\pm0.00003$ \\   
Dec $(\degree)$ & $66.758\pm0.002$ & $42.22615\pm0.00002$ \\				
\hline
\end{tabular}
\caption{\label{tab:CI}Right ascension (R.A.) and declination (Dec) of Ceres \citep{park2016} and Vesta  \citep{konopliv2014} at the epoch J2000 in the ICRF frame.}
\end{table}

\section{Orbital and rotational solutions obtained with the symplectic integration \label{SEC:orbobliquite}}

This section is dedicated to the long-term solutions \La{} for the orbital motion and \Ce{} for the rotational motion, which is obtained with the symplectic integration of the angular momentum described in section \ref{sec:oblisymp}.
The time origin of the solutions is the epoch J2000.

We analyze the solutions with the method of the frequency map analysis \citep{laskar1988,laskar1990,laskar1992,laskar1993,laskar2003}, which decomposes a discrete temporal function in a quasi-periodic approximation. The precision of the obtained frequencies is estimated by performing a frequency analysis of the solution rebuilt from the frequency decomposition with a temporal offset \citep{laskar1990}. The differences between the frequencies of the two decompositions give an estimate of the accuracy on the determination of the frequencies.

\begin{table*}
\centering
\begin{tabular}{|c|c|c|c|c|c|c|c|}
\hline
 & $k_2$ & $Q$ & $\omega$ $(\degday)$ & $R$ $(\km)$ & $\C$ & $r$ $(\AU)$ & $\Gamma/(C\omega)$ $(\yr^{-1})$\\
 \hline
Mars 	& $0.149$ & $92$ & $350.89198521$ & $3396$ & $0.3654$ & $1.5237$ & $\sim3\times10^{-13}$ \\
Ceres  	& $10^{-3}$ & $10$ & $952.1532$ & $470$ & $ 0.393$ & $2.7665$ & $\sim4\times10^{-16}$ \\
Vesta  	& $10^{-3}$ & $100$ & $1617.3331235$ & $265$ & $ 0.409$ & $2.3615$ & $\sim3\times10^{-17}$ \\
\hline
\end{tabular}
\caption{\label{tab:tides}Estimation of the solar tidal torque $\Gamma/(C\omega)$ on Mars \citep{laskar2004a,konopliv2006}, Ceres \citep{rambaux2011,park2016}, and Vesta \citep{bills2011,konopliv2014}.}
\end{table*}

\subsection{Perturbations on the rotation axis}

We investigate and estimate some effects that can affect the long-term rotation in addition to the torques exerted by the Sun and the planets.

\subsubsection{Tidal dissipation}

The torque exerted on a celestial body for the solar tides is \citep{mignard1979}
\begin{equation}
\mathbf{\Gamma} = 3\frac{k_{2}\mathcal{G}\Msol^2R^5}{Cr^8}\Delta t\left[
\left(\mathbf{r}.\mathbf{G}\right)\mathbf{r}-r^2\mathbf{G}+C\mathbf{r}\times\mathbf{v}\right]
\end{equation}
with $\Delta t$ the time delay between the stress exerted by the Sun and the response of the body, $k_2$ the Love number, $R$ the radius of the body, $\mathbf{r}$ and $\mathbf{v}$ the heliocentric position and velocity of the body, and $r$ the norm of $\mathbf{r}$. For a circular and equatorial orbit, \cite{mignard1979} writes
\begin{equation}
\Gamma = 3\frac{k_{2}\mathcal{G}\Msol^2R^5}{2r^6}|\sin\left(2\delta\right)|
\end{equation}
with $\delta=\left(\omega-n\right)\Delta t$ the phase lag and $n$ the mean motion. The phase lag $\delta$ is related to the effective specific tidal dissipation function $Q$ by $1/Q=\tan(2\delta)$ \citep{macdonald1964}.

Because of the dependence in $r^{-6}$, the torque decreases strongly with the distance to the Sun. \cite{laskar2004a} concluded that the tidal dissipation in the long-term rotation of Mars has an effect on the obliquity inferior to $0.002\degree$ in $10\MYR$. We estimate this torque for Ceres and Vesta and compare these values with that for Mars in table \ref{tab:tides}. The values of $k_2$ and $Q$ used for the estimation of the torque are those used by \cite{rambaux2011} for Ceres and by \cite{bills2011} for Vesta. The ratio of the torque on the rotation angular momentum is respectively about $1000$ and $10000$ times weaker for Ceres and Vesta than for Mars, for which the effect can already be considered weak \citep{laskar2004a}. Therefore, the solar tidal dissipation for Ceres and Vesta was not considered.

\subsubsection{Close encounters\label{sec:closeencounters}}

\begin{table*}
\centering
\begin{tabular}{ccccccc}
\hline
perturbed body & perturbing body & R ($\km$) & $N_c$ ($10^{-3}\times\Gyr^{-1}$) & $A$ ($10^8\times\AU^{-2}.\Gyr^{-1}$) & $B$ ($10^{-10}\times\AU^{3/2}$) & $V$ ($\Gyr^{-1}$) \\ 
\hline
(1) 				& (4) 	& $256$ 	& $2.0$ 	& $1.7$		& $6.3$		& $1.3\times10^{-5}$\\
$R_1=476\,\km$ 	& (2) 	& $252$ 	& $0.9$ 	& $0.76$		& $5.3$		& $4.4\times10^{-6}$\\
 				& (7) 	& $112$ 	& $1.3$ 	& $1.7$		& $0.41$		& $7.2\times10^{-8}$\\
 				& (324) & $102$	& $1.0$ 	& $1.3$		& $0.25$		& $2.1\times10^{-8}$\\
\hline
(4) 				& (1) 	& $476$	& $2.0$ 	& $1.7$		& $34$ 		& $3.9\times10^{-4}$\\
$R_4=256\,\km$ 	& (2) 	& $252$	& $1.0$ 	& $1.7$		& $13$ 		& $8.0\times10^{-5}$\\
 				& (7) 	& $112$	& $1.4$ 	& $4.6$		& $1.2$ 		& $2.5\times10^{-6}$\\
 				& (324) & $102$	& $0.5$ 	& $1.7$		& $0.71$		& $3.6\times10^{-7}$\\
\hline
\end{tabular}
\caption{\label{tab:closeencounters}Radii $R$ used by \cite{laskar2011b} to compute the collision probabilities $N_c$ between Ceres and Vesta and some bodies. The values of $R$ for Ceres, Vesta, and the other bodies and the values of $N_c$ are extracted from table 3 in \cite{laskar2011b}. The coefficients $A$ computed from these values, the coefficient $B$, and the variance $V$ are also indicated.}
\end{table*}

In the long-term solution La2010 \citep{laskar2011a}, the five bodies of the asteroid belt (1) Ceres, (2) Pallas, (4) Vesta, (7) Iris, and (324) Bamberga are considered planets and there are mutual gravitational interactions between them. \cite{laskar2011a} considered these bodies because Ceres, Vesta, and Pallas are the three main bodies of the main asteroid belt and because Iris and Bamberga significantly influence the orbital motion of Mars.
\cite{laskar2011b} studied the close encounters between these bodies and showed that they are responsible for their chaotic behavior. If a body comes close to Ceres or Vesta, it can exert a significant torque during the encounter. The effects of close encounters on the rotation axes of the giant planets have been studied by \cite{lee2007}. For a body of mass $m$ with no satellites, the maximum difference $\|\Delta \mathbf{w}\|$ between the angular momentum before an encounter with a perturbing body of mass $m_{pert}$ and the angular momentum after
is \citep{lee2007}
\begin{equation}
\|\Delta \mathbf{w}\|=\frac{\pi}{2}\alpha\frac{m_{pert}}{\Msol}\frac{ a^3}{r_p^2v_p}
\end{equation}
with $\alpha$ the precession constant (Eq. (\ref{eq:prec_const})), $a$ the semi-major axis, and $r_p$ and $v_p=\sqrt{2\mathcal{G}(m+m_{pert})/r_p}$ respectively the distance and the relative speed between the two bodies at the closest distance. We can write this formula as
\begin{equation}
\|\Delta \mathbf{w}\|=B r_p^{-3/2}
\end{equation}
with 
\begin{equation}
B=\frac{3\pi}{4}m_{pert}\sqrt{\frac{\mathcal{G}}{2\left(m+m_{pert}\right)}}\frac{J_2}{\C\omega}.
\end{equation}
The values of the coefficient $B$ have been computed in table \ref{tab:closeencounters} for close encounters considered in \cite{laskar2011b}. Therefore, a close encounter changes the orientation of the rotation axis at the most of the angle
\begin{equation}
\theta=\arccos \left(1-\frac{B^2}{2r_p^3}\right).
\end{equation}

\cite{laskar2011b} studied the probability of close encounters between the five bodies of the asteroid belt considered in the solution La2010 \citep{laskar2011a} and determined that the probability density $\rho(r_p)$ per unit of time of an encounter with a distance $r_p$ at the closest approach can be fitted by a linear function of $r_p$ for $r_p\leq1\times10^{-3}\,\AU$
\begin{equation}
\rho\left(r_p\right)=Ar_p,
\end{equation}
with
\begin{equation}
A=\frac{2N_c}{\left(R_1+R_2\right)^{2}}.
\end{equation}
Here $N_c$ is the collision probability per unit of time between two bodies of radii $R_1$ and $R_2$. Table \ref{tab:closeencounters} gives the radii, the collision probability $N_c$ extracted from table 3 in \cite{laskar2011b}, and the deduced coefficient $A$ between each considered pair on $1\,\Gyr$ for the five bodies considered in \cite{laskar2011b}.

We suppose that each close encounter moves the angular momentum in a random direction. The motion of the rotation axis is described by a random walk on a sphere of distribution \citep{perrin1928,roberts1960}
\begin{equation}
\rho_S\left(\theta\right)=\sum_{k=0}^{\infty}\frac{2k+1}{4\pi}e^{-\frac{k\left(k+1\right)}{4}V}P_k\left(\cos\theta\right)
\end{equation}
with the variance $V$, $\int_0^{2\pi}\int_0^\pi\rho_S\left(\theta\right)\sin\theta d\theta d\phi=1$, and $P_k$ the Legendre polynomial of order $k$. For a random walk of $N$ steps with a large value of $N$, where each step causes a small change $\beta$ in the orientation, the variance $V$ is \citep{roberts1960}
\begin{equation}
V=\sum_{k=1}^N\int_0^\pi\beta^2dp_k\left(\beta\right)
\end{equation}
with $dp_k\left(\beta\right)$ the probability to have a change of angle $\beta$ for the step $k$. For Ceres, we consider the close encounters with the bodies considered in \cite{laskar2011b} for which we have the probability of close encounters and we can write
\begin{equation}
V_1=\sum_{k\in \{2,4,7,324\}}N_k\int_{\beta_{kmin}}^{\beta_{kmax}}\beta^2dp_k\left(\beta\right)
\end{equation}
with $k$ the number of the body for which we consider the close encounter with Ceres, $\beta_{kmin}$ the minimum change of orientation at the distance $1\times10^{-3}\,\AU$, $\beta_{kmax}$ the maximum change of orientation for a grazing encounter at the distance $R_1+R_k$, $N_k$ the number of close encounters, and $dp_k\left(\beta\right)$ the probability distribution to have the change $\beta$ for a close encounter with the body $k$. As $|dp_k\left(\beta\right)|=|A_{1k}r_pdr_p|/N_k$, the variance $V_1$ verifies
\begin{equation}
V_1=\sum_{k\in \{2,4,7,324\}}A_{1k}\int_{R_1+R_k}^{10^{-3}\,\AU}\arccos^2 \left(1-\frac{B_{1k}^2}{2r_p^3}\right)r_pdr_p.
\end{equation}
We compute the standard deviation of the distribution of the rotation axis of Ceres on $1\,\Gyr$ under the effects of the close encounters with the bodies (2) Pallas, (4) Vesta, (7) Iris, and (324) Bamberga with the formula
\begin{equation}
\theta_{1sd}=\sqrt{\sum_{k=0}^{\infty}\frac{2k+1}{4\pi}e^{-\frac{k\left(k+1\right)}{4}V_1}\int_0^{2\pi}\int_0^\pi \theta^2P_k\left(\cos\theta\right)\sin\theta d\theta d\phi}.
\end{equation}
We obtain with the intermediary quantities in table \ref{tab:closeencounters} about
\begin{equation}
\theta_{1sd}=0.24\degree.
\end{equation}
We realize a similar computation for Vesta in table \ref{tab:closeencounters} to obtain
\begin{equation}
\theta_{4sd}=1.3\degree.
\end{equation}
The time interval $\left[-100:0\right]\MYR$ is smaller than $1\,\Gyr$ and the effects of close encounters are then weaker on this interval. Moreover, these standard deviations are computed for close encounters, which cause a maximum effect on the rotation axis.

Although their effects are weak, we consider for the long-term integration of the rotation the torques exerted on the angular momenta of Ceres and Vesta by the five bodies of the main asteroid belt considered in \cite{laskar2011b}.

\begin{table}
\centering
\begin{tabular}{cccc}
\hline
$g_i$ & ($\arc$) & $s_i$ & ($\arc$) \\
\hline
$g_1$ & $5.59$ & $s_1$ & $-5.61$ \\
$g_2$ & $7.453$ & $s_2$ & $-7.06$ \\
$g_3$ & $17.368$ & $s_3$ & $-18.848$ \\
$g_4$ & $17.916$ & $s_4$ & $-17.751$ \\
$g_5$ & $4.257492$ & & \\
$g_6$ & $28.2452$ & $s_6$ & $-26.347856$ \\
$g_7$ & $3.087927$ & $s_7$ & $-2.9925254$ \\
$g_8$ & $0.673022$ & $s_8$ & $-0.691742$ \\
$g_9$ & $-0.35019$ & $s_9$ & $-0.35012$ \\
\hline
\end{tabular}
\caption{\label{tab:freqref} Principal secular frequencies of the solution La2011 $g_i$, $s_i$ determined on $\left[-20:0\right]\MYR$ for the four inner planets and on $\left[-50:0\right]\MYR$ for the four giant planets and Pluto.}
\end{table}

\begin{figure}
\centering
\input{figures/tex_figure2.tex}
\caption{\label{fig:paleo11ceres}Eccentricity (a) and inclination (b) of Ceres for the solution La2011.}
\end{figure}
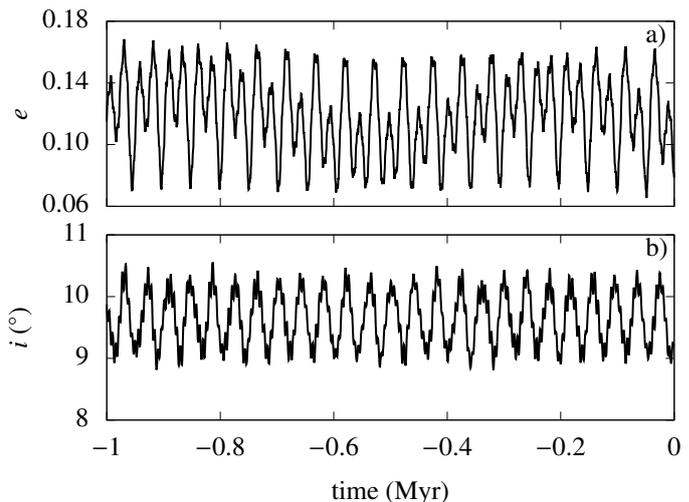

\begin{figure}
\centering
\input{figures/tex_figure3.tex}
\caption{\label{fig:paleo11vesta}Eccentricity (a) and inclination (b) of Vesta for the solution La2011.}
\end{figure}
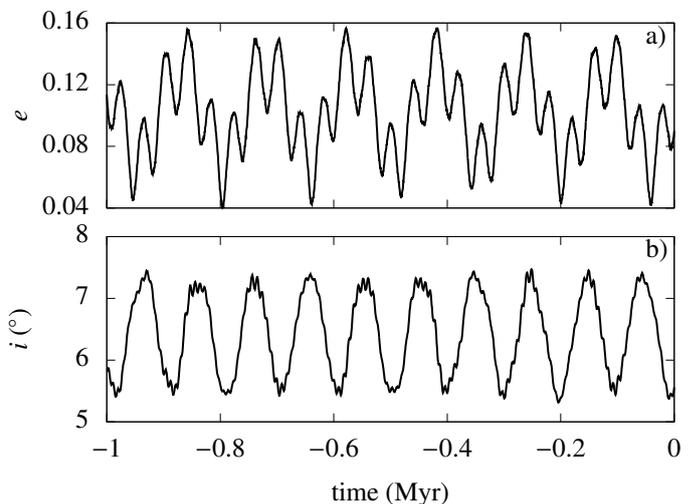

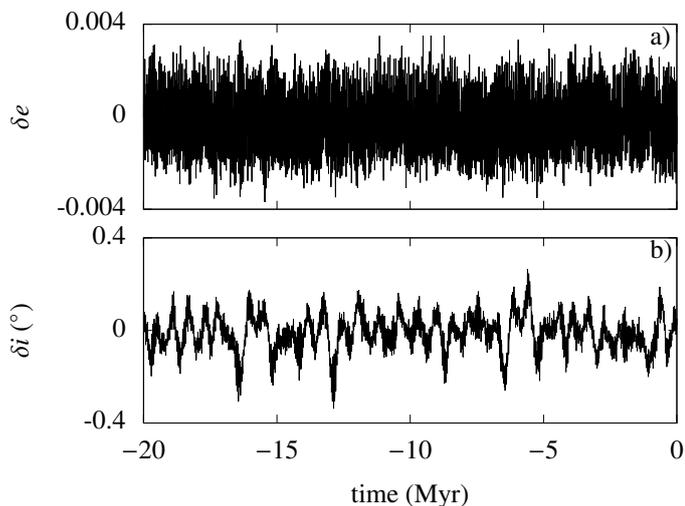
\begin{figure}
\centering
\input{figures/tex_figure4.tex}
\caption{\label{fig:compmodsecceres}Difference for Ceres in eccentricity (a) and inclination (b) between the solution La2011 and the secular solution.}
\end{figure}

\begin{figure}
\centering
\input{figures/tex_figure5.tex}
\caption{\label{fig:compmodsecvesta}Difference for Vesta in eccentricity (a) and inclination (b) between the solution La2011 and the secular solution.}
\end{figure}
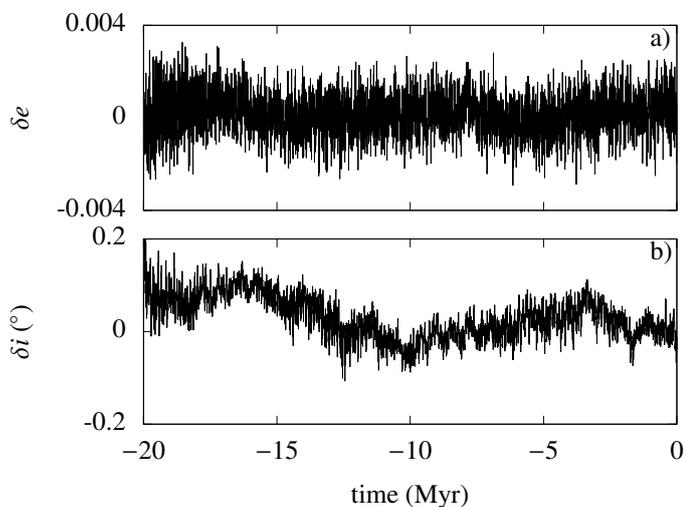

\subsection{Orbital motion La2011 \label{sec:resultorb}}

The orbital solution La2011 is computed on $\left[-250:250\right]\MYR$ in a frame associated with the invariable plane \citep{laskar2011a}. Two successive rotations allow us to pass from this frame to the ICRF as explained in Appendix \ref{sec:planinv}. The variables $z=e \exp(i\varpi)$ and $\zeta=\sin\left(i/2\right)\exp(i\Omega)$ are computed from the noncanonical elliptical elements $(a,\lambda,e,\varpi,i,\Omega)$, where $a$ is the semi-major axis, $\lambda$ the mean longitude, $e$ the eccentricity ,$\varpi$ the longitude of the perihelion, $i$ the inclination with respect to the invariable plane, and $\Omega$ the longitude of the ascending node. These elements are computed from the heliocentric positions and velocities.

As made for the solution La2004 of \cite{laskar2004b}, we perform a frequency analysis of the quantities $z_i$ and $\zeta_i$ on $\left[-20:0\right]\MYR$ for the four inner planets and on $\left[-50:0\right]\MYR$ for the four giant planets and Pluto to obtain the proper perihelion precession frequencies $g_i$ and ascending node precession frequencies $s_i$ in table \ref{tab:freqref}.

The evolutions of the eccentricity and the inclination are represented for Ceres and Vesta on $\left[-1:0\right]\MYR$ respectively in figures \ref{fig:paleo11ceres} and \ref{fig:paleo11vesta}. For Ceres, the eccentricity oscillates between $0.0629$ and $0.169$ and the inclination between $8.77$ and $10.6\degree$ on $\left[-20:0\right]\MYR$. For Vesta, the eccentricity varies between $0.0392$ and $0.160$ and the inclination between $5.21$ and $7.56\degree$ on $\left[-20:0\right]\MYR$. The amplitudes of the variations have the same order of magnitude for Ceres and Vesta on $\left[-250:250\right]\MYR$.

For Ceres and Vesta, we perform a frequency analysis of $z$ and $\zeta$ on the time interval $\left[-25:5\right]\MYR$.
We consider the 50 secular terms with the highest amplitudes which have a frequency in the interval $\left[-300:300\right]\arc$ in tables \ref{tab:freqorbiceres} and \ref{tab:freqorbivesta} (Appendix \ref{sec:freqdecomp}).
The frequency decompositions of tables \ref{tab:freqorbiceres} and \ref{tab:freqorbivesta} allow us to obtain a secular solution, which reproduces the secular evolution of the solution \La{} on $\left[-20:0\right]\MYR$ in figures \ref{fig:compmodsecceres} and \ref{fig:compmodsecvesta}, where the differences with the solution \La{} correspond to the short-period terms excluded from the secular solution. It is not the case for those obtained with a frequency analysis on the time interval $\left[-20:0\right]\MYR$.

These frequency decompositions are used in section \ref{SEC:stab} to compute the orbital quantities in Eq. (\ref{eq:integsec}) needed for the secular integration of the rotation. The terms of weak amplitude can play a role in the long-term rotation in the case of secular resonances. For instance, the passage through the resonance with the frequency $s_6+g_5-g_6$ is responsible for a decrease in the obliquity of about $0.4\degree$ for the Earth \citep{laskarjoutelboudin1993,laskar2004b}. Therefore, we add to the frequency decompositions of the variable $\zeta$ for Ceres the 100 following terms in the interval $\left[-45:60\right]\arc$ and for Vesta the 100 following terms in the interval $\left[-34:60\right]\arc$. These boundaries have been chosen such that they select the frequencies, which can play a role in the long-term rotation without all the terms close to the principal frequencies $s$.

For Ceres, the proper secular frequencies are $g_C=54.2525\pm0.0006\arc$ and $s_C=-59.254\pm0.002\arc$ with the respectively associated periods $23.888\KYR$ and $21.872\KYR$. The first 50 secular terms of the frequency decompositions do not include proper frequencies of the inner planets. Their perturbations on the orbital motions are then much weaker than those of the giant planets. We observe the proximity of the frequencies $2g_6-g_5\approx52.23\arc$ and $2g_6-g_7\approx53.40\arc$ with $g_C$. Resonances with these two frequencies could affect the orbital motion of Ceres.

For Vesta, the proper secular frequencies are $g_V=36.895\pm0.003\arc$ and $s_V=-39.609\pm0.003\arc$ with respectively associated periods $35.13\KYR$ and $32.72\KYR$. The proper frequencies of the inner planets are not present except perhaps for the frequency $-17.74\arc$, which could correspond to the node frequency of Mars $s_4$. Vesta has a shorter semi-major axis, and the planetary perturbations of Mars are then more important than for Ceres, which could explain the presence of this frequency with a higher amplitude.

\begin{figure}
\centering
\input{figures/tex_figure6.tex}
\caption{\label{fig:gserrceres}Secular frequencies $g_C$ (a) and $s_C$ (b) of Ceres on $\left[-250:250\right]\MYR$ computed with a step of $5\MYR$ with the frequency map analysis on an interval of $30\MYR$. The error bars are given by the precision of the frequency map analysis.}
\end{figure}
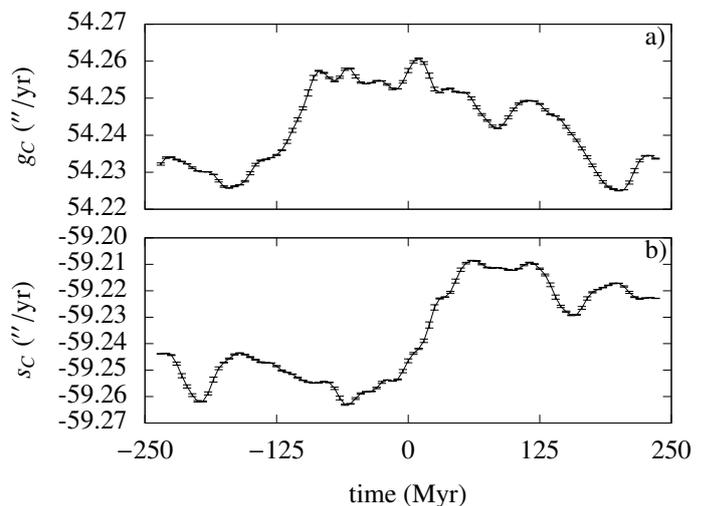

\begin{figure}
\centering
\input{figures/tex_figure7.tex}
\caption{\label{fig:gserrvesta}Secular frequencies $g_V$ (a) and $s_V$ (b) of Vesta on $\left[-250:250\right]\MYR$, computed with a step of $5\MYR$ with the frequency map analysis on an interval of $30\MYR$. The error bars are given by the precision of the frequency map analysis.}
\end{figure}
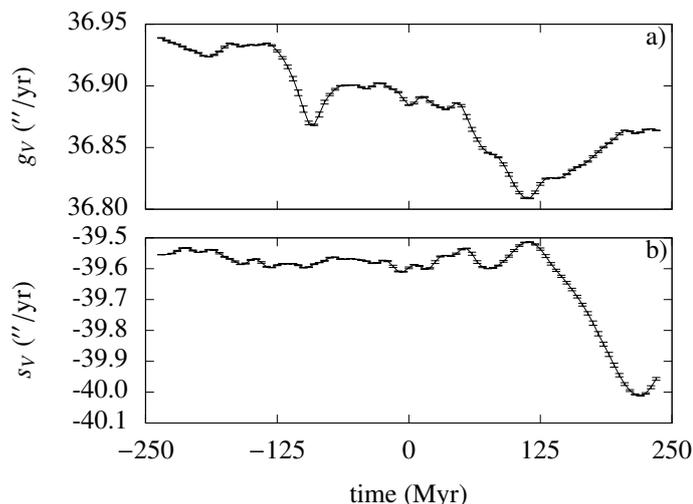

As in \cite{laskar1990}, we estimate the size of the chaotic zones and perform a frequency analysis of the solution La2011 on sliding intervals of $30\MYR$ over $\left[-250:250\right]\MYR$ with a $5\MYR$ step size. The evolutions of the proper frequencies of Ceres and Vesta are respectively in figures \ref{fig:gserrceres} and \ref{fig:gserrvesta}. The values of frequencies $g_C$ and $s_C$ vary respectively in about $\left[54.225:54.261\right]\arc$ and $\left[-59.263:-59.209\right]\arc$, and $g_V$ and $s_V$ respectively in about $\left[36.809:36.939\right]\arc$ and $\left[-40.011:-39.514\right]\arc$. The secular frequencies vary because of the chaotic diffusion, which is then higher for Vesta than for Ceres. The frequency $s_V$ has the highest diffusion with a decrease of about $0.50\arc$ on $\left[115:220\right]\MYR$.

\subsection{Rotational motion Ceres2017\label{sec:resultoblisymp}}

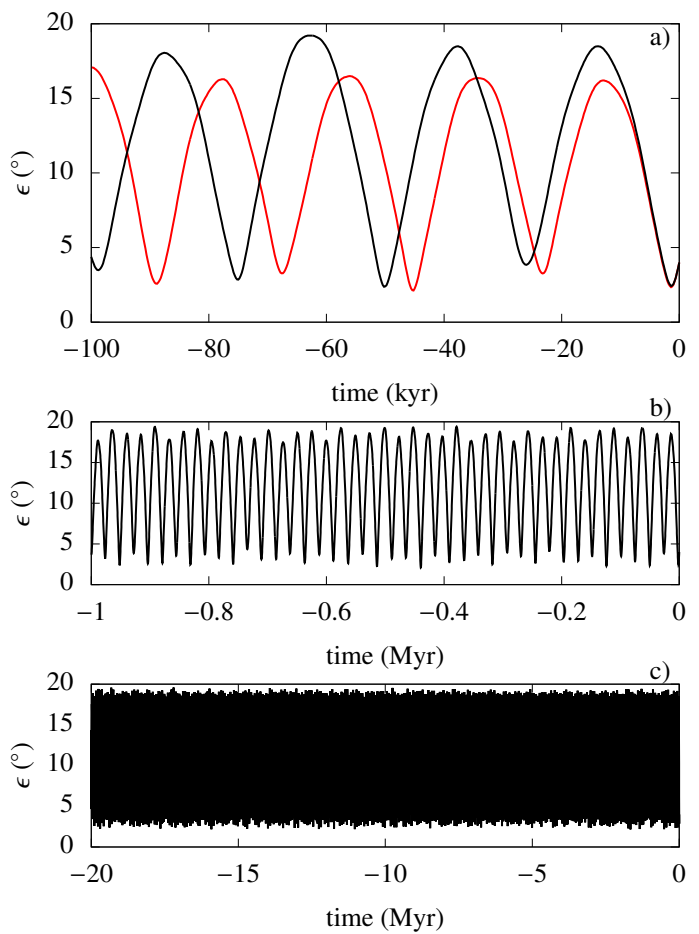
\begin{figure}
\centering
\input{figures/tex_figure8.tex}
\caption{\label{fig:obliceres}Obliquity of Ceres on $\left[-100:0\right]\KYR$ (a), $\left[-1:0\right]\MYR$ (b) and $\left[-20:0\right]\MYR$ (c). In (a) the obliquity caused only by the change in orientation of the orbit is represented by the red curve.}
\end{figure}
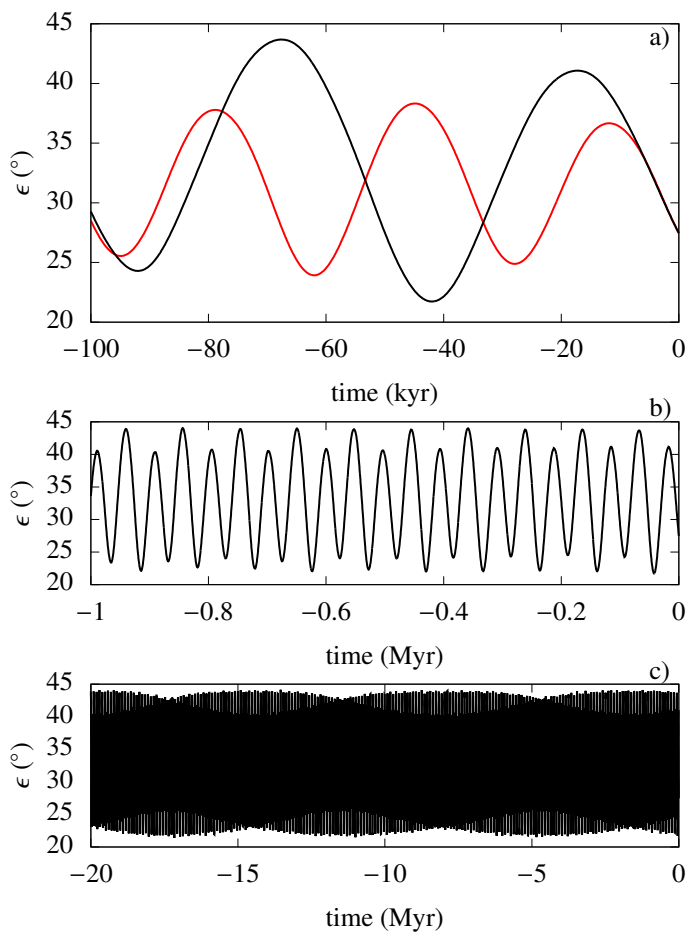
\begin{figure}
\centering
\input{figures/tex_figure9.tex}
\caption{\label{fig:oblivesta}Obliquity of Vesta on $\left[-100:0\right]\KYR$ (a), $\left[-1:0\right]\MYR$ (b) and $\left[-20:0\right]\MYR$ (c). In (a) the obliquity caused only by the change in orientation of the orbit is represented by the red curve.}
\end{figure}

\begin{table}
\centering
\begin{tabular}{ccc}
								& Ceres 					& Vesta \\
\hline
$J_2$ 							& $2.6499\times10^{-2}$				& $7.1060892\times10^{-2}$ \\
$R$ ($\km$) 					& $470$					& $265$ \\
$\omega$ ($\mathrm{rad}.s^{-1}$) 	& $1.923403741\times10^{-4}$ & $3.26710510494\times10^{-4}$ \\
$\C$ 							& $0.393$             	& $0.409$ \\
\end{tabular}
\caption{\label{tab:phycha}Physical characteristics of Ceres and Vesta used for the computation of the long-term rotation.}
\end{table}

The solution La2011 does not include the integration of the rotation axes of Ceres and Vesta. We compute then the solution Ceres2017 where the spin axes of Ceres and Vesta are integrated with the symplectic method presented in section \ref{sec:oblisymp}. We consider the interactions between the orbital and rotational motions and the torques exerted by the Sun and the planets on Ceres and Vesta. As they are in La2011, Ceres, Vesta, Pallas, Iris, and Bamberga are considered planets and exert a torque on Ceres and Vesta. We use the same initial condition for the orbital motion as in La2011. To integrate the long-term rotation, we use the parameters listed in table \ref{tab:phycha} and the initial conditions for the rotation axis in table \ref{tab:CI}. The integration is realized on $\left[-100:100\right]\MYR$ in extended precision with a time step of $0.005\,\yr$.
We use the integrator $\mathcal{SABA}_{C3}$ developed for perturbed Hamiltonians by \cite{laskar2001}.
A symmetric composition of this integrator with the method of \cite{suzuki1990} allows one to obtain a higher order integrator, as indicated in \cite{laskar2001}.

The differences between La2011 and Ceres2017 for the eccentricity and inclination of Ceres and Vesta oscillate around zero. The amplitudes on $\left[-20:0\right]\MYR$ are about $0.008$ and $0.1\degree$ for the eccentricity and the inclination of Ceres and about $0.02$ and $0.2\degree$ for Vesta. These differences have similar amplitudes to those observed for a small change ($1\times10^{-10}\,\rad$) of the initial mean longitude $\lambda$ of Ceres and Vesta. Therefore, they come from the chaotic behavior for the orbital motions of Ceres and Vesta \citep{laskar2011b} and are thus not significant.

The evolution of the obliquity is represented on the time intervals $\left[-100:0\right]\KYR$, $\left[-1:0\right]\MYR$ and $\left[-20:0\right]\MYR$ in figures \ref{fig:obliceres} and \ref{fig:oblivesta} respectively for Ceres and Vesta. For Ceres, we obtain similar results to \cite{bills2017} and \cite{ermakov2017a} with oscillations between about $2.06$ and $19.6\degree$ on $\left[-20:0\right]\MYR$. For Vesta, we observe oscillations between $21.4$ and $44.1\degree$. The amplitudes of the oscillations of the obliquities of Ceres and Vesta are similar on $\left[-100:100\right]\MYR$.

We perform the frequency analysis of the solution Ceres2017 on the time interval $\left[-20:0\right]\MYR$. The frequency decompositions of the quantity $w_x+iw_y$, where $w_x$ and $w_y$ are the coordinates in the invariant frame of the component parallel to the invariable plane of the normalized angular momentum, are in tables \ref{tab:freqobliceres} and \ref{tab:freqoblivesta} (Appendix \ref{sec:freqdecomp}) respectively for Ceres and Vesta. For Ceres, the precession frequency of the rotation axis is then $f_C=-6.1588\pm0.0002\arc$, which corresponds to a precession period of about $210.43\KYR$ and is consistent with the precession period of $210\KYR$ determined by \cite{ermakov2017a}. For Vesta, the precession frequency of the rotation axis is $f_V=-12.882\pm0.002\arc$, which corresponds to a period of precession of about $100.61\KYR$.

\cite{skoglov1996} noticed that bodies like Ceres and Vesta, which have a high inclination and a precession frequency of the ascending node higher than the precession frequency of the rotation axis, could have strong variations in the obliquity.
Indeed, the obliquity is given by
\begin{equation}
\cos \epsilon=\mathbf{n}.\mathbf{w}=\cos i \cos l + \sin i \sin l \cos\left(\Omega-L\right)
\end{equation}
with $(l,L)$ the inclination and the longitude of the ascending node of the equatorial plane and $(i,\Omega)$ those of the orbital plane in the frame of the invariable plane.
Then the precession of the ascending node causes obliquity variations if the inclination of the orbital plane is not null.
The inclination of the orbit plane with respect to the initial equatorial plane is represented by a red curve in figures \ref{fig:obliceres} and \ref{fig:oblivesta} respectively for Ceres and Vesta. For Ceres, a large part of the amplitude of the obliquity is caused by the precession of the ascending node, which creates oscillations between $2.1$ and $17.1\degree$ on $\left[-100:0\right]\KYR$. For Vesta, the contribution is less important.

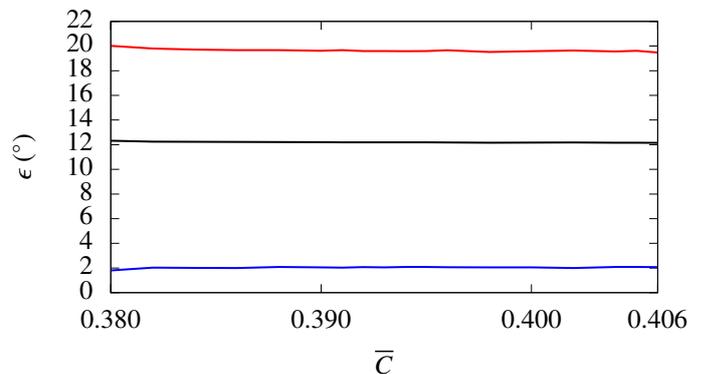
\begin{figure}
\centering
\input{figures/tex_figure10.tex}
\caption{\label{fig:CCeres}Maximum, mean, and minimum obliquities respectively in red, black, and blue for Ceres on $\left[-20:0\right]\MYR$ with respect to the normalized polar moment of inertia.}
\end{figure}

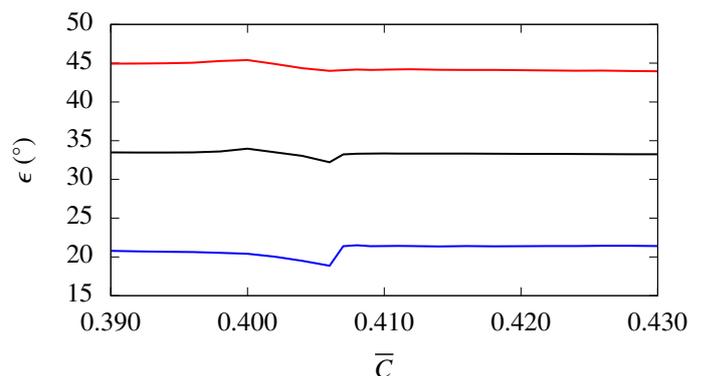
\begin{figure}
\centering
\input{figures/tex_figure11.tex}
\caption{\label{fig:CVesta}Maximum, mean, and minimum obliquities respectively in red, black, and blue for Vesta on $\left[-20:0\right]\MYR$ with respect to the normalized polar moment of inertia.}
\end{figure}

We have integrated for the time interval $\left[-100:0\right]\MYR$ the rotation of Ceres and Vesta for different normalized polar moments of inertia respectively in the intervals $\left[0.380:0.406\right]$ and $\left[0.390:0.430\right]$. For Ceres, all the different normalized polar moments of inertia give solutions for the obliquity with oscillations of similar amplitude (Fig. \ref{fig:CCeres}), as noticed by \cite{ermakov2017a}. The mean differences come from the precession frequency, which depends on the normalized polar moment of inertia. A small difference on the precession frequency causes a phase difference, which grows when the time increases. For Vesta, the obliquity solutions of the different normalized polar moments of inertia have all oscillations of similar amplitude (Fig. \ref{fig:CVesta}) except for the solution obtained for $\C=0.406$. Because of a secular resonance with the orbital frequency $2s_6-s_V$ (see section \ref{sec:closereson}), the obliquity can decrease to $18.9\degree$ on $\left[-20:0\right]\MYR$ for $\C=0.406$.

\section{Secular model for the orbital motion\label{SEC:secularmodels}}

In section \ref{sec:resultorb}, we have observed the proximity of Ceres with the resonances of the frequencies $2g_6-g_5\approx52.23\arc$ and $2g_6-g_7\approx53.40\arc$. If Ceres is close to these two resonances and if these resonances overlap, it could affect its orbital motion and therefore the rotational motion. We have especially seen in section \ref{sec:resultoblisymp} that the values of the inclination have direct consequences on the variations of the obliquity. Moreover, as noted by \cite{laskarrobutel1993}, the chaotic behavior of the orbital motion can widen by diffusion the possible chaotic zones of the rotation axis. 

A secular model can be obtained from the secular Hamiltonian of Ceres and Vesta to get secular equations, which are integrated much faster than full equations. From the development of the secular Hamiltonian of \cite{laskarrobutel1995}, we build a secular model of Ceres and Vesta perturbed only by Jupiter and Saturn, which allows us to identify the important terms of the planetary perturbations and to study the close secular resonances.

\subsection{Hamiltonian secular model\label{sec:modsecham}}

\cite{laskarrobutel1995} computed the development of the Hamiltonian of the planetary perturbations. 
We consider the case of a body only perturbed by Jupiter and Saturn. 
From \cite{laskarrobutel1995}, the Hamiltonian is
\begin{equation}
H= \sum_{i=5}^{6}\sum_{k,k'}\sum_{\mathcal{N}}\Gamma_{\mathcal{N}}\left(\Lambda,\Lambda_{i}\right) X^{n}X_{i}^{n'}\overline{X}^{\overline{n}}\overline{X}_{i}^{\overline{n}'}Y^{m}Y_{i}^{m'}\overline{Y}^{\overline{m}}\overline{Y}_{i}^{\overline{m}'}e^{i\left( k\lambda+k'\lambda_{i}\right)}\label{eq:devham}
\end{equation}
with $\mathcal{N}=(n,n',\overline{n},\overline{n}',m,m',\overline{m},\overline{m}')$ and the coefficients $\Gamma_{N}\left(\Lambda,\Lambda_{i}\right)$, which depend only on the ratio of the semi-major axes. The Poincaré rectangular canonical coordinates $\left(\Lambda,\lambda,x,-i\overline{x},y,-i\overline{y} \right)$ are defined by
\begin{equation}
\Lambda=\beta\sqrt{\mu a},
\end{equation}
\begin{equation}
x=\sqrt{\Lambda\left( 1-\sqrt{1-e^{2}} \right)}e^{i\varpi},
\end{equation}
\begin{equation}
y=\sqrt{\Lambda\sqrt{1-e^{2}} \left(1-\cos i \right)}e^{i\Omega},
\end{equation}
with $\beta=m\Msol/(m+\Msol)$, $\mu=\mathcal{G}(m+\Msol)$ and $m$ the mass of the perturbed body. The variables $X$ and $Y$ are given by $X=x\sqrt{2/\Lambda}$ and $Y=y/\sqrt{2\Lambda}$ \citep{laskarrobutel1995}. We select the terms verifying the secular inequality $\left(0,0\right)$ for $\left(k,k'\right)$ to obtain the secular part of the Hamiltonian (Eq. (\ref{eq:devham})) and we consider the case of a massless perturbed body.

The secular interaction Hamiltonian has been computed for the order 1 in mass and the degree 4 in eccentricity and inclination. We perform a frequency analysis of the solution La2011 on $\left[-20:0\right]\MYR$ and conserve only the main secular terms to create a secular solution of Jupiter and Saturn, which we inject in the Hamiltonian. The Hamiltonian depends then only on time and on $X$, $Y$. The equations of the motion are
\begin{equation}
\frac{dX}{dt}=-\frac{2i}{\Lambda}\frac{\partial H}{\partial \overline{X}}\label{eq:secX}
\end{equation}
\begin{equation}
\frac{dY}{dt}=-\frac{i}{2\Lambda}\frac{\partial H}{\partial \overline{Y}}\label{eq:secY}.
\end{equation}

\subsection{Adjustment of the secular model\label{sec:modsecadj}}

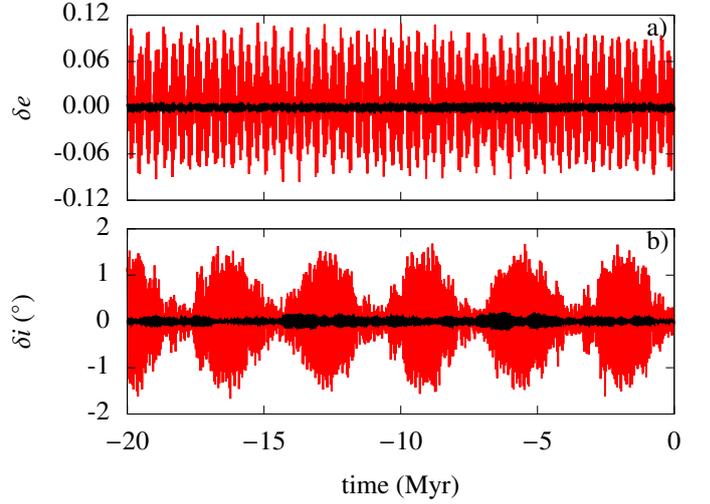
\begin{figure}
\centering
\input{figures/tex_figure12.tex}
\caption{\label{fig:compLaXsecceres}Difference in eccentricity (a) and inclination (b) for Ceres between the solution La2011 and the Hamiltonian secular model with adjustment of the frequencies in black, and without in red.}
\end{figure}
\begin{figure}
\centering
\input{figures/tex_figure13.tex}
\caption{\label{fig:compLaXsecvesta}Difference in eccentricity (a) and inclination (b) for Vesta between the solution La2011 and the Hamiltonian secular model with adjustment of the frequencies in black, and without in red.}
\end{figure}
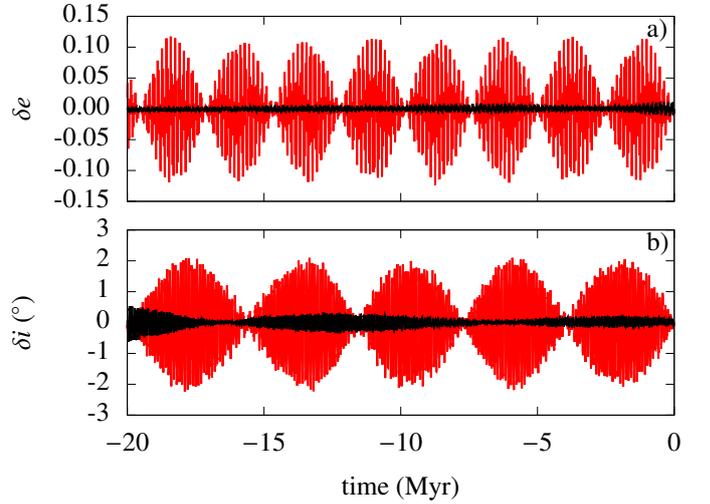

Equations (\ref{eq:secX}, \ref{eq:secY}) are integrated with a step size of $100$ years with a numerical integrator Runge-Kutta 8(7) on $\left[-20:0\right]\MYR$. The obtained solution allows us to reproduce the amplitudes of the oscillations of the eccentricity and the inclination of the solution La2011.
However, there are differences in the proper frequencies $g$ and $s$ with the solution La2011.
These differences of frequency cause phase differences between the perihelion and ascending node longitudes, which grow approximately linearly with time.
The secular model does not allow us then to reproduce the solution La2011 (Figs. \ref{fig:compLaXsecceres} and \ref{fig:compLaXsecvesta}).
To increase the precision of the secular model, it is possible to increase the order of the secular Hamiltonian, but this only allows us to partly reduce the differences in the frequencies.

As was done by \cite{laskar1990}, we adjust the secular frequencies of the model.
The differences in the perihelion and ascending node longitudes between the solution La2011 and the secular model are fitted by the affine functions $\mathcal{A} t + d \varpi_0$ and $\mathcal{B} t + d \Omega_0$. The frequencies of the secular model are adjusted by applying the following procedure to obtain the Hamiltonian $H'$
\begin{equation}
H'=H-\frac{\mathcal{A}\Lambda}{2} X\overline{X}-2\mathcal{B}\Lambda Y\overline{Y}.
\end{equation}
The initial conditions for the perihelion longitudes and ascending node longitudes are also respectively corrected by the quantities $d \varpi_0$ and $d \Omega_0$. The initial conditions for the eccentricity and the inclination are also slightly corrected. We iterate this procedure until we obtain a difference between the solution La2011 and the secular model, which has a mean close to zero for the four quantities $e$, $i$, $\varpi$, and $\Omega$. The adjustment of the frequencies is then about $\mathcal{A}\approx4.1\arc$ and $\mathcal{B}\approx0.20\arc$ for Ceres and $\mathcal{A}\approx0.51\arc$ and $\mathcal{B}\approx-0.41\arc$ for Vesta. The differences for the eccentricity and the inclination between the two solutions then oscillate around zero and correspond to short-period terms, which are not reproduced by the secular Hamiltonian (Figs. \ref{fig:compLaXsecceres} and \ref{fig:compLaXsecvesta}). On $\left[-20:0\right]\MYR$, the maximum differences in absolute value between the solution La2011 and the adjusted secular model are then $0.0082$ and $0.23\degree$ for the eccentricity and the inclination of Ceres and $0.013$ and $0.65\degree$ for the eccentricity and the inclination of Vesta.

This Hamiltonian model with the adjustment of the frequencies $g$ and $s$ allows us to reproduce the variations in the eccentricity and the inclination of Ceres and Vesta on $\left[-20:0\right]\MYR$. Therefore, the long-term orbital dynamics of Ceres and Vesta is given for the most part by the planetary perturbations of Jupiter and Saturn, as noticed by \cite{skoglov1996} for Ceres and Vesta and \cite{ermakov2017a} for Ceres.

\begin{figure}
\centering
\input{figures/tex_figure14.tex}
\caption{\label{fig:alphaceres}Eccentricity (a), inclination (b), and frequency $g_C$ (c) with respect to $\mathcal{A}$ for Ceres. For (a) and (b), the red, black, and blue curves correspond respectively to the maximum, mean, and minimum values. The vertical red line represents the value of $\mathcal{A}$ for the secular model.}
\end{figure}
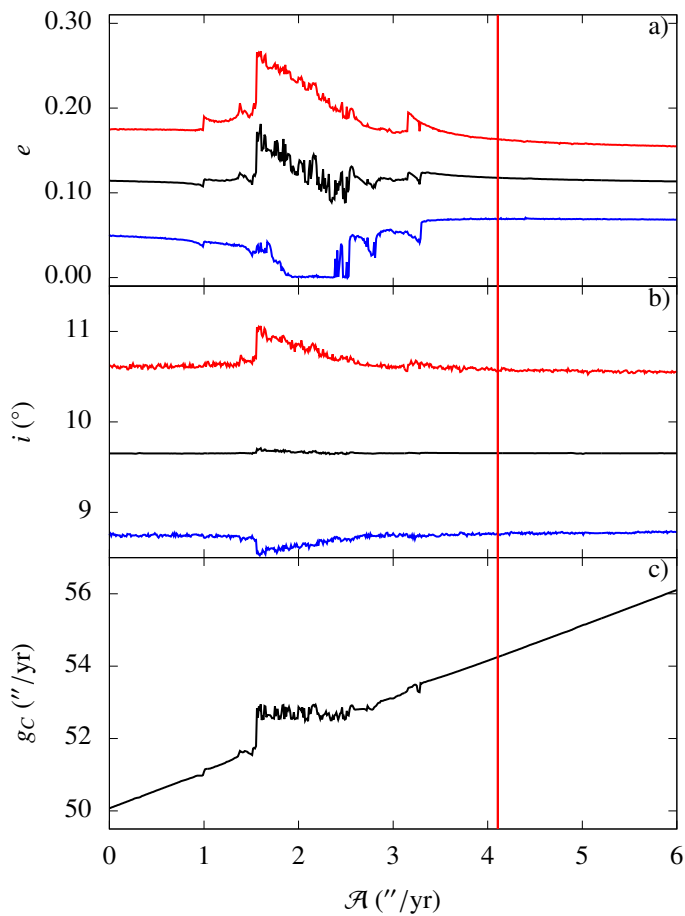

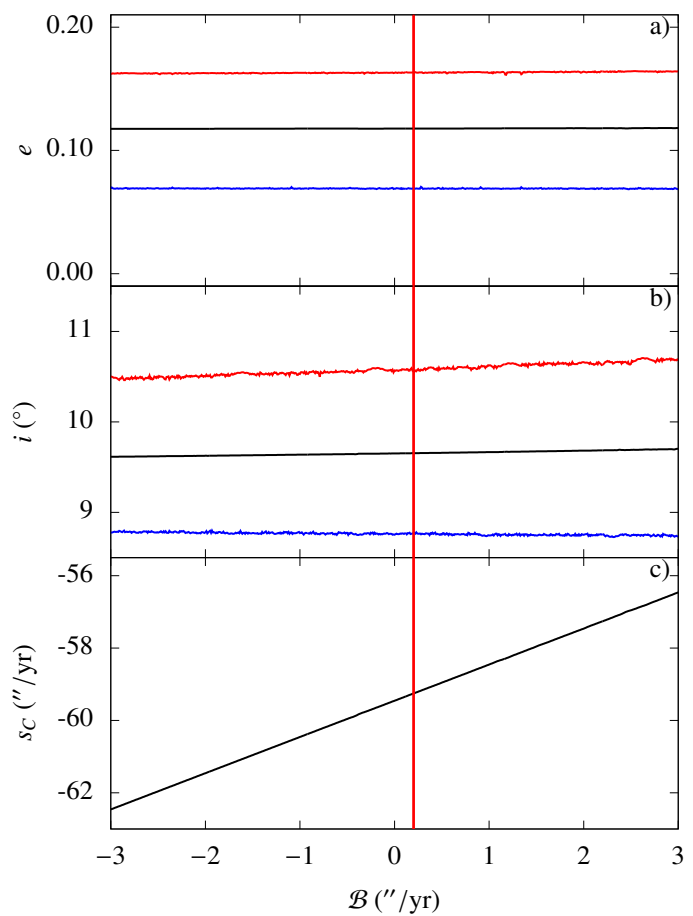
\begin{figure}
\centering
\input{figures/tex_figure15.tex}
\caption{\label{fig:betaceres}Eccentricity (a), inclination (b), and frequency $s_C$ (c) with respect to $\mathcal{B}$ for Ceres. For (a) and (b), the red, black, and blue curves correspond respectively to the maximum, mean, and minimum values. The vertical red line represents the value of $\mathcal{B}$ for the secular model.}
\end{figure}

\begin{figure}
\centering
\input{figures/tex_figure16.tex}
\caption{\label{fig:alphavesta}Eccentricity (a), inclination (b), and frequency $g_V$ (c) with respect to $\mathcal{A}$ for Vesta. For (a) and (b), the red, black, and blue curves correspond respectively to the maximum, mean, and minimum values. The vertical red line represents the value of $\mathcal{A}$ for the secular model.}
\end{figure}
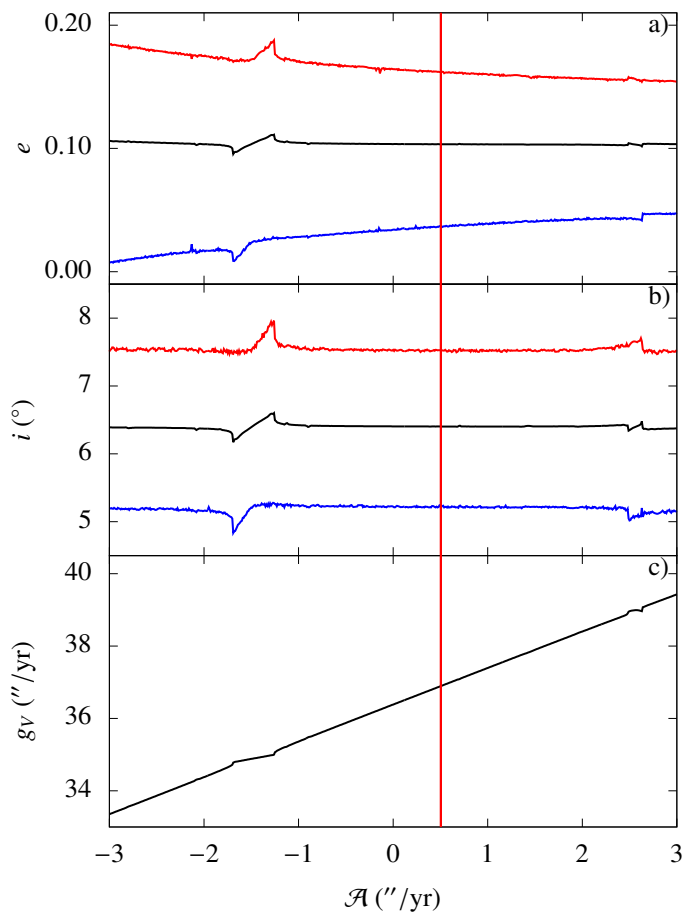

\begin{figure}
\centering
\input{figures/tex_figure17.tex}
\caption{\label{fig:betavesta}Eccentricity (a), inclination (b), and frequency $s_V$ (c) with respect to $\mathcal{B}$ for Vesta. For (a) and (b), the red, black, and blue curves correspond respectively to the maximum, mean, and minimum values. The vertical red line represents the value of $\mathcal{B}$ for the secular model.}
\end{figure}
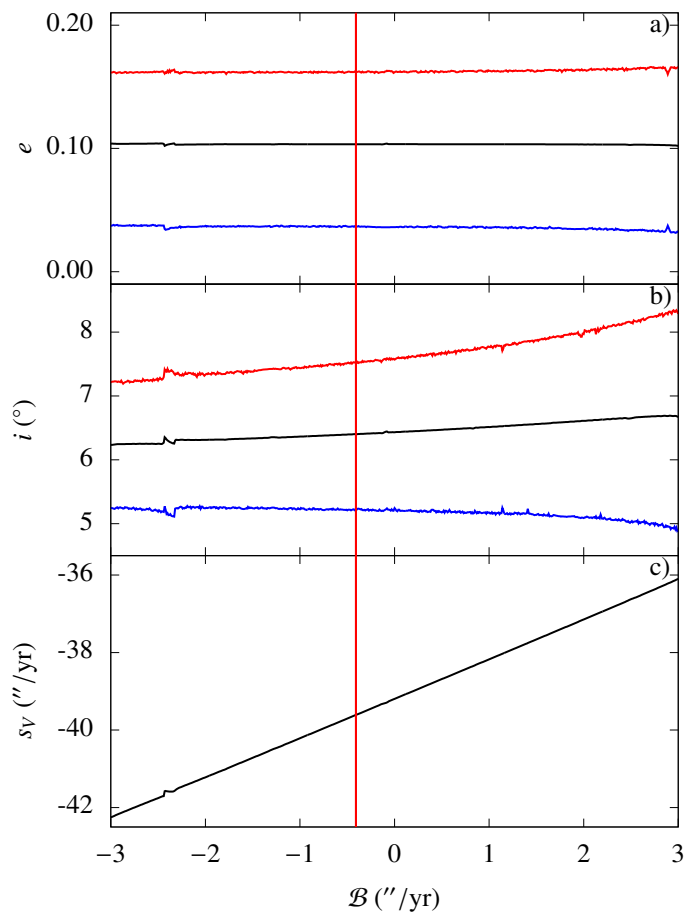

\subsection{Study of the close resonances}

This model allows us to study the resonances close to Ceres and Vesta. The integration of the secular Hamiltonian is about $10^4$ faster than the complete integration and allows us to proceed to many integrations with different parameters $\mathcal{A}$ and $\mathcal{B}$ near the values used for the models to see the effects of the close secular resonances. For each value of these parameters, Eqs. (\ref{eq:secX}, \ref{eq:secY}) are integrated on $\left[-20:0\right]\MYR$ and the secular frequencies $g$ and $s$ are determined with the frequency analysis.

For Ceres, the evolutions of the eccentricity, the inclination, and the frequency $g_C$ are in figure \ref{fig:alphaceres} for $\mathcal{A}\in \left[0:6\right]\arc$. The resonance with the frequency $2g_6-g_5\approx52.23\arc$, present in the secular motion of Jupiter and Saturn, acts for about $g_C\in\left[51.32:53.16\right]\arc$. The maximum and minimum eccentricities vary respectively from $0.18$ to $0.27$ and from $0.04$ to $0.0002$. The maximum inclination rises from $10.6$ to $11.1\degree$, which would increase the variations in the obliquity, as noticed in section \ref{sec:resultoblisymp}. For about $g_C\in\left[53.21:53.60\right]\arc$, there is a resonance with the frequency $2g_6-g_7\approx53.40\arc$ present in the secular motion of Jupiter and Saturn, and the maximum eccentricity increases from $0.17$ to $0.19$. Therefore, the resonance with the frequency $2g_6-g_7$ has a weaker chaotic nature than that with $2g_6-g_5$. In section \ref{sec:resultorb} we have given the interval $\left[54.225:54.261\right]\arc$ as an estimation of the variation in the frequency $g_C$ because of the chaotic diffusion on $\left[-250:250\right]\MYR$. Therefore, the chaotic diffusion of Ceres is too weak to put Ceres in resonance with the frequencies $2g_6-g_5$ and $2g_6-g_7$ on $\left[-250:250\right]\MYR$. The frequency $g_7+2g_6-2g_5\approx51.06\arc$ in the motion of Jupiter and Saturn causes a resonance with weaker but observable effects on the eccentricity for $g_C\in\left[50.97:51.20\right]\arc$. In the secular model of the motions of Jupiter and Saturn, we find the frequency $3g_6-2g_5+s_6-s_7\approx52.87\arc$ with a smaller amplitude and it is then difficult to distinguish its effects from those of the resonance with the frequency $2g_6-g_5$. The evolutions of the eccentricity, the inclination, and the frequency $s_C$ are in figure \ref{fig:betaceres} for $\mathcal{B}\in\left[-3:3\right]\arc$ and have slight irregularities for $s_C\in\left[-59.95:-59.73\right]\arc$. This frequency interval does not correspond to a term used for the secular motion of Jupiter and Saturn.

For Vesta, the evolutions of the eccentricity, the inclination, and the frequency $g_V$ are in figure \ref{fig:alphavesta} for $\mathcal{A}\in\left[-3:3\right]\arc$. For $g_V\in\left[34.56:35.25\right]\arc$, there is a resonance with the frequency $2g_5-s_6\approx34.86\arc$ where the maximum eccentricity increases from $0.17$ to $0.19$ and the maximum inclination from $7.5$ to $8.0\degree$. For $g_V\in\left[38.86:39.12\right]\arc$, the maximum inclination increases from $7.6$ to $7.7\degree$. This area does not correspond to terms used for the secular motion of Jupiter and Saturn. The evolutions of the eccentricity, the inclination, and the frequency $s_V$ are in figure \ref{fig:betavesta} for $\mathcal{B}\in\left[-3:3\right]\arc$. For $s_V\in\left[-41.71:-41.49\right]\arc$, the inclination can increase from $7.3$ to $7.4\degree$. This resonance does not match any term used for the secular motion of Jupiter and Saturn.

\section{Stability of the rotation axes\label{SEC:stab}}

In this section we are interested in the study of the long-term stability of the rotation axis. Like in \cite{laskarjoutelrobutel1993} and \cite{laskarrobutel1993}, the stability of the rotation axis can be estimated using frequency analysis. We determine the precession frequency $f_1$ on the interval $\left[-20:0\right]\MYR$ and the precession frequency $f_2$ on the interval $\left[-40:-20\right]\MYR$. The quantity $\sigma=|(f_1-f_2)/f_1|$ gives an estimate of the diffusion of the precession frequency \citep{laskar1993,dumaslaskar1993}. For an integrable system, this quantity must stay null. For a weakly perturbed system, this quantity is small, but increases if the system becomes chaotic.

\subsection{Secular solution for the obliquity}

\begin{figure}
\centering
\input{figures/tex_figure18.tex}
\caption{Difference between the solution Ceres2017 and the secular solution for the obliquity of Ceres (a) and of Vesta  (b) on $\left[-20:0\right]\MYR$.\label{fig:dobliLaXmodsec}}
\end{figure}
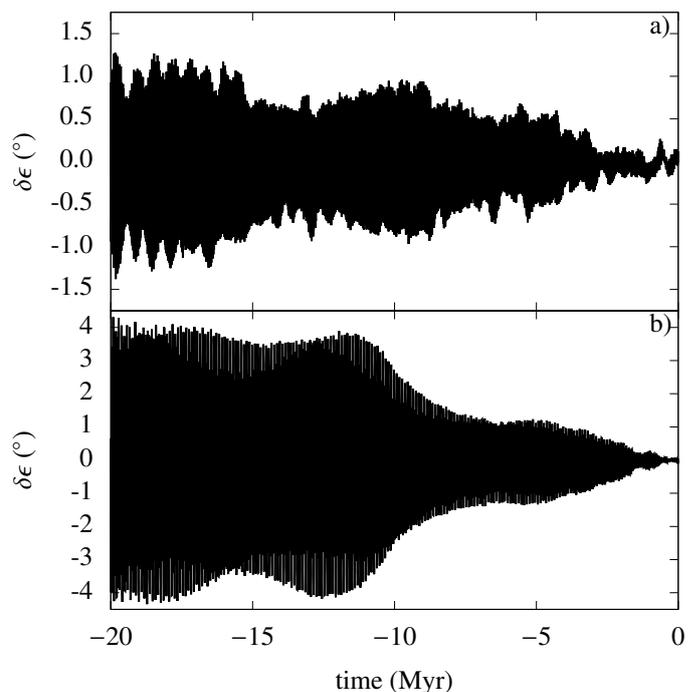

We integrate the secular equation (\ref{eq:integsec}) with an Adams integrator and a step size of $100$ years. The normal to the orbit $\bf{n}$ and the eccentricity $e$ are computed from the secular orbital solution obtained from the secular frequency decompositions in section \ref{sec:resultorb}. We use the initial conditions for the rotation axis of table \ref{tab:CI}. The secular solutions for the obliquities are compared to the nonsecular ones for Ceres and Vesta in figure \ref{fig:dobliLaXmodsec} for the time interval $\left[-20:0\right]\MYR$. The secular computation of the obliquity, which is about 1 million times faster, allows us then to reproduce correctly the evolution of the obliquity.

The secular orbital solution has initial conditions different from those of solution La2011 because we have removed the short-period variations in figures \ref{fig:compmodsecceres} and \ref{fig:compmodsecvesta}. This modifies the initial obliquities of Ceres and Vesta of about $0.02\degree$ for Ceres and $-0.05\degree$ for Vesta, and could explain the differences observed in figure \ref{fig:dobliLaXmodsec}.

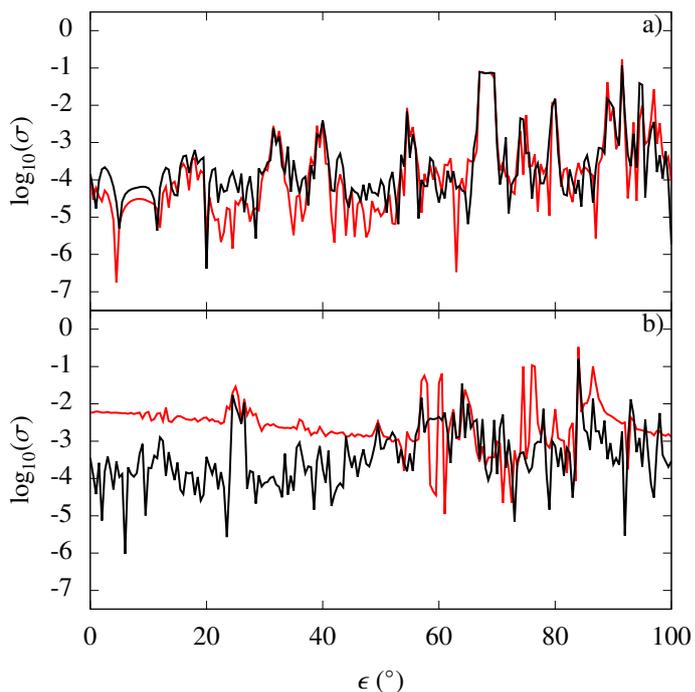
\begin{figure}
\centering
\input{figures/tex_figure19.tex}
\caption{\label{fig:epsstab}Diffusion of the precession frequency with respect to the initial obliquity for the complete solution (in black) and for the secular solution (in red) for Ceres (a) and Vesta (b).}
\end{figure}

We integrate on $\left[-40:0\right]\MYR$ the rotation axis with the symplectic method of the section \ref{sec:oblisymp} and the secular equation (\ref{eq:integsec}). For both integrations the initial obliquities vary from $0$ to $100\degree$ with a step of $0.5\degree$. The diffusion of the precession frequency  is represented in figure \ref{fig:epsstab} with respect to the initial obliquity. For Ceres, the diffusion is quite similar with close amplitude and evolution and the areas with a strong increase in $\sigma$ allow us to recognize resonances with the orbital frequencies for the two cases. For Vesta, the diffusion $\sigma$ is higher for the secular solution. However, the areas with high values of the diffusion $\sigma$ correspond. The secular and nonsecular solutions of the obliquity thus have close stability properties.

\subsection{Study of the close resonances}

We integrate the secular equation (\ref{eq:integsec}) on $\left[-40:0\right]\MYR$ for different precession constants in an interval with a step of $0.01\arc$ to determine the effects of the resonances.

\subsubsection{Ceres}

The precession frequency, its diffusion, and the variations in the obliquity are represented for Ceres in figure \ref{fig:alphastabceres} with respect to the precession constant in the interval $\left[0.01:12\right]\arc$. We observe areas with strong variations in the diffusion, specified in table \ref{tab:areadifceres}, which correspond to resonances with orbital frequencies. Most of these frequencies are already present in the frequency decompositions of the variables $z$ and $\zeta$ used for the construction of the secular solution. The quantities $\mathbf{n}$ and $e$, which appear in the secular equation (\ref{eq:integsec}) and which are used to obtain the angular momentum $\mathbf{w}$, are computed from the secular solution of the variables $z$ and $\zeta$, and can include additional frequencies. To identify the remaining frequencies in table \ref{tab:areadifceres}, we then perform a frequency analysis of the quantities $n_x+in_y$ and $(n_x+in_y)/(1-e^2)^{3/2}$, where $n_x$ and $n_y$ are the coordinates in the invariant frame of the component parallel to the invariable plane of the normal to the orbit $\mathbf{n}$. We find no trace of the remaining frequencies in the first 4000 terms of the frequency analysis of $n_x+in_y$. In the first 4000 terms of the frequency analysis of $(n_x+in_y)/(1-e^2)^{3/2}$, we find the missing frequencies of the areas identified in table \ref{tab:areadifceres}. The variations in the eccentricity are then responsible for the apparition of additional secular resonances between the orbital and the rotational motions.

We note in particular the appearance of the resonance with the frequency $s_C+2(g_C-g_6)+(g_5-g_7)\approx-6.07\arc$, which is included in the interval of uncertainty of the precession constant. Therefore, Ceres could be in resonance with this frequency. However, this effect on the obliquity is very limited. In the vicinity of the interval of uncertainty, we observe a narrow area with a small decrease up to $1\degree$ of the minimum obliquity because of the resonance with the frequency $s_C+(3g_C-4g_6+g_7)\approx-6.39\arc$. More distant resonances have stronger effects on the obliquity of Ceres. The resonance with the frequency $s_C+(g_C-g_5)\approx-9.26\arc$ causes variations in the obliquity in the interval between $0$ and almost $40\degree$ and that with $s_7\approx-2.99\arc$ variations between $0$ and almost $30\degree$. Ceres is closer to a less important resonance with $s_{C}+2(g_C-g_6)\approx-7.24\arc$, where there are variations in the obliquity between $0$ and $26\degree$. However Ceres should have a precession constant between about $7.15\arc$ and $7.85\arc$ to be inside this resonance.

In figure \ref{fig:alphastabceres}, we see the diffusion for the precession constants computed with a rotation rate that is $7\%$ higher \citep{mao2018}, as discussed in section \ref{sec:earlyceresalpha}. If the early Ceres were in hydrostatic equilibrium, as supposed by \cite{mao2018}, it could be in resonance with the frequencies $s_1\approx-5.61\arc$, $s_C+2(g_C-g_6)+(g_5-g_7)\approx-6.07\arc$ and $s_C+(3g_C-4g_6+g_7)\approx-6.39\arc$, which have weak effects on the obliquity as seen in figure \ref{fig:alphastabceres} and the amplitudes of the oscillations of the obliquity would be similar. The events or phenomena, which would have changed its rotation rate, would not have significantly changed the interval of variation in the obliquity.

As discussed in section \ref{sec:earlyceresalpha}, if the early Ceres were in hydrosatic equilibrium and the shape and the internal structure had not changed as supposed by \cite{mao2018}, the present Ceres would have a precession constant in the interval $\left[6.58:6.98\right]\arc$ for a normalized polar moment of inertia of $\C=0.371$. With these precession constants, Ceres could be in resonance with the frequency $s_C+(3g_C-4g_6+g_7)\approx-6.39\arc$ (table \ref{tab:areadifceres}) with no significant changes in the obliquity.

\begin{figure}
\centering
\input{figures/tex_figure20.tex}
\caption{\label{fig:alphastabceres}Obliquity (a), precession frequency (b), and diffusion of the precession frequency (c) for Ceres on $\left[-40:0\right]\MYR$ with respect to the precession constant. In (a) the maximum, mean, and minimum obliquities are respectively in red, black, and blue. In (c) the rectangle A represents the precession constants for $\C\in[0.380:0.406]$ with a vertical red line for $\C=0.393$. B corresponds to the precession constants in $[5.91:6.77]\arc$ with a vertical red line for $\alpha=6.34\arc$ computed in section \ref{sec:earlyceresalpha} for a spin rate $7\%$ higher \citep{mao2018}.}
\end{figure}

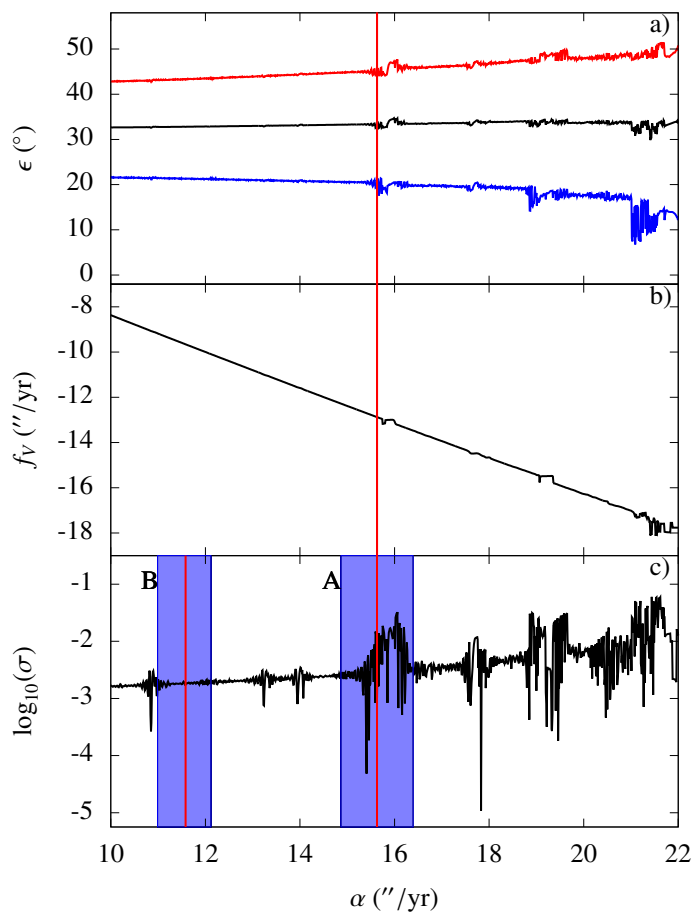
\begin{figure}
\centering
\input{figures/tex_figure21.tex}
\caption{\label{fig:alphastabvesta}Obliquity (a), precession frequency (b), and diffusion of the precession frequency (c) for Vesta on $\left[-40:0\right]\MYR$ with respect to the precession constant. In (a) the maximum, mean, and minimum obliquities are respectively in red, black, and blue. In (c) the rectangle A represents the precession constants for $\C\in[0.390:0.430]$ with a vertical red line for $\C=0.409$ and B the same but before the two giant impacts.}
\end{figure}

\subsubsection{Vesta\label{sec:closereson}}

The precession frequency, its diffusion, and the variations in the obliquity are represented for Vesta in figure \ref{fig:alphastabvesta} with respect to the precession constant in the interval $\left[10:22\right]\arc$. The frequencies of the resonances are in table \ref{tab:areadifvesta}. We identify the frequencies $2s_6-s_V\approx-13.09\arc$, $s_V-g_5+g_6\approx-15.62\arc$, and $-17.74\arc$, which are among the frequencies of the secular model. As for Ceres, we perform a frequency analysis of the quantities $n_x+in_y$ and $(n_x+in_y)/(1-e^2)^{3/2}$ to identify the remaining frequencies. We do not find them in the frequency analysis of $n_x+in_y$, but in the frequency analysis of $(n_x+in_y)/(1-e^2)^{3/2}$ we find the frequencies $-9.09\arc$, $s_7-(g_V-g_6)\approx-11.65\arc$, $s_V+(g_6-g_7)\approx-14.46\arc$, which can correspond to the areas identified in table \ref{tab:areadifvesta}. As for Ceres, the variations of the eccentricity are responsible for the appearance of some resonances. The interval of frequency $\left[-11.16:-10.93\right]\arc$ in table \ref{tab:areadifvesta} does not correspond to any term of the frequency analysis.

The resonance which has the most important effect in the vicinity of Vesta is that with the frequency $-17.74\arc$. If the maximum obliquity increases by about $3\degree$, the minimum obliquity decreases by about $10\degree$. The domain of the resonance with the frequency $2s_6-s_V\approx-13.09\arc$ is included in the uncertainty interval for the precession constant. We have observed in section \ref{sec:resultoblisymp} that for the value $\C=0.406$ of the normalized polar moment of inertia, the minimum obliquity decreases compared to the evolution of the obliquity for the other normalized polar moments of inertia. We can see here that it is an effect of the resonance with the frequency $2s_6-s_V$. In this resonance, the minimum obliquity can decrease to $17.6\degree$ and the maximum obliquity can increase to $47.7\degree$.

In figure \ref{fig:alphastabvesta}, we observe the diffusion for the precession constants computed with the paleorotation rate of the early Vesta determined by \cite{fu2014} and the physical parameters discussed in section \ref{sec:earlyvestaalpha}. The two giant impacts could then have put Vesta closer to the resonance with the frequency $2s_6-s_V$, which involves the crossing of two small resonances and could also have slightly increased the interval of variation of the obliquity.

\begin{table*}
\centering
\begin{tabular}{ccccc}
$\alpha$ $(\arc)$	& frequency $(\arc)$ & identification & approximate value & \\
\hline
$\left[0.52:1.10\right]$ & $\left[-1.07:-0.51\right]$ & $s_8$ & $-0.69\arc$ & * \\
$\left[1.77:2.00\right]$ & $\left[-1.94:-1.72\right]$ & $s_7+(g_5-g_7)$ & $-1.83\arc$ & * \\
$\left[2.19:2.60\right]$ & $\left[-2.52:-2.12\right]$ & $s_6-(g_5-g_6)$ & $-2.36\arc$ & * \\
$\left[2.76:3.49\right]$ & $\left[-3.38:-2.67\right]$ & $s_7$ & $-2.99\arc$ & * \\
$\left[4.21:4.54\right]$ & $\left[-4.38:-4.07\right]$ & $s_7-(g_5-g_7)$ & $-4.16\arc$ & * \\
$\left[5.14:5.34\right]$ & $\left[-5.15:-4.96\right]$ & $s_7-(g_C+g_5-2g_6)$ & $-5.01\arc$ & \\
$\left[5.36:5.60\right]$ & $\left[-5.39:-5.16\right]$ & $s_C+(3g_C+g_5-4g_6)$ & $-5.22\arc$ & *  \\
$\left[5.75:6.04\right]$ & $\left[-5.81:-5.54\right]$ & $s_1$ & $-5.61\arc$ & *  \\
$\left[6.21:6.48\right]$ & $\left[-6.23:-5.97\right]$ & $s_C+2(g_C-g_6)+(g_5-g_7)$ & $-6.07\arc$ & \\
$\left[6.52:6.79\right]$ & $\left[-6.52:-6.27\right]$ & $s_C+(3g_C-4g_6+g_7)$ & $-6.39\arc$ & *  \\
$\left[7.15:7.85\right]$ & $\left[-7.53:-6.86\right]$ & $s_C+2(g_C-g_6)$ & $-7.24\arc$ & *  \\
$\left[8.10:8.29\right]$ & $\left[-7.94:-7.76\right]$ & $s_C-(s_6-s_7-2g_C-g_5+3g_6)$ & $-7.88\arc$ & \\
$\left[8.34:8.51\right]$ & $\left[-8.16:-7.99\right]$ & $s_C+(g_C-g_7)$ & $-8.09\arc$ & *  \\
$\left[8.63:8.95\right]$ & $\left[-8.56:-8.27\right]$ & $s_C+2(g_C-g_6)-(g_5-g_7)$ & $-8.41\arc$ & *  \\
$\left[9.48:10.19\right]$ & $\left[-9.66:-9.04\right]$ & $s_C+(g_C-g_5)$ & $-9.26\arc$ & *  \\
$\left[10.48:10.75\right]$ & $\left[-10.19:-9.95\right]$ & $s_C-(g_5-2g_6+g_7)$ & $-10.11\arc$ & *  \\
$\left[10.93:11.12\right]$ & $\left[-10.55:-10.37\right]$ & $s_C+(g_C-2g_5+g_7)$ & $-10.43\arc$ & * \\
$\left[11.15:11.37\right]$ & $\left[-10.78:-10.58\right]$ & $s_C+(s_6-s_7-3(g_5-g_6))$ & $-10.65\arc$ & \\
$\left[11.51:11.66\right]$ & $\left[-11.03:-10.90\right]$ & $s_C-(g_C+g_5-4g_6+2g_7)$ & $-10.96\arc$ & \\
$\left[11.75:11.99\right]$ & $\left[-11.40:-11.11\right]$ & $s_C-2(g_5-g_6)$ & $-11.28\arc$ & * \\
\end{tabular}
\caption{\label{tab:areadifceres}Areas with strong variations in the diffusion in figure \ref{fig:alphastabceres} for Ceres. The sign * indicates the frequencies used to construct the secular orbital solution in section \ref{sec:resultorb}.}
\end{table*}

\begin{table*}
\centering
\begin{tabular}{ccccc}
$\alpha$ $(\arc)$	& frequency $(\arc)$ & identification & approximate value & \\
\hline
$\left[10.71:11.07\right]$ & $\left[-9.24:-8.95\right]$  & & $-9.09\arc$ \\
$\left[13.16:13.45\right]$ & $\left[-11.16:-10.93\right]$  & & \\
$\left[13.84:14.18\right]$ & $\left[-11.74:-11.47\right]$  & $s_7-(g_V-g_6)$ & $-11.65\arc$ & \\
$\left[14.89:16.48\right]$ & $\left[-13.54:-12.30\right]$  & $2s_6-s_V$ & $-13.09\arc$ & * \\
$\left[17.36:18.18\right]$ & $\left[-14.82:-14.22\right]$  & $s_V+(g_6-g_7)$ & $-14.46\arc$\\
$\left[18.77:19.67\right]$ & $\left[-16.02:-15.25\right]$  & $s_V-(g_5-g_6)$ & $-15.62\arc$ & * \\
$\left[20.12:21.98\right]$ & $\left[-18.12:-16.34\right]$  & & $-17.74\arc$ & * \\
\end{tabular}
\caption{\label{tab:areadifvesta}Areas with strong variations in the diffusion in figure \ref{fig:alphastabvesta} for Vesta. The sign * indicates the frequencies used to construct the secular orbital solution in section \ref{sec:resultorb}.}
\end{table*}

\subsection{Global stability of the rotation axis}

As in \cite{laskarjoutelrobutel1993} and in \cite{laskarrobutel1993}, we look for the long-term stability of the rotation axis. We integrate the rotation axis on $\left[-40:0\right]\MYR$ with the secular equation (\ref{eq:integsec}) on a grid of $24120$ points for initial obliquities from $0$ to $100\degree$ with a step of $0.5\degree$ and for precession constants from $0.5$ to $60\arc$ with a step of $0.5\arc$.

The precession frequency corresponds in the frequency analysis to the frequency with the largest amplitude. However, in the case of an important resonance, the frequency with the largest amplitude can correspond to the resonance frequency. We also consider as precession frequency the one with the largest amplitude, for which the difference with the frequencies $s$ and $s_6$ is larger than $5\times10^{-3}\arc$. 

\subsubsection{Ceres}

\begin{figure*}
\centering
\input{figures/tex_figure22a.tex}
\input{figures/tex_figure22b.tex}
\caption{\label{fig:stab}Stability of the rotation axis (a,d), amplitude of the obliquity on $\left[-40:0\right]\MYR$  (b,e), precession frequency on $\left[-20:0\right]\MYR$ (c,f) respectively for Ceres and Vesta with respect to the initial obliquity and the precession constant for a grid of $24120$ points. The white circles represent Ceres and Vesta for the epoch J2000. The color scale of the diffusion is represented on $[-6:-1]$, although $\log_{10}\left(\sigma\right)$ takes values outside this interval. For (a) and (d), red points correspond to $\log_{10}(\sigma)\geq -1$ and black points to $\log_{10}(\sigma)\leq -6$. For (d,e,f), the white squares represent Vesta before the two giant impacts.}
\end{figure*}
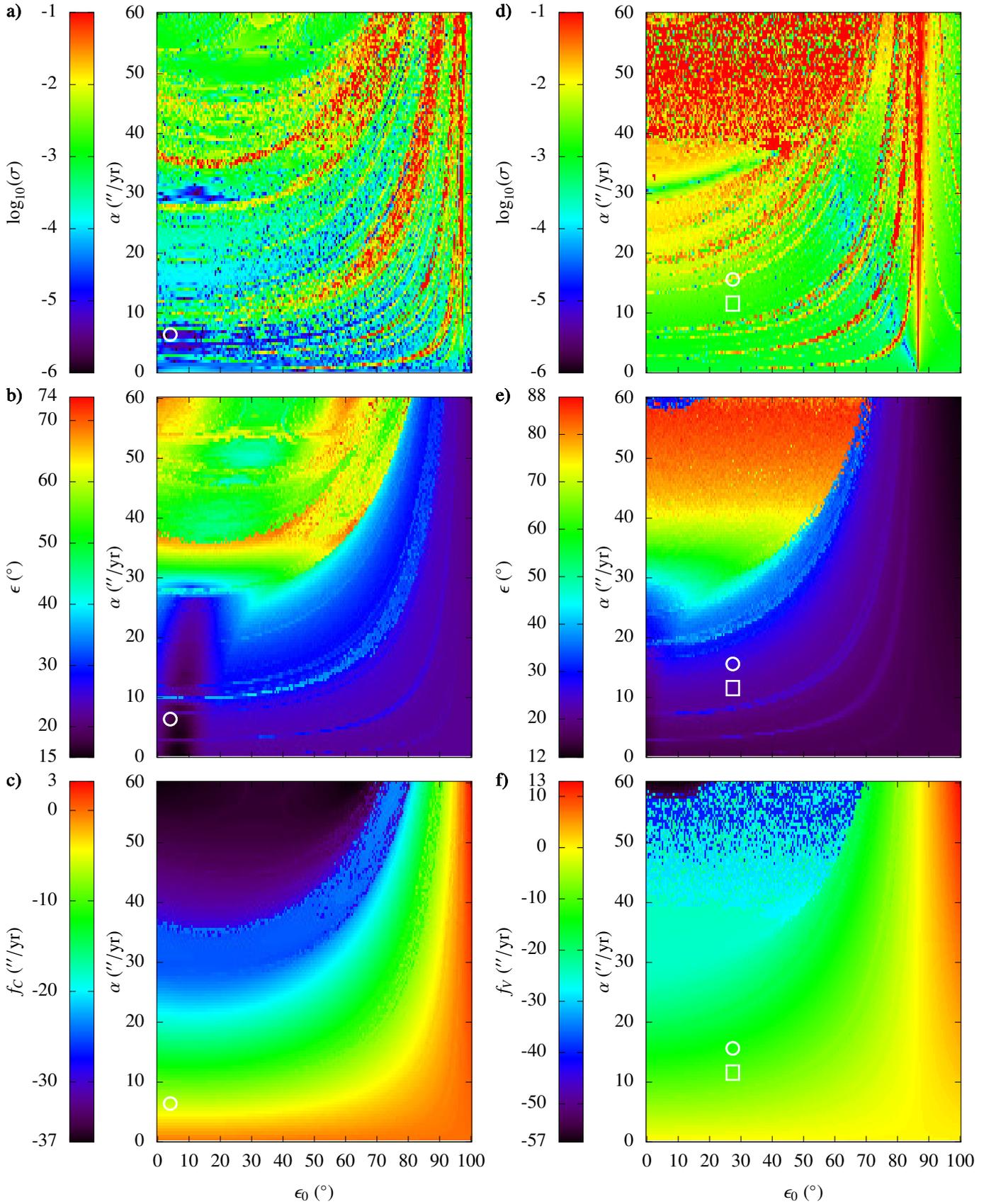

In figure \ref{fig:stab}, we see the value of the quantity $\log_{10}\left(\sigma\right)$, the maximum amplitude of the obliquity on $\left[-40:0\right]\MYR$, which corresponds to the difference between the minimum and maximum obliquities on $\left[-40:0\right]\MYR$ and the precession frequency of Ceres obtained by frequency analysis on $\left[-20:0\right]\MYR$. The position of Ceres for the epoch J2000 is indicated with a white circle. 

Ceres is in a quite stable zone and is far from the most chaotic zones, which correspond to the resonance with the frequencies $s_C$, $s_6$ and $s_C+\left(g_C-g_6\right)$. The motion of its rotation axis is relatively stable although Ceres has a precession frequency $f_C=-6.1588\arc$ (table \ref{tab:freqobliceres}) close to the node precession frequencies of the inner planets, Mercury $s_1=-5.61\arc$ and Venus $s_2=-7.06\arc$ (table \ref{tab:freqref}). The secular orbital motion of Ceres is almost entirely determined by the planetary perturbations of Jupiter and Saturn (section \ref{sec:modsecadj}), and the amplitudes of frequencies of the inner planets are small in the motion of the ascending node of Ceres. Therefore, the width of these resonances is small and they do not overlap, contrary to the case of the inner planets \citep{laskarrobutel1993}.

The resonances with the frequencies $s_7$ and $s_8$ have more important effects than those with the inner planets. These resonances can increase the amplitude of the obliquity of several degrees. With the present precession constant, we see that the resonance with $s_7$ can affect the rotation axis of Ceres only if Ceres has an initial obliquity of about $70\degree$ and the resonance with $s_8$ can affect Ceres if the initial obliquity is about $90\degree$. The resonance with the frequency $s_6$ has large amplitudes of the obliquity, but does not correspond to that of the most chaotic zone except when it overlaps with the resonance at the frequency $s_C+\left(g_C-g_6\right)$. As for the case of the planets, as noted by \cite{laskarrobutel1993}, the resonance with the frequency $s_6$ is isolated.

The most important nearby resonance is with the frequency $s_C+\left(g_C-g_5\right)\approx-9.24\arc$, which has a significant effect on the amplitude of the obliquity. For an initial obliquity between $0$ and $10\degree$, the amplitude of the obliquity passes from about $20$ to $40\degree$. However, this resonance has a limited width of about $1\arc$ and it has no influence on Ceres.

\subsubsection{Vesta}

The diffusion $\log_{10}\left(\sigma\right)$, the maximum amplitude of the obliquity, and the precession frequency are represented for Vesta in figure \ref{fig:stab}. Vesta for the epoch J2000 is indicated with a white circle. 

Vesta is at the boundary of a relatively stable region. Vesta is far from the chaotic zone created by the resonances with the frequencies $s_V$ and $s_6$ and is close to the resonance with the orbital frequency $2s_6-s_V$. The most important resonance in the vicinity is the one with the frequency $-17.74\arc$, which could correspond to the frequency $s_4$. In this resonance, the amplitude of the obliquity passes about from $30$ to $40\degree$. Because of this limited width, it has no influence on Vesta. The effect of the resonances with the frequencies $s_7$ and $s_8$ are less important than for Ceres. The resonance with $s_7$ increases the amplitude of the obliquity only by a few degrees. As for Ceres, the resonance with the frequency $s_V+\left(g_V-g_5\right)\approx-6.97\arc$ still has an important effect on the amplitude of the obliquity, which increases from about $20$ to $30\degree$ in the resonance. 

Like Ceres, Vesta has a precession frequency $f_V=-12.882\arc$ (table \ref{tab:freqoblivesta}) close to the node precession frequencies of the inner planets, the Earth $s_3=-18.848\arc$, and Mars $s_4=-17.751\arc$ (table \ref{tab:freqref}), but their perturbations on the orbit of Vesta are too weak to have significant consequences on the stability of the rotation axis for the present Vesta.

The early Vesta is represented in figure \ref{fig:stab} by a white square for the precession constant computed from the supposed rotational parameters before the two giant impacts. The early Vesta would also be in a more stable region. As seen in section \ref{sec:closereson}, the two giant impacts could have put Vesta closer to the resonance with the orbital frequency $2s_6-s_V$.

\section{Conclusion}

We applied the method of \cite{farago2009} to realize a symplectic integration of the rotation axes only averaged over the fast proper rotation. The obliquity variations of Ceres have been obtained between $2$ and $20\degree$ for the last $20\MYR$ in agreement with the results of \cite{bills2017} and \cite{ermakov2017a}. If we use for Ceres the value of the normalized polar moment of inertia $\C=0.395$ (Eq. (\ref{eq:Cnorm})), which takes into account the nonspherical form of Ceres, we obtain obliquity variations in the same interval and the frequency precession decreases in absolute value of about $0.5\%$ with respect to the value obtained for $\C=0.393$. For Vesta, the obliquity variations are between $21$ and $45\degree$ for the last $20\MYR$. As noted by \cite{skoglov1996}, these large variations in the obliquity are due to the significant inclinations of Ceres and Vesta with respect to the invariable plane.

The secular orbital model in section \ref{SEC:secularmodels} has allowed us to show that the chaotic diffusion of the secular frequency $g_C$ of Ceres does not seem sufficiently important to put Ceres in a secular orbital resonance with the frequencies $2g_6-g_5$ and $2g_6-g_7$. For Vesta, the chaotic diffusion of the secular frequencies is more important especially for $s_V$. This model has also allowed us to show that a secular model of Ceres and Vesta only perturbed by Jupiter and Saturn could entirely reproduce their secular orbital motions. The secular orbital dynamics of Ceres and Vesta is then dominated by the perturbations of Jupiter and Saturn, as noted by \cite{skoglov1996} for Ceres and Vesta and confirmed by \cite{ermakov2017a} for Ceres.

Ceres and Vesta have precession frequencies close to the secular orbital frequencies of the terrestrial planets, as is the case for Mars. The precession frequency of Ceres is close to secular orbital frequencies of Mercury and Venus and that of Vesta to secular orbital frequencies of the Earth and Mars. However, their long-term rotations are relatively stable. They are in an orbital region where the perturbations of Jupiter and Saturn dominate the secular orbital dynamics and the perturbations of the inner planets are relatively weak. The secular resonances with the inner planets have smaller widths and do not overlap, contrary to the case of the inner planets.

This is an illustration that the stability of the long-term rotation depends strongly on the orbital motion. For Ceres and Vesta, there is a chaotic zone with large oscillations of the obliquity as for the inner planets, but it is caused by the overlapping of resonances due to their proper secular frequencies with other resonances due to the perturbations of Jupiter and Saturn. We also note for Ceres and Vesta that the evolution of the eccentricity is responsible for the appearance of secular resonances for the spin axis. However, their effects on the obliquity and the stability are modest.

The two giant impacts suffered by Vesta modified the precession constant and could have put Vesta closer to the resonance with the orbital frequency $2s_6-s_V$. Given the uncertainty on the polar moment of inertia, the present Vesta could be in resonance with the frequency $2s_6-s_V$, where the obliquity can decrease to about $17\degree$ and increase to about $48\degree$.

\begin{acknowledgements}
T. Vaillant thanks Nathan Hara for the fruitful discussions and references about the random walk on a sphere. The authors thank Anton Ermakov for the useful comments on this work.
\end{acknowledgements}

\bibliographystyle{aa} 
\bibliography{biblio} 

\begin{appendix}
\section{Passage from the invariable plane frame to the ICRF\label{sec:planinv} }

We consider a vector $\mathbf{x}$ in the frame associated with the invariable plane. The coordinates in the ICRF become
\begin{equation}
\mathbf{x}'=R_z\left(\theta_3\right)R_x\left(\theta_1\right)\mathbf{x}
\end{equation}
with $R_x$ the rotation of axis $(1,0,0)$ and $R_z$ the rotation of axis $(0,0,1)$. The angles $\theta_1$ and $\theta_3$ are given by
\begin{equation}
\theta_1=0.4015807829125271\,\rad
\end{equation}
being about $\theta_1\approx23.01\degree$ and
\begin{equation}
\theta_3=0.06724103544220839\,\rad
\end{equation}
being about $\theta_3\approx3.85\degree$.

\section{Frequency decompositions\label{sec:freqdecomp}}

\begin{table*}
\subfloat[$z=e \exp(i\varpi)$]{
\begin{tabular}{@{}C{3.65cm}R{1.65cm}R{1.1cm}r@{}}
\hline
 & $\nu_k$ ($\arc$) & $10^6\times A_k$ & $\phi_k$ ($\degree$) \\
\hline
$g_C$ & 54.25253 & 114938 & 158.975\\
$g_5$ & 4.25750 & 30684 & 27.077\\
$g_6$ & 28.24513 & 19564 & -55.734\\
$2g_6-g_5$ & 52.23276 & 8023 & 41.516\\
$2s_C-g_C$ & -172.75959 & 3548 & 176.528\\
& 54.17786 & 1655 & -83.503\\
& 54.32863 & 1623 & -134.203\\
$2g_6-g_7$ & 53.40275 & 1283 & -46.914\\
$g_7$ & 3.08802 & 1241 & 117.284\\
$2s_C-\left(2g_C-\left(2g_6-g_5\right)\right)$ & -174.77941 & 1127 & 58.984\\
$2g_C-\left(2g_6-g_5\right)$ & 56.27217 & 905 & -84.214\\
$2s_C-\left(2g_6-g_5\right)$ & -170.73981 & 840 & 113.990\\
$g_C-\left(s_C-s_6\right)$ & 87.15851 & 764 & -154.007\\
$g_C+\left(s_C-s_6\right)$ & 21.34721 & 742 & -66.583\\
$2g_C-g_5$ & 104.24751 & 634 & 110.562\\
$g_C-\left(g_5-g_7\right)$ & 53.08393 & 630 & 70.370\\
& 54.21717 & 616 & -84.317\\
$g_C+\left(g_5-g_7\right)$ & 55.42147 & 614 & 67.809\\
& 54.28472 & 558 & -129.243\\
& 54.08927 & 501 & -65.955\\
$2g_C-\left(2g_6-g_7\right)$ & 55.10340 & 497 & 9.247\\
& 54.41912 & 489 & -152.236\\
$2s_C-\left(2g_C-\left(2g_6-g_7\right)\right)$ & -173.60964 & 486 & -30.478\\
$2s_C-\left(2g_6-g_7\right)$ & -171.90928 & 480 & -155.596\\
$2s_C-g_5$ & -122.76450 & 446 & -51.407\\
& 54.15524 & 399 & -49.259\\
$2s_C-g_6$ & -146.75209 & 360 & 31.354\\
& 54.36418 & 355 & -158.902\\
$2g_C-g_6$ & 80.26030 & 353 & -165.663\\
& 54.31987 & 328 & -159.069\\
& 54.24404 & 319 & -112.835\\
& 54.18020 & 313 & -79.241\\
& 54.28220 & 270 & -132.179\\
$s_C+s_6-g_C$ & -139.85387 & 268 & 42.730\\
$g_6+\left(g_5-g_7\right)$ & 29.41435 & 259 & -146.243\\
$g_5+2\left(g_6-g_7\right)$ & 54.57100 & 256 & 48.097\\
& 54.12303 & 246 & -42.931\\
$g_6-\left(g_5-g_7\right)$ & 27.07620 & 238 & -143.927\\
& 54.38335 & 235 & 135.976\\
& 54.42899 & 226 & 142.071\\
& 54.35117 & 224 & -167.422\\
$2s_C-\left(g_C-\left(g_5-g_7\right)\right)$ & -171.59009 & 219 & -93.331\\
$2g_C-g_6+s_C-s_6$ & 47.35451 & 205 & -32.146\\
& 54.08255 & 184 & 10.421\\
& 53.93389 & 172 & 91.974\\
& 54.13501 & 157 & -16.590\\
& 54.51732 & 145 & 49.156\\
& 54.80607 & 130 & -157.283\\
& 53.98072 & 121 & 85.795\\
& 54.02290 & 103 & 63.359\\
\hline
\end{tabular}}\subfloat[$\zeta=\sin\left(i/2\right)\exp(i\Omega)$]{
\begin{tabular}{@{}C{4.3cm}R{1.45cm}R{1.1cm}r@{}}
\hline
 & $\nu_k$ ($\arc$) & $10^6\times A_k$ & $\phi_k$ ($\degree$) \\
\hline
$s_C$ & -59.25351 & 81688 & 78.182\\
$s_C-\left(g_C-\left(2g_6-g_5\right)\right)$ & -61.27328 & 12344 & -39.650\\
$s_C+\left(g_C-\left(2g_6-g_5\right)\right)$ & -57.23374 & 12196 & 15.144\\
$s_C+\left(g_C-\left(2g_6-g_7\right)\right)$ & -58.40327 & 5834 & 105.370\\
$s_C-\left(g_C-\left(2g_6-g_7\right)\right)$ & -60.10380 & 5635 & -130.022\\
$s_6$ & -26.34785 & 5229 & -56.215\\
& -59.31728 & 2065 & -158.140\\
& -59.18992 & 2006 & 137.687\\
$s_C+\left(g_5-g_7\right)$ & -58.08408 & 1633 & 167.789\\
$2g_C-s_C$ & 167.75857 & 1349 & -120.054\\
& -59.46608 & 1244 & -36.536\\
& -59.04171 & 1188 & 10.471\\
& -59.11983 & 1018 & 136.500\\
$s_C-2\left(g_C-\left(2g_6-g_5\right)\right)$ & -63.29327 & 962 & -157.935\\
& -59.38279 & 958 & -151.270\\
$s_C-\left(2g_C+g_5-4g_6+g_7\right)$ & -62.12350 & 895 & 112.305\\
$s_C-\left(g_C-g_6\right)$ & -85.26094 & 861 & -136.937\\
$s_C+2\left(g_C-\left(2g_6-g_5\right)\right)$ & -55.21408 & 861 & -47.864\\
& -59.21440 & 839 & 178.291\\
$s_C+(g_C-g_5-2g_6+2g_7)$ & -59.57496 & 833 & -175.559\\
$s_C+\left(2g_C+g_5-4g_6+g_7\right)$ & -56.38355 & 827 & 42.285\\
& -59.15776 & 815 & 145.609\\
$s_C+\left(g_C-g_6\right)$ & -33.24610 & 811 & 112.398\\
& -59.27631 & 754 & -158.149\\
$s_C-(g_C-g_5-2g_6+2g_7)$ & -58.93431 & 748 & 145.231\\
& -59.34031 & 658 & -136.208\\
$s_8$ & -0.69175 & 578 & 20.281\\
& -59.13924 & 537 & 69.232\\
& -59.09718 & 502 & 110.955\\
$s_7$ & -2.99254 & 498 & 136.651\\
& -59.18243 & 494 & 169.976\\
& -57.30115 & 464 & 140.185\\
& -59.40588 & 432 & -117.262\\
& -57.16365 & 430 & 81.300\\
$s_C+\left(s_6-s_7-g_C-2g_5+3g_6\right)$ & -60.64119 & 421 & 42.144\\
$s_C-\left(s_6-s_7-g_C-2g_5+3g_6\right)$ & -57.86597 & 410 & -68.694\\
$g_C+g_5-s_C$ & 117.76361 & 352 & 108.466\\
$s_C-2\left(g_C-\left(2g_6-g_7\right)\right)$ & -60.95377 & 347 & 25.982\\
& -59.08710 & 335 & 44.985\\
$g_C+g_6-s_C$ & 141.75125 & 332 & 25.633\\
& -59.18321 & 317 & 61.105\\
& -59.29432 & 306 & -156.136\\
$s_C+2\left(g_C-\left(2g_6-g_7\right)\right)$ & -57.55339 & 297 & 130.966\\
$s_C+\left(g_C-g_5\right)$ & -9.25848 & 291 & -150.471\\
$s_C+\left(g_C+2g_5-2g_6-g_7\right)$ & -56.06436 & 273 & 105.958\\
& -59.22602 & 268 & 176.849\\
$s_C+\left(g_5-g_6\right)$ & -83.24112 & 267 & 160.665\\
& -59.16970 & 238 & 106.185\\
& -59.24087 & 234 & 154.444\\
& -57.08095 & 199 & 104.126\\
\hline
\end{tabular}}
\caption{\label{tab:freqorbiceres}First 50 terms of the frequency decomposition $\sum_{k=1}^{50}A_ke^{i\left(\nu_k t+\phi_k\right)}$ of $z$ (a) and $\zeta$ (b) for Ceres on $\left[-25:5\right]\MYR$.}
\end{table*}

\begin{table*}
\centering
\subfloat[$z=e \exp(i\varpi)$]{
\begin{tabular}{@{}C{3.65cm}R{1.65cm}R{1.1cm}r@{}}
\hline
 & $\nu_k$ ($\arc$) & $10^6\times A_k$ & $\phi_k$ ($\degree$) \\
\hline
$g_V$ & 36.89490 & 98564 & -122.569\\
$g_6$ & 28.24512 & 31697 & -55.781\\
$g_5$ & 4.25749 & 26156 & 27.041\\
& 36.97140 & 4721 & 115.878\\
& 36.81313 & 4554 & 163.495\\
& 36.84862 & 2759 & 166.234\\
& 36.78643 & 2558 & 179.195\\
& 36.99679 & 2525 & 85.970\\
& 36.93803 & 1731 & 122.690\\
& 36.89072 & 1663 & 127.326\\
& 36.72857 & 1647 & -140.285\\
& 37.05697 & 1550 & 34.707\\
& 36.89173 & 1376 & 154.679\\
$g_V-\left(s_V-s_6\right)$ & 50.15557 & 1349 & -107.631\\
$g_V+\left(s_V-s_6\right)$ & 23.63377 & 1300 & 41.845\\
& 36.93341 & 1190 & 74.762\\
$2s_V-g_V$ & -116.11448 & 1102 & 154.886\\
& 36.84000 & 1064 & -179.914\\
& 36.84716 & 1047 & 162.744\\
$g_7$ & 3.08795 & 1045 & 117.063\\
& 36.78526 & 926 & -157.751\\
& 36.88950 & 802 & 69.901\\
$2g_6-g_5$ & 52.23282 & 676 & -138.360\\
& 36.94612 & 652 & 120.228\\
$g_V+\left(g_5-g_7\right)$ & 38.06152 & 640 & 140.591\\
$g_V-\left(g_5-g_7\right)$ & 35.72795 & 626 & 156.535\\
$2g_V-g_5$ & 69.53245 & 611 & -91.314\\
& 36.98925 & 581 & 79.842\\
$g_6-\left(s_V-s_6\right)$ & 41.50591 & 494 & -40.729\\
& 36.74614 & 477 & -116.184\\
& 36.69132 & 471 & -78.481\\
$g_6+\left(g_5-g_7\right)$ & 29.41516 & 430 & -143.914\\
$2g_V-g_6$ & 45.54421 & 422 & -9.384\\
& 36.93934 & 416 & 67.015\\
$2g_V-g_6+s_V-s_6$ & 32.28376 & 406 & 155.536\\
& 36.80741 & 403 & -139.857\\
& 36.94137 & 381 & 100.853\\
& 36.99255 & 377 & 27.377\\
& 37.04189 & 373 & 33.547\\
& 36.88943 & 372 & 119.603\\
$g_6-\left(g_5-g_7\right)$ & 27.07591 & 370 & -144.919\\
& 37.44533 & 358 & -86.385\\
& 36.34301 & 352 & 16.697\\
& 36.72254 & 349 & -62.084\\
& 36.79011 & 347 & -135.492\\
& 37.10060 & 344 & -39.502\\
& 36.65369 & 310 & -37.489\\
& 36.84687 & 268 & 177.786\\
& 37.14608 & 266 & -64.218\\
& 36.84949 & 237 & -174.862\\
\hline
\end{tabular}}\subfloat[$\zeta=\sin\left(i/2\right)\exp(i\Omega)$]{
\begin{tabular}{@{}C{3.65cm}R{1.65cm}R{1.1cm}r@{}}
\hline
 & $\nu_k$ ($\arc$) & $10^6\times A_k$ & $\phi_k$ ($\degree$) \\
\hline
$s_V$ & -39.60884 & 53659 & 107.187\\
$s_6$ & -26.34784 & 8415 & -56.192\\
& -39.67443 & 7943 & -20.319\\
& -39.53472 & 7659 & 76.357\\
& -39.49150 & 2826 & -33.655\\
& -39.57146 & 2652 & -140.255\\
& -39.62997 & 2198 & -137.520\\
& -39.52960 & 2098 & 112.525\\
$s_V-\left(g_V-g_6\right)$ & -48.25866 & 2079 & 174.744\\
$s_V+\left(g_V-g_6\right)$ & -30.95913 & 1985 & -138.459\\
& -39.58772 & 1560 & 105.743\\
& -39.47439 & 1467 & 103.816\\
& -39.69832 & 1399 & 162.185\\
& -39.65518 & 1248 & 58.214\\
& -39.61420 & 945 & -56.211\\
& -39.73436 & 860 & -67.914\\
& -39.50462 & 851 & -93.953\\
& -39.69100 & 816 & -167.379\\
& -39.39015 & 772 & -106.169\\
& -39.43391 & 770 & -42.332\\
& -39.46848 & 707 & 124.045\\
& -39.55488 & 697 & -45.834\\
& -39.77783 & 682 & 29.955\\
& -39.65303 & 651 & 66.999\\
$s_8$ & -0.69175 & 584 & 20.276\\
& -39.43395 & 574 & -2.148\\
& -39.73011 & 563 & -49.221\\
$2g_V-s_V$ & 113.39865 & 533 & 7.177\\
$s_7$ & -2.99254 & 514 & 136.656\\
& -39.42303 & 488 & 46.722\\
& -39.69201 & 466 & -179.610\\
& -39.76367 & 464 & 94.984\\
& -39.35408 & 379 & 139.039\\
$s_6-\left(g_V-g_6\right)$ & -34.99780 & 363 & -169.922\\
& -39.81461 & 357 & 121.250\\
& -39.79370 & 336 & -103.462\\
& -48.33380 & 327 & -0.018\\
& -48.17492 & 326 & -164.608\\
& -31.01869 & 309 & 126.019\\
& -30.89244 & 301 & 144.122\\
& -40.15440 & 284 & 86.579\\
$g_V+g_6-s_V$ & 104.74876 & 277 & 73.310\\
& -39.39361 & 274 & -103.823\\
& -39.05991 & 271 & -42.700\\
$s_V-\left(g_5-g_7\right)$ & -40.77220 & 263 & -145.014\\
$2s_V-s_6$ & -52.87189 & 259 & 88.498\\
$s_V+\left(g_5-g_7\right)$ & -38.44674 & 249 & -177.002\\
& -17.73694 & 212 & -34.193\\
$s_V+\left(g_V-g_5\right)$ & -6.97330 & 136 & -50.263\\
$g_V+g_5-s_V$ & 80.76118 & 133 & 156.327\\
\hline
\end{tabular}}
\caption{\label{tab:freqorbivesta}First 50 terms of the frequency decomposition $\sum_{k=1}^{50}A_ke^{i\left(\nu_k t+\phi_k\right)}$ of $z$ (a) and $\zeta$ (b) for Vesta on $\left[-25:5\right]\MYR$.}
\end{table*}

\begin{table}
\centering
\begin{tabular}{@{}C{3.65cm}R{1.65cm}R{1.1cm}r@{}}
\hline
 & $\nu_k$ ($\arc$) & $10^6\times A_k$ & $\phi_k$ ($\degree$) \\
\hline
$f_C$ & -6.15875 & 132796 & 4.534\\
$s_C$ & -59.25393 & 19264 & 162.921\\
$s_6$ & -26.34785 & 3150 & 33.785\\
$s_C+(g_C-g_5)$ & -9.25982 & 3022 & -68.345\\
$s_C+\left(g_C-\left(2g_6-g_5\right)\right)$ & -57.23494 & 2952 & 97.518\\
$s_C-\left(g_C-\left(2g_6-g_5\right)\right)$ & -61.27289 & 2832 & 48.975\\
$s_7$ & -2.99104 & 1915 & 50.007\\
$s_C+\left(g_C-\left(2g_6-g_7\right)\right)$ & -58.40439 & 1392 & -172.293\\
$s_C-\left(g_C-\left(2g_6-g_7\right)\right)$ & -60.10345 & 1312 & -41.211\\
$s_8$ & -0.69160 & 1303 & -69.308\\
$s_C+2(g_C-g_6)$ & -7.23883 & 1170 & -127.942\\
$f_C-\left(g_C-\left(2g_6-g_5\right)\right)$ & -8.17557 & 668 & -100.922\\
$f_C+\left(g_C-\left(2g_6-g_5\right)\right)$ & -4.14024 & 658 & -57.248\\
$s_C+2(g_C-g_6)+(g_5-g_7)$ & -6.06275 & 573 & 2.238\\
$s_C-2(g_5-g_6)$ & -11.27863 & 420 & 177.727\\
$s_C+(g_5-g_7)$ & -58.08387 & 391 & -105.808\\
$f_C+\left(g_C-\left(2g_6-g_7\right)\right)$ & -5.30451 & 313 & 48.054\\
& -59.15554 & 305 & -28.298\\
$f_C-\left(g_C-\left(2g_6-g_7\right)\right)$ & -6.99815 & 290 & -168.326\\
& -59.34498 & 280 & -166.997\\
$s_C+\left(3g_C-4g_6+g_7\right)$ & -6.39098 & 276 & -113.654\\
& -59.45168 & 267 & 113.695\\
$s_C+2(g_C-g_6)-(g_5-g_7)$ & -8.41092 & 266 & -45.638\\
& -6.27076 & 251 & 121.424\\
& -59.04637 & 229 & 60.256\\
$s_C+(g_C-g_6)$ & -33.24727 & 227 & -165.136\\
& -6.15880 & 225 & -88.261\\
$s_C-(g_5-2g_6+g_7)$ & -10.10925 & 219 & 86.998\\
$s_C+2\left(g_C-\left(2g_6-g_5\right)\right)$ & -55.21618 & 216 & 30.997\\
$s_C-2\left(g_C-\left(2g_6-g_5\right)\right)$ & -63.29223 & 215 & -66.897\\
$s_C-(2g_C+g_5-4g_6+g_7)$ & -62.12250 & 204 & -155.831\\
$s_C+(2g_C+g_5-4g_6+g_7)$ & -56.38576 & 203 & 121.140\\
& -59.25884 & 202 & -106.222\\
$2f_C-s_C$ & 46.93652 & 195 & -153.986\\
$s_C+(g_C-g_5-2g_6+2g_7)$ & -59.57497 & 183 & -77.254\\
$s_C-(g_C-g_6)$ & -85.26054 & 183 & -48.320\\
$s_C-(g_C-g_5-2g_6+2g_7)$ & -58.93920 & 173 & -154.076\\
$s_1$ & -5.61671 & 159 & -118.819\\
$2s_C-f_C$ & -112.34910 & 140 & 141.919\\
$s_C+(g_C-g_7)$ & -8.08029 & 127 & -125.830\\
$f_C+(g_C-g_6)$ & 19.84789 & 105 & 36.075\\
$f_C-(g_C-g_6)$ & -32.16614 & 104 & 150.147\\
$s_C-(s_6-s_7-g_C-2g_5+3g_6)$ & -57.86704 & 96 & 15.532\\
$s_C+(s_6-s_7-g_C-2g_5+3g_6)$ & -60.64206 & 94 & 127.961\\
$2g_C-s_C$ & 167.75745 & 90 & 148.483\\
$f_C+(s_C-s_6)$ & -39.06485 & 90 & -46.170\\
& -56.11382 & 85 & 12.387\\
& -5.91789 & 82 & 112.898\\
$s_C+\left(3g_C+g_5-4g_6\right)$ & -5.21225 & 78 & 10.716\\
& -59.37412 & 68 & 159.704\\
\hline
\end{tabular}
\caption{\label{tab:freqobliceres}First 50 terms of the frequency decomposition $\sum_{k=1}^{50}A_ke^{i\left(\nu_k t+\phi_k\right)}$ of $w_x+iw_y$ for Ceres on $\left[-20:0\right]\MYR$.}
\end{table}

\begin{table}
\centering
\begin{tabular}{@{}C{3.65cm}R{1.65cm}R{1.1cm}r@{}}
\hline
 & $\nu_k$ ($\arc$) & $10^6\times A_k$ & $\phi_k$ ($\degree$) \\
\hline
$f_V$ & -12.88235 & 536537 & -32.774\\
$2f_V-\left(2s_6-s_V\right)$ & -12.68720 & 53372 & -129.004\\
$2s_6-s_V$ & -13.07751 & 49031 & -114.858\\
$s_V$ & -39.61376 & 31572 & -172.011\\
& -12.77160 & 17140 & 52.745\\
& -12.99626 & 14654 & 48.485\\
$2f_V-s_V$ & 13.84895 & 13649 & 106.512\\
& -12.67225 & 10137 & -159.973\\
$s_6$ & -26.34823 & 7832 & 32.968\\
& -13.10016 & 7321 & -115.847\\
& -12.55303 & 6177 & -19.085\\
& -12.92955 & 5260 & 14.327\\
$2f_V-s_6$ & 0.58288 & 4646 & -99.913\\
& -12.73796 & 4601 & -121.794\\
$f_V-\left(s_V-s_6\right)$ & 0.38433 & 4002 & 174.403\\
$f_V-\left(g_V-g_6\right)$ & -21.53224 & 3891 & 27.961\\
& -12.82979 & 3878 & 130.392\\
$f_V+\left(g_V-g_6\right)$ & -4.23214 & 3863 & 87.770\\
$f_V+\left(s_V-s_6\right)$ & -26.14817 & 3364 & -58.177\\
& -12.83327 & 3345 & 116.714\\
& -13.06876 & 3064 & 61.903\\
$3f_V-2s_6$ & 14.04715 & 2933 & 13.044\\
$f_V-2(s_V-s_6)$ & 13.65255 & 2838 & 24.381\\
& -12.44726 & 2791 & -125.619\\
& -13.16296 & 2581 & -44.359\\
& -39.71053 & 2303 & -17.854\\
& -12.99531 & 2220 & 41.091\\
& -39.51364 & 2149 & -138.939\\
$2s_V-f_V$ & -66.34537 & 2086 & -131.843\\
$s_V+(g_V-g_5)$ & -6.97876 & 1971 & -142.636\\
& -13.25676 & 1924 & -122.046\\
& -12.86902 & 1919 & -78.108\\
& -12.61890 & 1819 & 99.975\\
& -12.53596 & 1587 & 159.537\\
& -13.11606 & 1508 & 156.028\\
$s_V+(g_V-g_6)$ & -30.96333 & 1357 & -51.344\\
& -12.78299 & 1342 & -14.556\\
& -12.93541 & 1319 & -23.383\\
$s_7$ & -2.99285 & 1201 & 45.632\\
$s_V+s_6-f_V$ & -53.07896 & 1116 & 75.117\\
$s_V-(g_V-g_6)$ & -48.26419 & 1101 & -112.768\\
$s_8$ & -0.69187 & 1056 & -69.630\\
& -12.37138 & 1049 & -141.990\\
& -13.18382 & 1020 & -117.994\\
& 0.77870 & 1013 & 161.465\\
& -39.43894 & 953 & -156.160\\
$f_V-\left(g_V-g_5\right)$ & -45.52000 & 900 & 110.360\\
$f_V+\left(g_V-g_5\right)$ & 19.75530 & 886 & 4.257\\
& -12.46117 & 863 & 169.774\\
$f_V+\left(g_V-g_6+s_V-s_6\right)$ & -17.49709 & 595 & 62.754\\
\hline
\end{tabular}
\caption{\label{tab:freqoblivesta}First 50 terms of the frequency decomposition $\sum_{k=1}^{50}A_ke^{i\left(\nu_k t+\phi_k\right)}$ of $w_x+iw_y$ for Vesta on $\left[-20:0\right]\MYR$.}
\end{table}
\end{appendix}

\end{document}

%% file: figures/tex_figure2.tex
\begingroup
  \makeatletter
  \providecommand\color[2][]{%
    \GenericError{(gnuplot) \space\space\space\@spaces}{%
      Package color not loaded in conjunction with
      terminal option `colourtext'%
    }{See the gnuplot documentation for explanation.%
    }{Either use 'blacktext' in gnuplot or load the package
      color.sty in LaTeX.}%
    \renewcommand\color[2][]{}%
  }%
  \providecommand\includegraphics[2][]{%
    \GenericError{(gnuplot) \space\space\space\@spaces}{%
      Package graphicx or graphics not loaded%
    }{See the gnuplot documentation for explanation.%
    }{The gnuplot epslatex terminal needs graphicx.sty or graphics.sty.}%
    \renewcommand\includegraphics[2][]{}%
  }%
  \providecommand\rotatebox[2]{#2}%
  \@ifundefined{ifGPcolor}{%
    \newif\ifGPcolor
    \GPcolorfalse
  }{}%
  \@ifundefined{ifGPblacktext}{%
    \newif\ifGPblacktext
    \GPblacktextfalse
  }{}%
  \let\gplgaddtomacro\g@addto@macro
  \gdef\gplbacktext{}%
  \gdef\gplfronttext{}%
  \makeatother
  \ifGPblacktext
    \def\colorrgb#1{}%
    \def\colorgray#1{}%
  \else
    \ifGPcolor
      \def\colorrgb#1{\color[rgb]{#1}}%
      \def\colorgray#1{\color[gray]{#1}}%
      \expandafter\def\csname LTw\endcsname{\color{white}}%
      \expandafter\def\csname LTb\endcsname{\color{black}}%
      \expandafter\def\csname LTa\endcsname{\color{black}}%
      \expandafter\def\csname LT0\endcsname{\color[rgb]{1,0,0}}%
      \expandafter\def\csname LT1\endcsname{\color[rgb]{0,1,0}}%
      \expandafter\def\csname LT2\endcsname{\color[rgb]{0,0,1}}%
      \expandafter\def\csname LT3\endcsname{\color[rgb]{1,0,1}}%
      \expandafter\def\csname LT4\endcsname{\color[rgb]{0,1,1}}%
      \expandafter\def\csname LT5\endcsname{\color[rgb]{1,1,0}}%
      \expandafter\def\csname LT6\endcsname{\color[rgb]{0,0,0}}%
      \expandafter\def\csname LT7\endcsname{\color[rgb]{1,0.3,0}}%
      \expandafter\def\csname LT8\endcsname{\color[rgb]{0.5,0.5,0.5}}%
    \else
      \def\colorrgb#1{\color{black}}%
      \def\colorgray#1{\color[gray]{#1}}%
      \expandafter\def\csname LTw\endcsname{\color{white}}%
      \expandafter\def\csname LTb\endcsname{\color{black}}%
      \expandafter\def\csname LTa\endcsname{\color{black}}%
      \expandafter\def\csname LT0\endcsname{\color{black}}%
      \expandafter\def\csname LT1\endcsname{\color{black}}%
      \expandafter\def\csname LT2\endcsname{\color{black}}%
      \expandafter\def\csname LT3\endcsname{\color{black}}%
      \expandafter\def\csname LT4\endcsname{\color{black}}%
      \expandafter\def\csname LT5\endcsname{\color{black}}%
      \expandafter\def\csname LT6\endcsname{\color{black}}%
      \expandafter\def\csname LT7\endcsname{\color{black}}%
      \expandafter\def\csname LT8\endcsname{\color{black}}%
    \fi
  \fi
    \setlength{\unitlength}{0.0500bp}%
    \ifx\gptboxheight\undefined%
      \newlength{\gptboxheight}%
      \newlength{\gptboxwidth}%
      \newsavebox{\gptboxtext}%
    \fi%
    \setlength{\fboxrule}{0.5pt}%
    \setlength{\fboxsep}{1pt}%
\begin{picture}(5102.00,3570.00)%
    \gplgaddtomacro\gplbacktext{%
      \csname LTb\endcsname%
      \put(633,499){\makebox(0,0)[r]{\strut{}   8}}%
      \put(633,963){\makebox(0,0)[r]{\strut{}   9}}%
      \put(633,1427){\makebox(0,0)[r]{\strut{}  10}}%
      \put(633,1891){\makebox(0,0)[r]{\strut{}  11}}%
      \put(765,279){\makebox(0,0){\strut{}$-1$}}%
      \put(1612,279){\makebox(0,0){\strut{}$-0.8$}}%
      \put(2458,279){\makebox(0,0){\strut{}$-0.6$}}%
      \put(3305,279){\makebox(0,0){\strut{}$-0.4$}}%
      \put(4151,279){\makebox(0,0){\strut{}$-0.2$}}%
      \put(4998,279){\makebox(0,0){\strut{}$0$}}%
    }%
    \gplgaddtomacro\gplfronttext{%
      \csname LTb\endcsname%
      \put(127,1195){\rotatebox{-270}{\makebox(0,0){\strut{}$i$ $(\degree)$}}}%
      \put(2881,-51){\makebox(0,0){\strut{}time $(\Myr)$}}%
      \csname LTb\endcsname%
      \put(4786,1780){\makebox(0,0)[l]{\strut{}b)}}%
    }%
    \gplgaddtomacro\gplbacktext{%
      \csname LTb\endcsname%
      \put(633,2106){\makebox(0,0)[r]{\strut{}0.06}}%
      \put(633,2570){\makebox(0,0)[r]{\strut{}0.10}}%
      \put(633,3033){\makebox(0,0)[r]{\strut{}0.14}}%
      \put(633,3497){\makebox(0,0)[r]{\strut{}0.18}}%
      \put(765,1886){\makebox(0,0){\strut{}}}%
      \put(1612,1886){\makebox(0,0){\strut{}}}%
      \put(2458,1886){\makebox(0,0){\strut{}}}%
      \put(3305,1886){\makebox(0,0){\strut{}}}%
      \put(4151,1886){\makebox(0,0){\strut{}}}%
      \put(4998,1886){\makebox(0,0){\strut{}}}%
    }%
    \gplgaddtomacro\gplfronttext{%
      \csname LTb\endcsname%
      \put(127,2801){\rotatebox{-270}{\makebox(0,0){\strut{}$e$}}}%
      \put(2881,1820){\makebox(0,0){\strut{}}}%
      \csname LTb\endcsname%
      \put(4786,3386){\makebox(0,0)[l]{\strut{}a)}}%
    }%
    \gplbacktext
    \put(0,0){\includegraphics{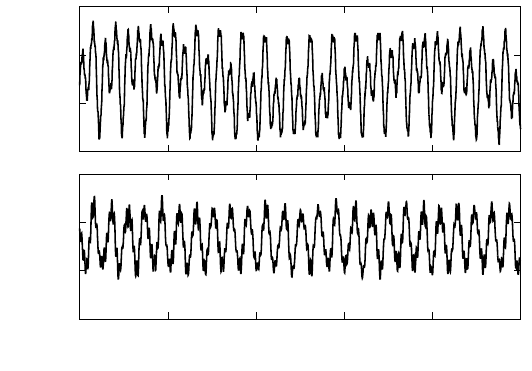}}%
    \gplfronttext
  \end{picture}%
\endgroup

%% file: figures/tex_figure3.tex
\begingroup
  \makeatletter
  \providecommand\color[2][]{%
    \GenericError{(gnuplot) \space\space\space\@spaces}{%
      Package color not loaded in conjunction with
      terminal option `colourtext'%
    }{See the gnuplot documentation for explanation.%
    }{Either use 'blacktext' in gnuplot or load the package
      color.sty in LaTeX.}%
    \renewcommand\color[2][]{}%
  }%
  \providecommand\includegraphics[2][]{%
    \GenericError{(gnuplot) \space\space\space\@spaces}{%
      Package graphicx or graphics not loaded%
    }{See the gnuplot documentation for explanation.%
    }{The gnuplot epslatex terminal needs graphicx.sty or graphics.sty.}%
    \renewcommand\includegraphics[2][]{}%
  }%
  \providecommand\rotatebox[2]{#2}%
  \@ifundefined{ifGPcolor}{%
    \newif\ifGPcolor
    \GPcolorfalse
  }{}%
  \@ifundefined{ifGPblacktext}{%
    \newif\ifGPblacktext
    \GPblacktextfalse
  }{}%
  \let\gplgaddtomacro\g@addto@macro
  \gdef\gplbacktext{}%
  \gdef\gplfronttext{}%
  \makeatother
  \ifGPblacktext
    \def\colorrgb#1{}%
    \def\colorgray#1{}%
  \else
    \ifGPcolor
      \def\colorrgb#1{\color[rgb]{#1}}%
      \def\colorgray#1{\color[gray]{#1}}%
      \expandafter\def\csname LTw\endcsname{\color{white}}%
      \expandafter\def\csname LTb\endcsname{\color{black}}%
      \expandafter\def\csname LTa\endcsname{\color{black}}%
      \expandafter\def\csname LT0\endcsname{\color[rgb]{1,0,0}}%
      \expandafter\def\csname LT1\endcsname{\color[rgb]{0,1,0}}%
      \expandafter\def\csname LT2\endcsname{\color[rgb]{0,0,1}}%
      \expandafter\def\csname LT3\endcsname{\color[rgb]{1,0,1}}%
      \expandafter\def\csname LT4\endcsname{\color[rgb]{0,1,1}}%
      \expandafter\def\csname LT5\endcsname{\color[rgb]{1,1,0}}%
      \expandafter\def\csname LT6\endcsname{\color[rgb]{0,0,0}}%
      \expandafter\def\csname LT7\endcsname{\color[rgb]{1,0.3,0}}%
      \expandafter\def\csname LT8\endcsname{\color[rgb]{0.5,0.5,0.5}}%
    \else
      \def\colorrgb#1{\color{black}}%
      \def\colorgray#1{\color[gray]{#1}}%
      \expandafter\def\csname LTw\endcsname{\color{white}}%
      \expandafter\def\csname LTb\endcsname{\color{black}}%
      \expandafter\def\csname LTa\endcsname{\color{black}}%
      \expandafter\def\csname LT0\endcsname{\color{black}}%
      \expandafter\def\csname LT1\endcsname{\color{black}}%
      \expandafter\def\csname LT2\endcsname{\color{black}}%
      \expandafter\def\csname LT3\endcsname{\color{black}}%
      \expandafter\def\csname LT4\endcsname{\color{black}}%
      \expandafter\def\csname LT5\endcsname{\color{black}}%
      \expandafter\def\csname LT6\endcsname{\color{black}}%
      \expandafter\def\csname LT7\endcsname{\color{black}}%
      \expandafter\def\csname LT8\endcsname{\color{black}}%
    \fi
  \fi
    \setlength{\unitlength}{0.0500bp}%
    \ifx\gptboxheight\undefined%
      \newlength{\gptboxheight}%
      \newlength{\gptboxwidth}%
      \newsavebox{\gptboxtext}%
    \fi%
    \setlength{\fboxrule}{0.5pt}%
    \setlength{\fboxsep}{1pt}%
\begin{picture}(5102.00,3570.00)%
    \gplgaddtomacro\gplbacktext{%
      \csname LTb\endcsname%
      \put(633,499){\makebox(0,0)[r]{\strut{}   5}}%
      \put(633,963){\makebox(0,0)[r]{\strut{}   6}}%
      \put(633,1427){\makebox(0,0)[r]{\strut{}   7}}%
      \put(633,1891){\makebox(0,0)[r]{\strut{}   8}}%
      \put(765,279){\makebox(0,0){\strut{}$-1$}}%
      \put(1612,279){\makebox(0,0){\strut{}$-0.8$}}%
      \put(2458,279){\makebox(0,0){\strut{}$-0.6$}}%
      \put(3305,279){\makebox(0,0){\strut{}$-0.4$}}%
      \put(4151,279){\makebox(0,0){\strut{}$-0.2$}}%
      \put(4998,279){\makebox(0,0){\strut{}$0$}}%
    }%
    \gplgaddtomacro\gplfronttext{%
      \csname LTb\endcsname%
      \put(127,1195){\rotatebox{-270}{\makebox(0,0){\strut{}$i$ $(\degree)$}}}%
      \put(2881,-51){\makebox(0,0){\strut{}time $(\Myr)$}}%
      \csname LTb\endcsname%
      \put(4786,1780){\makebox(0,0)[l]{\strut{}b)}}%
    }%
    \gplgaddtomacro\gplbacktext{%
      \csname LTb\endcsname%
      \put(633,2106){\makebox(0,0)[r]{\strut{}0.04}}%
      \put(633,2570){\makebox(0,0)[r]{\strut{}0.08}}%
      \put(633,3033){\makebox(0,0)[r]{\strut{}0.12}}%
      \put(633,3497){\makebox(0,0)[r]{\strut{}0.16}}%
      \put(765,1886){\makebox(0,0){\strut{}}}%
      \put(1612,1886){\makebox(0,0){\strut{}}}%
      \put(2458,1886){\makebox(0,0){\strut{}}}%
      \put(3305,1886){\makebox(0,0){\strut{}}}%
      \put(4151,1886){\makebox(0,0){\strut{}}}%
      \put(4998,1886){\makebox(0,0){\strut{}}}%
    }%
    \gplgaddtomacro\gplfronttext{%
      \csname LTb\endcsname%
      \put(127,2801){\rotatebox{-270}{\makebox(0,0){\strut{}$e$}}}%
      \put(2881,1820){\makebox(0,0){\strut{}}}%
      \csname LTb\endcsname%
      \put(4786,3386){\makebox(0,0)[l]{\strut{}a)}}%
    }%
    \gplbacktext
    \put(0,0){\includegraphics{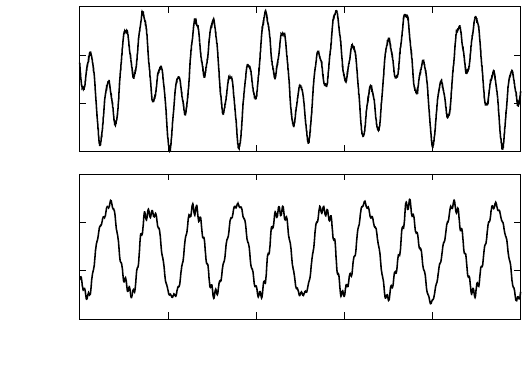}}%
    \gplfronttext
  \end{picture}%
\endgroup

%% file: figures/tex_figure4.tex
\begingroup
  \makeatletter
  \providecommand\color[2][]{%
    \GenericError{(gnuplot) \space\space\space\@spaces}{%
      Package color not loaded in conjunction with
      terminal option `colourtext'%
    }{See the gnuplot documentation for explanation.%
    }{Either use 'blacktext' in gnuplot or load the package
      color.sty in LaTeX.}%
    \renewcommand\color[2][]{}%
  }%
  \providecommand\includegraphics[2][]{%
    \GenericError{(gnuplot) \space\space\space\@spaces}{%
      Package graphicx or graphics not loaded%
    }{See the gnuplot documentation for explanation.%
    }{The gnuplot epslatex terminal needs graphicx.sty or graphics.sty.}%
    \renewcommand\includegraphics[2][]{}%
  }%
  \providecommand\rotatebox[2]{#2}%
  \@ifundefined{ifGPcolor}{%
    \newif\ifGPcolor
    \GPcolorfalse
  }{}%
  \@ifundefined{ifGPblacktext}{%
    \newif\ifGPblacktext
    \GPblacktextfalse
  }{}%
  \let\gplgaddtomacro\g@addto@macro
  \gdef\gplbacktext{}%
  \gdef\gplfronttext{}%
  \makeatother
  \ifGPblacktext
    \def\colorrgb#1{}%
    \def\colorgray#1{}%
  \else
    \ifGPcolor
      \def\colorrgb#1{\color[rgb]{#1}}%
      \def\colorgray#1{\color[gray]{#1}}%
      \expandafter\def\csname LTw\endcsname{\color{white}}%
      \expandafter\def\csname LTb\endcsname{\color{black}}%
      \expandafter\def\csname LTa\endcsname{\color{black}}%
      \expandafter\def\csname LT0\endcsname{\color[rgb]{1,0,0}}%
      \expandafter\def\csname LT1\endcsname{\color[rgb]{0,1,0}}%
      \expandafter\def\csname LT2\endcsname{\color[rgb]{0,0,1}}%
      \expandafter\def\csname LT3\endcsname{\color[rgb]{1,0,1}}%
      \expandafter\def\csname LT4\endcsname{\color[rgb]{0,1,1}}%
      \expandafter\def\csname LT5\endcsname{\color[rgb]{1,1,0}}%
      \expandafter\def\csname LT6\endcsname{\color[rgb]{0,0,0}}%
      \expandafter\def\csname LT7\endcsname{\color[rgb]{1,0.3,0}}%
      \expandafter\def\csname LT8\endcsname{\color[rgb]{0.5,0.5,0.5}}%
    \else
      \def\colorrgb#1{\color{black}}%
      \def\colorgray#1{\color[gray]{#1}}%
      \expandafter\def\csname LTw\endcsname{\color{white}}%
      \expandafter\def\csname LTb\endcsname{\color{black}}%
      \expandafter\def\csname LTa\endcsname{\color{black}}%
      \expandafter\def\csname LT0\endcsname{\color{black}}%
      \expandafter\def\csname LT1\endcsname{\color{black}}%
      \expandafter\def\csname LT2\endcsname{\color{black}}%
      \expandafter\def\csname LT3\endcsname{\color{black}}%
      \expandafter\def\csname LT4\endcsname{\color{black}}%
      \expandafter\def\csname LT5\endcsname{\color{black}}%
      \expandafter\def\csname LT6\endcsname{\color{black}}%
      \expandafter\def\csname LT7\endcsname{\color{black}}%
      \expandafter\def\csname LT8\endcsname{\color{black}}%
    \fi
  \fi
    \setlength{\unitlength}{0.0500bp}%
    \ifx\gptboxheight\undefined%
      \newlength{\gptboxheight}%
      \newlength{\gptboxwidth}%
      \newsavebox{\gptboxtext}%
    \fi%
    \setlength{\fboxrule}{0.5pt}%
    \setlength{\fboxsep}{1pt}%
\begin{picture}(5102.00,3570.00)%
    \gplgaddtomacro\gplbacktext{%
      \csname LTb\endcsname%
      \put(888,499){\makebox(0,0)[r]{\strut{}  -0.4}}%
      \put(888,1195){\makebox(0,0)[r]{\strut{}     0}}%
      \put(888,1891){\makebox(0,0)[r]{\strut{}   0.4}}%
      \put(1020,279){\makebox(0,0){\strut{}$-20$}}%
      \put(2015,279){\makebox(0,0){\strut{}$-15$}}%
      \put(3009,279){\makebox(0,0){\strut{}$-10$}}%
      \put(4004,279){\makebox(0,0){\strut{}$-5$}}%
      \put(4998,279){\makebox(0,0){\strut{}$0$}}%
    }%
    \gplgaddtomacro\gplfronttext{%
      \csname LTb\endcsname%
      \put(118,1195){\rotatebox{-270}{\makebox(0,0){\strut{}$\delta i$ $(\degree)$}}}%
      \put(3009,-51){\makebox(0,0){\strut{}time $(\Myr)$}}%
      \csname LTb\endcsname%
      \put(4799,1780){\makebox(0,0)[l]{\strut{}b)}}%
    }%
    \gplgaddtomacro\gplbacktext{%
      \csname LTb\endcsname%
      \put(888,2106){\makebox(0,0)[r]{\strut{}-0.004}}%
      \put(888,2802){\makebox(0,0)[r]{\strut{}0}}%
      \put(888,3497){\makebox(0,0)[r]{\strut{}0.004}}%
      \put(1020,1886){\makebox(0,0){\strut{}}}%
      \put(2015,1886){\makebox(0,0){\strut{}}}%
      \put(3009,1886){\makebox(0,0){\strut{}}}%
      \put(4004,1886){\makebox(0,0){\strut{}}}%
      \put(4998,1886){\makebox(0,0){\strut{}}}%
    }%
    \gplgaddtomacro\gplfronttext{%
      \csname LTb\endcsname%
      \put(118,2801){\rotatebox{-270}{\makebox(0,0){\strut{}$\delta e$}}}%
      \put(3009,1820){\makebox(0,0){\strut{}}}%
      \csname LTb\endcsname%
      \put(4799,3386){\makebox(0,0)[l]{\strut{}a)}}%
    }%
    \gplbacktext
    \put(0,0){\includegraphics{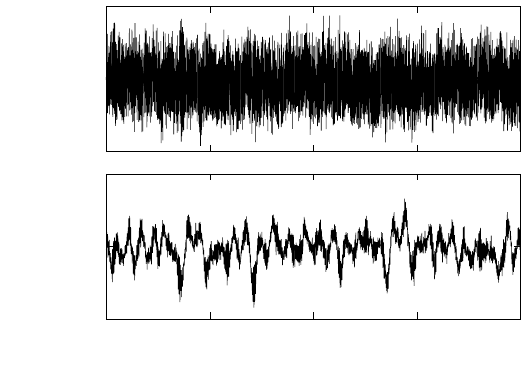}}%
    \gplfronttext
  \end{picture}%
\endgroup

%% file: figures/tex_figure5.tex
\begingroup
  \makeatletter
  \providecommand\color[2][]{%
    \GenericError{(gnuplot) \space\space\space\@spaces}{%
      Package color not loaded in conjunction with
      terminal option `colourtext'%
    }{See the gnuplot documentation for explanation.%
    }{Either use 'blacktext' in gnuplot or load the package
      color.sty in LaTeX.}%
    \renewcommand\color[2][]{}%
  }%
  \providecommand\includegraphics[2][]{%
    \GenericError{(gnuplot) \space\space\space\@spaces}{%
      Package graphicx or graphics not loaded%
    }{See the gnuplot documentation for explanation.%
    }{The gnuplot epslatex terminal needs graphicx.sty or graphics.sty.}%
    \renewcommand\includegraphics[2][]{}%
  }%
  \providecommand\rotatebox[2]{#2}%
  \@ifundefined{ifGPcolor}{%
    \newif\ifGPcolor
    \GPcolorfalse
  }{}%
  \@ifundefined{ifGPblacktext}{%
    \newif\ifGPblacktext
    \GPblacktextfalse
  }{}%
  \let\gplgaddtomacro\g@addto@macro
  \gdef\gplbacktext{}%
  \gdef\gplfronttext{}%
  \makeatother
  \ifGPblacktext
    \def\colorrgb#1{}%
    \def\colorgray#1{}%
  \else
    \ifGPcolor
      \def\colorrgb#1{\color[rgb]{#1}}%
      \def\colorgray#1{\color[gray]{#1}}%
      \expandafter\def\csname LTw\endcsname{\color{white}}%
      \expandafter\def\csname LTb\endcsname{\color{black}}%
      \expandafter\def\csname LTa\endcsname{\color{black}}%
      \expandafter\def\csname LT0\endcsname{\color[rgb]{1,0,0}}%
      \expandafter\def\csname LT1\endcsname{\color[rgb]{0,1,0}}%
      \expandafter\def\csname LT2\endcsname{\color[rgb]{0,0,1}}%
      \expandafter\def\csname LT3\endcsname{\color[rgb]{1,0,1}}%
      \expandafter\def\csname LT4\endcsname{\color[rgb]{0,1,1}}%
      \expandafter\def\csname LT5\endcsname{\color[rgb]{1,1,0}}%
      \expandafter\def\csname LT6\endcsname{\color[rgb]{0,0,0}}%
      \expandafter\def\csname LT7\endcsname{\color[rgb]{1,0.3,0}}%
      \expandafter\def\csname LT8\endcsname{\color[rgb]{0.5,0.5,0.5}}%
    \else
      \def\colorrgb#1{\color{black}}%
      \def\colorgray#1{\color[gray]{#1}}%
      \expandafter\def\csname LTw\endcsname{\color{white}}%
      \expandafter\def\csname LTb\endcsname{\color{black}}%
      \expandafter\def\csname LTa\endcsname{\color{black}}%
      \expandafter\def\csname LT0\endcsname{\color{black}}%
      \expandafter\def\csname LT1\endcsname{\color{black}}%
      \expandafter\def\csname LT2\endcsname{\color{black}}%
      \expandafter\def\csname LT3\endcsname{\color{black}}%
      \expandafter\def\csname LT4\endcsname{\color{black}}%
      \expandafter\def\csname LT5\endcsname{\color{black}}%
      \expandafter\def\csname LT6\endcsname{\color{black}}%
      \expandafter\def\csname LT7\endcsname{\color{black}}%
      \expandafter\def\csname LT8\endcsname{\color{black}}%
    \fi
  \fi
    \setlength{\unitlength}{0.0500bp}%
    \ifx\gptboxheight\undefined%
      \newlength{\gptboxheight}%
      \newlength{\gptboxwidth}%
      \newsavebox{\gptboxtext}%
    \fi%
    \setlength{\fboxrule}{0.5pt}%
    \setlength{\fboxsep}{1pt}%
\begin{picture}(5102.00,3570.00)%
    \gplgaddtomacro\gplbacktext{%
      \csname LTb\endcsname%
      \put(888,499){\makebox(0,0)[r]{\strut{}  -0.2}}%
      \put(888,1195){\makebox(0,0)[r]{\strut{}     0}}%
      \put(888,1891){\makebox(0,0)[r]{\strut{}   0.2}}%
      \put(1020,279){\makebox(0,0){\strut{}$-20$}}%
      \put(2015,279){\makebox(0,0){\strut{}$-15$}}%
      \put(3009,279){\makebox(0,0){\strut{}$-10$}}%
      \put(4004,279){\makebox(0,0){\strut{}$-5$}}%
      \put(4998,279){\makebox(0,0){\strut{}$0$}}%
    }%
    \gplgaddtomacro\gplfronttext{%
      \csname LTb\endcsname%
      \put(118,1195){\rotatebox{-270}{\makebox(0,0){\strut{}$\delta i$ $(\degree)$}}}%
      \put(3009,-51){\makebox(0,0){\strut{}time $(\Myr)$}}%
      \csname LTb\endcsname%
      \put(4799,1780){\makebox(0,0)[l]{\strut{}b)}}%
    }%
    \gplgaddtomacro\gplbacktext{%
      \csname LTb\endcsname%
      \put(888,2106){\makebox(0,0)[r]{\strut{}-0.004}}%
      \put(888,2802){\makebox(0,0)[r]{\strut{}0}}%
      \put(888,3497){\makebox(0,0)[r]{\strut{}0.004}}%
      \put(1020,1886){\makebox(0,0){\strut{}}}%
      \put(2015,1886){\makebox(0,0){\strut{}}}%
      \put(3009,1886){\makebox(0,0){\strut{}}}%
      \put(4004,1886){\makebox(0,0){\strut{}}}%
      \put(4998,1886){\makebox(0,0){\strut{}}}%
    }%
    \gplgaddtomacro\gplfronttext{%
      \csname LTb\endcsname%
      \put(118,2801){\rotatebox{-270}{\makebox(0,0){\strut{}$\delta e$}}}%
      \put(3009,1820){\makebox(0,0){\strut{}}}%
      \csname LTb\endcsname%
      \put(4799,3386){\makebox(0,0)[l]{\strut{}a)}}%
    }%
    \gplbacktext
    \put(0,0){\includegraphics{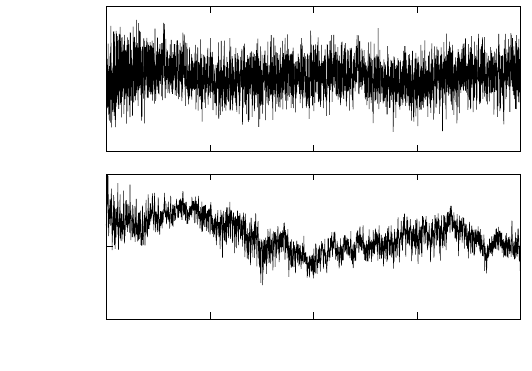}}%
    \gplfronttext
  \end{picture}%
\endgroup

%% file: figures/tex_figure6.tex
\begingroup
  \makeatletter
  \providecommand\color[2][]{%
    \GenericError{(gnuplot) \space\space\space\@spaces}{%
      Package color not loaded in conjunction with
      terminal option `colourtext'%
    }{See the gnuplot documentation for explanation.%
    }{Either use 'blacktext' in gnuplot or load the package
      color.sty in LaTeX.}%
    \renewcommand\color[2][]{}%
  }%
  \providecommand\includegraphics[2][]{%
    \GenericError{(gnuplot) \space\space\space\@spaces}{%
      Package graphicx or graphics not loaded%
    }{See the gnuplot documentation for explanation.%
    }{The gnuplot epslatex terminal needs graphicx.sty or graphics.sty.}%
    \renewcommand\includegraphics[2][]{}%
  }%
  \providecommand\rotatebox[2]{#2}%
  \@ifundefined{ifGPcolor}{%
    \newif\ifGPcolor
    \GPcolorfalse
  }{}%
  \@ifundefined{ifGPblacktext}{%
    \newif\ifGPblacktext
    \GPblacktextfalse
  }{}%
  \let\gplgaddtomacro\g@addto@macro
  \gdef\gplbacktext{}%
  \gdef\gplfronttext{}%
  \makeatother
  \ifGPblacktext
    \def\colorrgb#1{}%
    \def\colorgray#1{}%
  \else
    \ifGPcolor
      \def\colorrgb#1{\color[rgb]{#1}}%
      \def\colorgray#1{\color[gray]{#1}}%
      \expandafter\def\csname LTw\endcsname{\color{white}}%
      \expandafter\def\csname LTb\endcsname{\color{black}}%
      \expandafter\def\csname LTa\endcsname{\color{black}}%
      \expandafter\def\csname LT0\endcsname{\color[rgb]{1,0,0}}%
      \expandafter\def\csname LT1\endcsname{\color[rgb]{0,1,0}}%
      \expandafter\def\csname LT2\endcsname{\color[rgb]{0,0,1}}%
      \expandafter\def\csname LT3\endcsname{\color[rgb]{1,0,1}}%
      \expandafter\def\csname LT4\endcsname{\color[rgb]{0,1,1}}%
      \expandafter\def\csname LT5\endcsname{\color[rgb]{1,1,0}}%
      \expandafter\def\csname LT6\endcsname{\color[rgb]{0,0,0}}%
      \expandafter\def\csname LT7\endcsname{\color[rgb]{1,0.3,0}}%
      \expandafter\def\csname LT8\endcsname{\color[rgb]{0.5,0.5,0.5}}%
    \else
      \def\colorrgb#1{\color{black}}%
      \def\colorgray#1{\color[gray]{#1}}%
      \expandafter\def\csname LTw\endcsname{\color{white}}%
      \expandafter\def\csname LTb\endcsname{\color{black}}%
      \expandafter\def\csname LTa\endcsname{\color{black}}%
      \expandafter\def\csname LT0\endcsname{\color{black}}%
      \expandafter\def\csname LT1\endcsname{\color{black}}%
      \expandafter\def\csname LT2\endcsname{\color{black}}%
      \expandafter\def\csname LT3\endcsname{\color{black}}%
      \expandafter\def\csname LT4\endcsname{\color{black}}%
      \expandafter\def\csname LT5\endcsname{\color{black}}%
      \expandafter\def\csname LT6\endcsname{\color{black}}%
      \expandafter\def\csname LT7\endcsname{\color{black}}%
      \expandafter\def\csname LT8\endcsname{\color{black}}%
    \fi
  \fi
    \setlength{\unitlength}{0.0500bp}%
    \ifx\gptboxheight\undefined%
      \newlength{\gptboxheight}%
      \newlength{\gptboxwidth}%
      \newsavebox{\gptboxtext}%
    \fi%
    \setlength{\fboxrule}{0.5pt}%
    \setlength{\fboxsep}{1pt}%
\begin{picture}(5102.00,3570.00)%
    \gplgaddtomacro\gplbacktext{%
      \csname LTb\endcsname%
      \put(888,499){\makebox(0,0)[r]{\strut{}-59.27}}%
      \put(888,698){\makebox(0,0)[r]{\strut{}-59.26}}%
      \put(888,897){\makebox(0,0)[r]{\strut{}-59.25}}%
      \put(888,1096){\makebox(0,0)[r]{\strut{}-59.24}}%
      \put(888,1294){\makebox(0,0)[r]{\strut{}-59.23}}%
      \put(888,1493){\makebox(0,0)[r]{\strut{}-59.22}}%
      \put(888,1692){\makebox(0,0)[r]{\strut{}-59.21}}%
      \put(888,1891){\makebox(0,0)[r]{\strut{}-59.20}}%
      \put(1020,279){\makebox(0,0){\strut{}$-250$}}%
      \put(2002,279){\makebox(0,0){\strut{}$-125$}}%
      \put(2984,279){\makebox(0,0){\strut{}$0$}}%
      \put(3965,279){\makebox(0,0){\strut{}$125$}}%
      \put(4947,279){\makebox(0,0){\strut{}$250$}}%
    }%
    \gplgaddtomacro\gplfronttext{%
      \csname LTb\endcsname%
      \put(118,1195){\rotatebox{-270}{\makebox(0,0){\strut{}$s_C$ $(\arc)$}}}%
      \put(2983,-51){\makebox(0,0){\strut{}time $(\Myr)$}}%
      \csname LTb\endcsname%
      \put(4751,1780){\makebox(0,0)[l]{\strut{}b)}}%
    }%
    \gplgaddtomacro\gplbacktext{%
      \csname LTb\endcsname%
      \put(888,2106){\makebox(0,0)[r]{\strut{} 54.22}}%
      \put(888,2384){\makebox(0,0)[r]{\strut{} 54.23}}%
      \put(888,2662){\makebox(0,0)[r]{\strut{} 54.24}}%
      \put(888,2941){\makebox(0,0)[r]{\strut{} 54.25}}%
      \put(888,3219){\makebox(0,0)[r]{\strut{} 54.26}}%
      \put(888,3497){\makebox(0,0)[r]{\strut{} 54.27}}%
      \put(1020,1886){\makebox(0,0){\strut{}}}%
      \put(2002,1886){\makebox(0,0){\strut{}}}%
      \put(2984,1886){\makebox(0,0){\strut{}}}%
      \put(3965,1886){\makebox(0,0){\strut{}}}%
      \put(4947,1886){\makebox(0,0){\strut{}}}%
    }%
    \gplgaddtomacro\gplfronttext{%
      \csname LTb\endcsname%
      \put(118,2801){\rotatebox{-270}{\makebox(0,0){\strut{}$g_C$ $(\arc)$}}}%
      \put(2983,1820){\makebox(0,0){\strut{}}}%
      \csname LTb\endcsname%
      \put(4751,3386){\makebox(0,0)[l]{\strut{}a)}}%
    }%
    \gplbacktext
    \put(0,0){\includegraphics{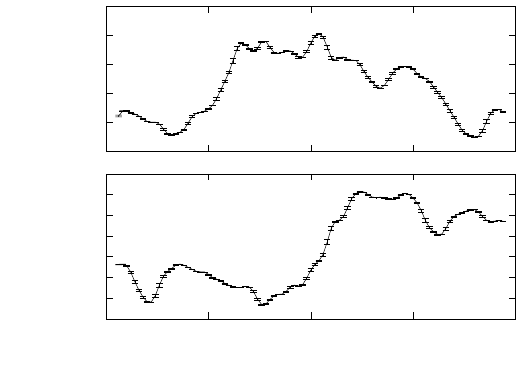}}%
    \gplfronttext
  \end{picture}%
\endgroup

%% file: figures/tex_figure7.tex
\begingroup
  \makeatletter
  \providecommand\color[2][]{%
    \GenericError{(gnuplot) \space\space\space\@spaces}{%
      Package color not loaded in conjunction with
      terminal option `colourtext'%
    }{See the gnuplot documentation for explanation.%
    }{Either use 'blacktext' in gnuplot or load the package
      color.sty in LaTeX.}%
    \renewcommand\color[2][]{}%
  }%
  \providecommand\includegraphics[2][]{%
    \GenericError{(gnuplot) \space\space\space\@spaces}{%
      Package graphicx or graphics not loaded%
    }{See the gnuplot documentation for explanation.%
    }{The gnuplot epslatex terminal needs graphicx.sty or graphics.sty.}%
    \renewcommand\includegraphics[2][]{}%
  }%
  \providecommand\rotatebox[2]{#2}%
  \@ifundefined{ifGPcolor}{%
    \newif\ifGPcolor
    \GPcolorfalse
  }{}%
  \@ifundefined{ifGPblacktext}{%
    \newif\ifGPblacktext
    \GPblacktextfalse
  }{}%
  \let\gplgaddtomacro\g@addto@macro
  \gdef\gplbacktext{}%
  \gdef\gplfronttext{}%
  \makeatother
  \ifGPblacktext
    \def\colorrgb#1{}%
    \def\colorgray#1{}%
  \else
    \ifGPcolor
      \def\colorrgb#1{\color[rgb]{#1}}%
      \def\colorgray#1{\color[gray]{#1}}%
      \expandafter\def\csname LTw\endcsname{\color{white}}%
      \expandafter\def\csname LTb\endcsname{\color{black}}%
      \expandafter\def\csname LTa\endcsname{\color{black}}%
      \expandafter\def\csname LT0\endcsname{\color[rgb]{1,0,0}}%
      \expandafter\def\csname LT1\endcsname{\color[rgb]{0,1,0}}%
      \expandafter\def\csname LT2\endcsname{\color[rgb]{0,0,1}}%
      \expandafter\def\csname LT3\endcsname{\color[rgb]{1,0,1}}%
      \expandafter\def\csname LT4\endcsname{\color[rgb]{0,1,1}}%
      \expandafter\def\csname LT5\endcsname{\color[rgb]{1,1,0}}%
      \expandafter\def\csname LT6\endcsname{\color[rgb]{0,0,0}}%
      \expandafter\def\csname LT7\endcsname{\color[rgb]{1,0.3,0}}%
      \expandafter\def\csname LT8\endcsname{\color[rgb]{0.5,0.5,0.5}}%
    \else
      \def\colorrgb#1{\color{black}}%
      \def\colorgray#1{\color[gray]{#1}}%
      \expandafter\def\csname LTw\endcsname{\color{white}}%
      \expandafter\def\csname LTb\endcsname{\color{black}}%
      \expandafter\def\csname LTa\endcsname{\color{black}}%
      \expandafter\def\csname LT0\endcsname{\color{black}}%
      \expandafter\def\csname LT1\endcsname{\color{black}}%
      \expandafter\def\csname LT2\endcsname{\color{black}}%
      \expandafter\def\csname LT3\endcsname{\color{black}}%
      \expandafter\def\csname LT4\endcsname{\color{black}}%
      \expandafter\def\csname LT5\endcsname{\color{black}}%
      \expandafter\def\csname LT6\endcsname{\color{black}}%
      \expandafter\def\csname LT7\endcsname{\color{black}}%
      \expandafter\def\csname LT8\endcsname{\color{black}}%
    \fi
  \fi
    \setlength{\unitlength}{0.0500bp}%
    \ifx\gptboxheight\undefined%
      \newlength{\gptboxheight}%
      \newlength{\gptboxwidth}%
      \newsavebox{\gptboxtext}%
    \fi%
    \setlength{\fboxrule}{0.5pt}%
    \setlength{\fboxsep}{1pt}%
\begin{picture}(5102.00,3570.00)%
    \gplgaddtomacro\gplbacktext{%
      \csname LTb\endcsname%
      \put(888,499){\makebox(0,0)[r]{\strut{} -40.1}}%
      \put(888,731){\makebox(0,0)[r]{\strut{} -40.0}}%
      \put(888,963){\makebox(0,0)[r]{\strut{} -39.9}}%
      \put(888,1195){\makebox(0,0)[r]{\strut{} -39.8}}%
      \put(888,1427){\makebox(0,0)[r]{\strut{} -39.7}}%
      \put(888,1659){\makebox(0,0)[r]{\strut{} -39.6}}%
      \put(888,1891){\makebox(0,0)[r]{\strut{} -39.5}}%
      \put(1020,279){\makebox(0,0){\strut{}$-250$}}%
      \put(2002,279){\makebox(0,0){\strut{}$-125$}}%
      \put(2984,279){\makebox(0,0){\strut{}$0$}}%
      \put(3965,279){\makebox(0,0){\strut{}$125$}}%
      \put(4947,279){\makebox(0,0){\strut{}$250$}}%
    }%
    \gplgaddtomacro\gplfronttext{%
      \csname LTb\endcsname%
      \put(118,1195){\rotatebox{-270}{\makebox(0,0){\strut{}$s_V$ $(\arc)$}}}%
      \put(2983,-51){\makebox(0,0){\strut{}time $(\Myr)$}}%
      \csname LTb\endcsname%
      \put(4751,1780){\makebox(0,0)[l]{\strut{}b)}}%
    }%
    \gplgaddtomacro\gplbacktext{%
      \csname LTb\endcsname%
      \put(888,2106){\makebox(0,0)[r]{\strut{} 36.80}}%
      \put(888,2570){\makebox(0,0)[r]{\strut{} 36.85}}%
      \put(888,3033){\makebox(0,0)[r]{\strut{} 36.90}}%
      \put(888,3497){\makebox(0,0)[r]{\strut{} 36.95}}%
      \put(1020,1886){\makebox(0,0){\strut{}}}%
      \put(2002,1886){\makebox(0,0){\strut{}}}%
      \put(2984,1886){\makebox(0,0){\strut{}}}%
      \put(3965,1886){\makebox(0,0){\strut{}}}%
      \put(4947,1886){\makebox(0,0){\strut{}}}%
    }%
    \gplgaddtomacro\gplfronttext{%
      \csname LTb\endcsname%
      \put(118,2801){\rotatebox{-270}{\makebox(0,0){\strut{}$g_V$ $(\arc)$}}}%
      \put(2983,1820){\makebox(0,0){\strut{}}}%
      \csname LTb\endcsname%
      \put(4751,3386){\makebox(0,0)[l]{\strut{}a)}}%
    }%
    \gplbacktext
    \put(0,0){\includegraphics{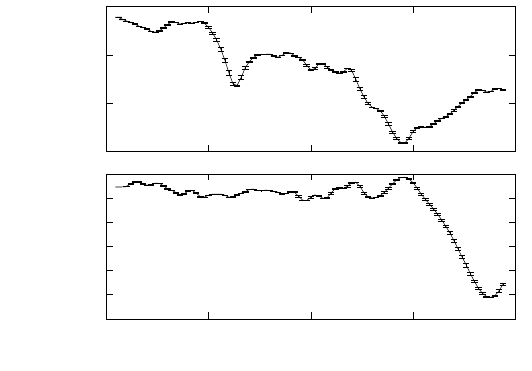}}%
    \gplfronttext
  \end{picture}%
\endgroup

%% file: figures/tex_figure8.tex
\begingroup
  \makeatletter
  \providecommand\color[2][]{%
    \GenericError{(gnuplot) \space\space\space\@spaces}{%
      Package color not loaded in conjunction with
      terminal option `colourtext'%
    }{See the gnuplot documentation for explanation.%
    }{Either use 'blacktext' in gnuplot or load the package
      color.sty in LaTeX.}%
    \renewcommand\color[2][]{}%
  }%
  \providecommand\includegraphics[2][]{%
    \GenericError{(gnuplot) \space\space\space\@spaces}{%
      Package graphicx or graphics not loaded%
    }{See the gnuplot documentation for explanation.%
    }{The gnuplot epslatex terminal needs graphicx.sty or graphics.sty.}%
    \renewcommand\includegraphics[2][]{}%
  }%
  \providecommand\rotatebox[2]{#2}%
  \@ifundefined{ifGPcolor}{%
    \newif\ifGPcolor
    \GPcolorfalse
  }{}%
  \@ifundefined{ifGPblacktext}{%
    \newif\ifGPblacktext
    \GPblacktextfalse
  }{}%
  \let\gplgaddtomacro\g@addto@macro
  \gdef\gplbacktext{}%
  \gdef\gplfronttext{}%
  \makeatother
  \ifGPblacktext
    \def\colorrgb#1{}%
    \def\colorgray#1{}%
  \else
    \ifGPcolor
      \def\colorrgb#1{\color[rgb]{#1}}%
      \def\colorgray#1{\color[gray]{#1}}%
      \expandafter\def\csname LTw\endcsname{\color{white}}%
      \expandafter\def\csname LTb\endcsname{\color{black}}%
      \expandafter\def\csname LTa\endcsname{\color{black}}%
      \expandafter\def\csname LT0\endcsname{\color[rgb]{1,0,0}}%
      \expandafter\def\csname LT1\endcsname{\color[rgb]{0,1,0}}%
      \expandafter\def\csname LT2\endcsname{\color[rgb]{0,0,1}}%
      \expandafter\def\csname LT3\endcsname{\color[rgb]{1,0,1}}%
      \expandafter\def\csname LT4\endcsname{\color[rgb]{0,1,1}}%
      \expandafter\def\csname LT5\endcsname{\color[rgb]{1,1,0}}%
      \expandafter\def\csname LT6\endcsname{\color[rgb]{0,0,0}}%
      \expandafter\def\csname LT7\endcsname{\color[rgb]{1,0.3,0}}%
      \expandafter\def\csname LT8\endcsname{\color[rgb]{0.5,0.5,0.5}}%
    \else
      \def\colorrgb#1{\color{black}}%
      \def\colorgray#1{\color[gray]{#1}}%
      \expandafter\def\csname LTw\endcsname{\color{white}}%
      \expandafter\def\csname LTb\endcsname{\color{black}}%
      \expandafter\def\csname LTa\endcsname{\color{black}}%
      \expandafter\def\csname LT0\endcsname{\color{black}}%
      \expandafter\def\csname LT1\endcsname{\color{black}}%
      \expandafter\def\csname LT2\endcsname{\color{black}}%
      \expandafter\def\csname LT3\endcsname{\color{black}}%
      \expandafter\def\csname LT4\endcsname{\color{black}}%
      \expandafter\def\csname LT5\endcsname{\color{black}}%
      \expandafter\def\csname LT6\endcsname{\color{black}}%
      \expandafter\def\csname LT7\endcsname{\color{black}}%
      \expandafter\def\csname LT8\endcsname{\color{black}}%
    \fi
  \fi
    \setlength{\unitlength}{0.0500bp}%
    \ifx\gptboxheight\undefined%
      \newlength{\gptboxheight}%
      \newlength{\gptboxwidth}%
      \newsavebox{\gptboxtext}%
    \fi%
    \setlength{\fboxrule}{0.5pt}%
    \setlength{\fboxsep}{1pt}%
\begin{picture}(5102.00,6802.00)%
    \gplgaddtomacro\gplbacktext{%
      \csname LTb\endcsname%
      \put(480,544){\makebox(0,0)[r]{\strut{}$0$}}%
      \put(480,850){\makebox(0,0)[r]{\strut{}$5$}}%
      \put(480,1156){\makebox(0,0)[r]{\strut{}$10$}}%
      \put(480,1462){\makebox(0,0)[r]{\strut{}$15$}}%
      \put(480,1768){\makebox(0,0)[r]{\strut{}$20$}}%
      \put(612,324){\makebox(0,0){\strut{}$-20$}}%
      \put(1709,324){\makebox(0,0){\strut{}$-15$}}%
      \put(2805,324){\makebox(0,0){\strut{}$-10$}}%
      \put(3902,324){\makebox(0,0){\strut{}$-5$}}%
      \put(4998,324){\makebox(0,0){\strut{}$0$}}%
    }%
    \gplgaddtomacro\gplfronttext{%
      \csname LTb\endcsname%
      \put(106,1156){\rotatebox{-270}{\makebox(0,0){\strut{}$\epsilon$ $(\degree)$}}}%
      \put(2805,-6){\makebox(0,0){\strut{}time $(\Myr)$}}%
      \csname LTb\endcsname%
      \put(4779,1854){\makebox(0,0)[l]{\strut{}c)}}%
    }%
    \gplgaddtomacro\gplbacktext{%
      \csname LTb\endcsname%
      \put(480,2516){\makebox(0,0)[r]{\strut{}$0$}}%
      \put(480,2822){\makebox(0,0)[r]{\strut{}$5$}}%
      \put(480,3128){\makebox(0,0)[r]{\strut{}$10$}}%
      \put(480,3434){\makebox(0,0)[r]{\strut{}$15$}}%
      \put(480,3740){\makebox(0,0)[r]{\strut{}$20$}}%
      \put(612,2296){\makebox(0,0){\strut{}$-1$}}%
      \put(1489,2296){\makebox(0,0){\strut{}$-0.8$}}%
      \put(2366,2296){\makebox(0,0){\strut{}$-0.6$}}%
      \put(3244,2296){\makebox(0,0){\strut{}$-0.4$}}%
      \put(4121,2296){\makebox(0,0){\strut{}$-0.2$}}%
      \put(4998,2296){\makebox(0,0){\strut{}$0$}}%
    }%
    \gplgaddtomacro\gplfronttext{%
      \csname LTb\endcsname%
      \put(106,3128){\rotatebox{-270}{\makebox(0,0){\strut{}$\epsilon$ $(\degree)$}}}%
      \put(2805,1966){\makebox(0,0){\strut{}time $(\Myr)$}}%
      \csname LTb\endcsname%
      \put(4779,3826){\makebox(0,0)[l]{\strut{}b)}}%
    }%
    \gplgaddtomacro\gplbacktext{%
      \csname LTb\endcsname%
      \put(480,4489){\makebox(0,0)[r]{\strut{}$0$}}%
      \put(480,5050){\makebox(0,0)[r]{\strut{}$5$}}%
      \put(480,5611){\makebox(0,0)[r]{\strut{}$10$}}%
      \put(480,6171){\makebox(0,0)[r]{\strut{}$15$}}%
      \put(480,6732){\makebox(0,0)[r]{\strut{}$20$}}%
      \put(612,4269){\makebox(0,0){\strut{}$-100$}}%
      \put(1489,4269){\makebox(0,0){\strut{}$-80$}}%
      \put(2366,4269){\makebox(0,0){\strut{}$-60$}}%
      \put(3244,4269){\makebox(0,0){\strut{}$-40$}}%
      \put(4121,4269){\makebox(0,0){\strut{}$-20$}}%
      \put(4998,4269){\makebox(0,0){\strut{}$0$}}%
    }%
    \gplgaddtomacro\gplfronttext{%
      \csname LTb\endcsname%
      \put(106,5610){\rotatebox{-270}{\makebox(0,0){\strut{}$\epsilon$ $(\degree)$}}}%
      \put(2805,3939){\makebox(0,0){\strut{}time $(\kyr)$}}%
      \csname LTb\endcsname%
      \put(4779,6620){\makebox(0,0)[l]{\strut{}a)}}%
    }%
    \gplbacktext
    \put(0,0){\includegraphics{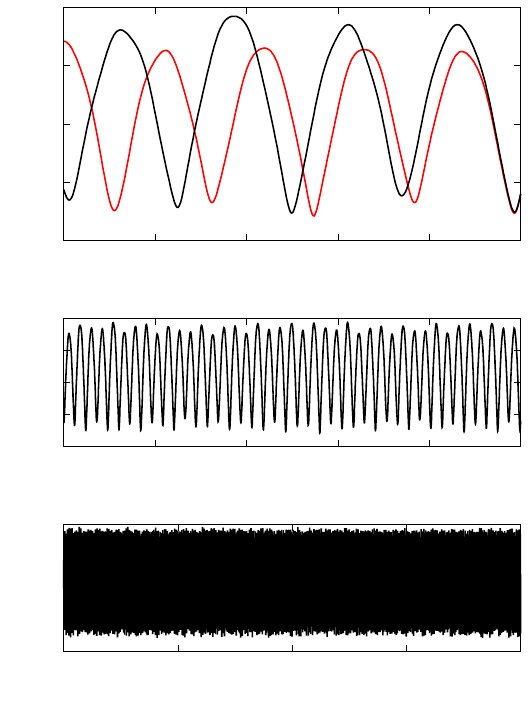}}%
    \gplfronttext
  \end{picture}%
\endgroup

%% file: figures/tex_figure9.tex
\begingroup
  \makeatletter
  \providecommand\color[2][]{%
    \GenericError{(gnuplot) \space\space\space\@spaces}{%
      Package color not loaded in conjunction with
      terminal option `colourtext'%
    }{See the gnuplot documentation for explanation.%
    }{Either use 'blacktext' in gnuplot or load the package
      color.sty in LaTeX.}%
    \renewcommand\color[2][]{}%
  }%
  \providecommand\includegraphics[2][]{%
    \GenericError{(gnuplot) \space\space\space\@spaces}{%
      Package graphicx or graphics not loaded%
    }{See the gnuplot documentation for explanation.%
    }{The gnuplot epslatex terminal needs graphicx.sty or graphics.sty.}%
    \renewcommand\includegraphics[2][]{}%
  }%
  \providecommand\rotatebox[2]{#2}%
  \@ifundefined{ifGPcolor}{%
    \newif\ifGPcolor
    \GPcolorfalse
  }{}%
  \@ifundefined{ifGPblacktext}{%
    \newif\ifGPblacktext
    \GPblacktextfalse
  }{}%
  \let\gplgaddtomacro\g@addto@macro
  \gdef\gplbacktext{}%
  \gdef\gplfronttext{}%
  \makeatother
  \ifGPblacktext
    \def\colorrgb#1{}%
    \def\colorgray#1{}%
  \else
    \ifGPcolor
      \def\colorrgb#1{\color[rgb]{#1}}%
      \def\colorgray#1{\color[gray]{#1}}%
      \expandafter\def\csname LTw\endcsname{\color{white}}%
      \expandafter\def\csname LTb\endcsname{\color{black}}%
      \expandafter\def\csname LTa\endcsname{\color{black}}%
      \expandafter\def\csname LT0\endcsname{\color[rgb]{1,0,0}}%
      \expandafter\def\csname LT1\endcsname{\color[rgb]{0,1,0}}%
      \expandafter\def\csname LT2\endcsname{\color[rgb]{0,0,1}}%
      \expandafter\def\csname LT3\endcsname{\color[rgb]{1,0,1}}%
      \expandafter\def\csname LT4\endcsname{\color[rgb]{0,1,1}}%
      \expandafter\def\csname LT5\endcsname{\color[rgb]{1,1,0}}%
      \expandafter\def\csname LT6\endcsname{\color[rgb]{0,0,0}}%
      \expandafter\def\csname LT7\endcsname{\color[rgb]{1,0.3,0}}%
      \expandafter\def\csname LT8\endcsname{\color[rgb]{0.5,0.5,0.5}}%
    \else
      \def\colorrgb#1{\color{black}}%
      \def\colorgray#1{\color[gray]{#1}}%
      \expandafter\def\csname LTw\endcsname{\color{white}}%
      \expandafter\def\csname LTb\endcsname{\color{black}}%
      \expandafter\def\csname LTa\endcsname{\color{black}}%
      \expandafter\def\csname LT0\endcsname{\color{black}}%
      \expandafter\def\csname LT1\endcsname{\color{black}}%
      \expandafter\def\csname LT2\endcsname{\color{black}}%
      \expandafter\def\csname LT3\endcsname{\color{black}}%
      \expandafter\def\csname LT4\endcsname{\color{black}}%
      \expandafter\def\csname LT5\endcsname{\color{black}}%
      \expandafter\def\csname LT6\endcsname{\color{black}}%
      \expandafter\def\csname LT7\endcsname{\color{black}}%
      \expandafter\def\csname LT8\endcsname{\color{black}}%
    \fi
  \fi
    \setlength{\unitlength}{0.0500bp}%
    \ifx\gptboxheight\undefined%
      \newlength{\gptboxheight}%
      \newlength{\gptboxwidth}%
      \newsavebox{\gptboxtext}%
    \fi%
    \setlength{\fboxrule}{0.5pt}%
    \setlength{\fboxsep}{1pt}%
\begin{picture}(5102.00,6802.00)%
    \gplgaddtomacro\gplbacktext{%
      \csname LTb\endcsname%
      \put(480,544){\makebox(0,0)[r]{\strut{}$20$}}%
      \put(480,789){\makebox(0,0)[r]{\strut{}$25$}}%
      \put(480,1034){\makebox(0,0)[r]{\strut{}$30$}}%
      \put(480,1278){\makebox(0,0)[r]{\strut{}$35$}}%
      \put(480,1523){\makebox(0,0)[r]{\strut{}$40$}}%
      \put(480,1768){\makebox(0,0)[r]{\strut{}$45$}}%
      \put(612,324){\makebox(0,0){\strut{}$-20$}}%
      \put(1709,324){\makebox(0,0){\strut{}$-15$}}%
      \put(2805,324){\makebox(0,0){\strut{}$-10$}}%
      \put(3902,324){\makebox(0,0){\strut{}$-5$}}%
      \put(4998,324){\makebox(0,0){\strut{}$0$}}%
    }%
    \gplgaddtomacro\gplfronttext{%
      \csname LTb\endcsname%
      \put(106,1156){\rotatebox{-270}{\makebox(0,0){\strut{}$\epsilon$ $(\degree)$}}}%
      \put(2805,-6){\makebox(0,0){\strut{}time $(\Myr)$}}%
      \csname LTb\endcsname%
      \put(4779,1854){\makebox(0,0)[l]{\strut{}c)}}%
    }%
    \gplgaddtomacro\gplbacktext{%
      \csname LTb\endcsname%
      \put(480,2516){\makebox(0,0)[r]{\strut{}$20$}}%
      \put(480,2761){\makebox(0,0)[r]{\strut{}$25$}}%
      \put(480,3006){\makebox(0,0)[r]{\strut{}$30$}}%
      \put(480,3250){\makebox(0,0)[r]{\strut{}$35$}}%
      \put(480,3495){\makebox(0,0)[r]{\strut{}$40$}}%
      \put(480,3740){\makebox(0,0)[r]{\strut{}$45$}}%
      \put(612,2296){\makebox(0,0){\strut{}$-1$}}%
      \put(1489,2296){\makebox(0,0){\strut{}$-0.8$}}%
      \put(2366,2296){\makebox(0,0){\strut{}$-0.6$}}%
      \put(3244,2296){\makebox(0,0){\strut{}$-0.4$}}%
      \put(4121,2296){\makebox(0,0){\strut{}$-0.2$}}%
      \put(4998,2296){\makebox(0,0){\strut{}$0$}}%
    }%
    \gplgaddtomacro\gplfronttext{%
      \csname LTb\endcsname%
      \put(106,3128){\rotatebox{-270}{\makebox(0,0){\strut{}$\epsilon$ $(\degree)$}}}%
      \put(2805,1966){\makebox(0,0){\strut{}time $(\Myr)$}}%
      \csname LTb\endcsname%
      \put(4779,3826){\makebox(0,0)[l]{\strut{}b)}}%
    }%
    \gplgaddtomacro\gplbacktext{%
      \csname LTb\endcsname%
      \put(480,4489){\makebox(0,0)[r]{\strut{}$20$}}%
      \put(480,4938){\makebox(0,0)[r]{\strut{}$25$}}%
      \put(480,5386){\makebox(0,0)[r]{\strut{}$30$}}%
      \put(480,5835){\makebox(0,0)[r]{\strut{}$35$}}%
      \put(480,6283){\makebox(0,0)[r]{\strut{}$40$}}%
      \put(480,6732){\makebox(0,0)[r]{\strut{}$45$}}%
      \put(612,4269){\makebox(0,0){\strut{}$-100$}}%
      \put(1489,4269){\makebox(0,0){\strut{}$-80$}}%
      \put(2366,4269){\makebox(0,0){\strut{}$-60$}}%
      \put(3244,4269){\makebox(0,0){\strut{}$-40$}}%
      \put(4121,4269){\makebox(0,0){\strut{}$-20$}}%
      \put(4998,4269){\makebox(0,0){\strut{}$0$}}%
    }%
    \gplgaddtomacro\gplfronttext{%
      \csname LTb\endcsname%
      \put(106,5610){\rotatebox{-270}{\makebox(0,0){\strut{}$\epsilon$ $(\degree)$}}}%
      \put(2805,3939){\makebox(0,0){\strut{}time $(\kyr)$}}%
      \csname LTb\endcsname%
      \put(4779,6620){\makebox(0,0)[l]{\strut{}a)}}%
    }%
    \gplbacktext
    \put(0,0){\includegraphics{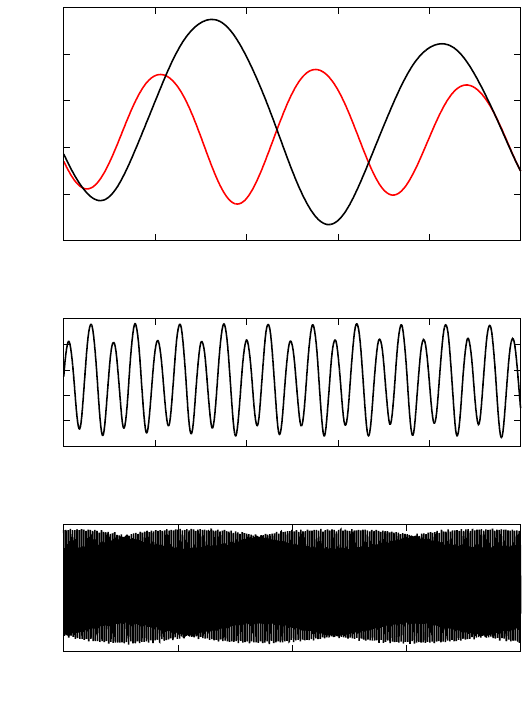}}%
    \gplfronttext
  \end{picture}%
\endgroup

%% file: figures/tex_figure10.tex
\begingroup
  \makeatletter
  \providecommand\color[2][]{%
    \GenericError{(gnuplot) \space\space\space\@spaces}{%
      Package color not loaded in conjunction with
      terminal option `colourtext'%
    }{See the gnuplot documentation for explanation.%
    }{Either use 'blacktext' in gnuplot or load the package
      color.sty in LaTeX.}%
    \renewcommand\color[2][]{}%
  }%
  \providecommand\includegraphics[2][]{%
    \GenericError{(gnuplot) \space\space\space\@spaces}{%
      Package graphicx or graphics not loaded%
    }{See the gnuplot documentation for explanation.%
    }{The gnuplot epslatex terminal needs graphicx.sty or graphics.sty.}%
    \renewcommand\includegraphics[2][]{}%
  }%
  \providecommand\rotatebox[2]{#2}%
  \@ifundefined{ifGPcolor}{%
    \newif\ifGPcolor
    \GPcolorfalse
  }{}%
  \@ifundefined{ifGPblacktext}{%
    \newif\ifGPblacktext
    \GPblacktextfalse
  }{}%
  \let\gplgaddtomacro\g@addto@macro
  \gdef\gplbacktext{}%
  \gdef\gplfronttext{}%
  \makeatother
  \ifGPblacktext
    \def\colorrgb#1{}%
    \def\colorgray#1{}%
  \else
    \ifGPcolor
      \def\colorrgb#1{\color[rgb]{#1}}%
      \def\colorgray#1{\color[gray]{#1}}%
      \expandafter\def\csname LTw\endcsname{\color{white}}%
      \expandafter\def\csname LTb\endcsname{\color{black}}%
      \expandafter\def\csname LTa\endcsname{\color{black}}%
      \expandafter\def\csname LT0\endcsname{\color[rgb]{1,0,0}}%
      \expandafter\def\csname LT1\endcsname{\color[rgb]{0,1,0}}%
      \expandafter\def\csname LT2\endcsname{\color[rgb]{0,0,1}}%
      \expandafter\def\csname LT3\endcsname{\color[rgb]{1,0,1}}%
      \expandafter\def\csname LT4\endcsname{\color[rgb]{0,1,1}}%
      \expandafter\def\csname LT5\endcsname{\color[rgb]{1,1,0}}%
      \expandafter\def\csname LT6\endcsname{\color[rgb]{0,0,0}}%
      \expandafter\def\csname LT7\endcsname{\color[rgb]{1,0.3,0}}%
      \expandafter\def\csname LT8\endcsname{\color[rgb]{0.5,0.5,0.5}}%
    \else
      \def\colorrgb#1{\color{black}}%
      \def\colorgray#1{\color[gray]{#1}}%
      \expandafter\def\csname LTw\endcsname{\color{white}}%
      \expandafter\def\csname LTb\endcsname{\color{black}}%
      \expandafter\def\csname LTa\endcsname{\color{black}}%
      \expandafter\def\csname LT0\endcsname{\color{black}}%
      \expandafter\def\csname LT1\endcsname{\color{black}}%
      \expandafter\def\csname LT2\endcsname{\color{black}}%
      \expandafter\def\csname LT3\endcsname{\color{black}}%
      \expandafter\def\csname LT4\endcsname{\color{black}}%
      \expandafter\def\csname LT5\endcsname{\color{black}}%
      \expandafter\def\csname LT6\endcsname{\color{black}}%
      \expandafter\def\csname LT7\endcsname{\color{black}}%
      \expandafter\def\csname LT8\endcsname{\color{black}}%
    \fi
  \fi
    \setlength{\unitlength}{0.0500bp}%
    \ifx\gptboxheight\undefined%
      \newlength{\gptboxheight}%
      \newlength{\gptboxwidth}%
      \newsavebox{\gptboxtext}%
    \fi%
    \setlength{\fboxrule}{0.5pt}%
    \setlength{\fboxsep}{1pt}%
\begin{picture}(5102.00,2550.00)%
    \gplgaddtomacro\gplbacktext{%
      \csname LTb\endcsname%
      \put(633,459){\makebox(0,0)[r]{\strut{}$0$}}%
      \put(633,644){\makebox(0,0)[r]{\strut{}$2$}}%
      \put(633,830){\makebox(0,0)[r]{\strut{}$4$}}%
      \put(633,1015){\makebox(0,0)[r]{\strut{}$6$}}%
      \put(633,1200){\makebox(0,0)[r]{\strut{}$8$}}%
      \put(633,1386){\makebox(0,0)[r]{\strut{}$10$}}%
      \put(633,1571){\makebox(0,0)[r]{\strut{}$12$}}%
      \put(633,1757){\makebox(0,0)[r]{\strut{}$14$}}%
      \put(633,1942){\makebox(0,0)[r]{\strut{}$16$}}%
      \put(633,2127){\makebox(0,0)[r]{\strut{}$18$}}%
      \put(633,2313){\makebox(0,0)[r]{\strut{}$20$}}%
      \put(633,2498){\makebox(0,0)[r]{\strut{}$22$}}%
      \put(765,239){\makebox(0,0){\strut{}0.380}}%
      \put(2334,239){\makebox(0,0){\strut{}0.390}}%
      \put(3903,239){\makebox(0,0){\strut{}0.400}}%
      \put(4845,239){\makebox(0,0){\strut{}0.406}}%
    }%
    \gplgaddtomacro\gplfronttext{%
      \csname LTb\endcsname%
      \put(127,1478){\rotatebox{-270}{\makebox(0,0){\strut{}$\epsilon$ $(\degree)$}}}%
      \put(2805,-91){\makebox(0,0){\strut{}$\C$}}%
    }%
    \gplbacktext
    \put(0,0){\includegraphics{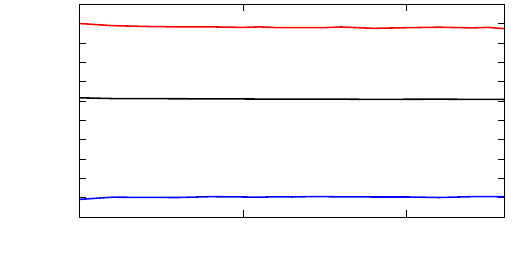}}%
    \gplfronttext
  \end{picture}%
\endgroup

%% file: figures/tex_figure11.tex
\begingroup
  \makeatletter
  \providecommand\color[2][]{%
    \GenericError{(gnuplot) \space\space\space\@spaces}{%
      Package color not loaded in conjunction with
      terminal option `colourtext'%
    }{See the gnuplot documentation for explanation.%
    }{Either use 'blacktext' in gnuplot or load the package
      color.sty in LaTeX.}%
    \renewcommand\color[2][]{}%
  }%
  \providecommand\includegraphics[2][]{%
    \GenericError{(gnuplot) \space\space\space\@spaces}{%
      Package graphicx or graphics not loaded%
    }{See the gnuplot documentation for explanation.%
    }{The gnuplot epslatex terminal needs graphicx.sty or graphics.sty.}%
    \renewcommand\includegraphics[2][]{}%
  }%
  \providecommand\rotatebox[2]{#2}%
  \@ifundefined{ifGPcolor}{%
    \newif\ifGPcolor
    \GPcolorfalse
  }{}%
  \@ifundefined{ifGPblacktext}{%
    \newif\ifGPblacktext
    \GPblacktextfalse
  }{}%
  \let\gplgaddtomacro\g@addto@macro
  \gdef\gplbacktext{}%
  \gdef\gplfronttext{}%
  \makeatother
  \ifGPblacktext
    \def\colorrgb#1{}%
    \def\colorgray#1{}%
  \else
    \ifGPcolor
      \def\colorrgb#1{\color[rgb]{#1}}%
      \def\colorgray#1{\color[gray]{#1}}%
      \expandafter\def\csname LTw\endcsname{\color{white}}%
      \expandafter\def\csname LTb\endcsname{\color{black}}%
      \expandafter\def\csname LTa\endcsname{\color{black}}%
      \expandafter\def\csname LT0\endcsname{\color[rgb]{1,0,0}}%
      \expandafter\def\csname LT1\endcsname{\color[rgb]{0,1,0}}%
      \expandafter\def\csname LT2\endcsname{\color[rgb]{0,0,1}}%
      \expandafter\def\csname LT3\endcsname{\color[rgb]{1,0,1}}%
      \expandafter\def\csname LT4\endcsname{\color[rgb]{0,1,1}}%
      \expandafter\def\csname LT5\endcsname{\color[rgb]{1,1,0}}%
      \expandafter\def\csname LT6\endcsname{\color[rgb]{0,0,0}}%
      \expandafter\def\csname LT7\endcsname{\color[rgb]{1,0.3,0}}%
      \expandafter\def\csname LT8\endcsname{\color[rgb]{0.5,0.5,0.5}}%
    \else
      \def\colorrgb#1{\color{black}}%
      \def\colorgray#1{\color[gray]{#1}}%
      \expandafter\def\csname LTw\endcsname{\color{white}}%
      \expandafter\def\csname LTb\endcsname{\color{black}}%
      \expandafter\def\csname LTa\endcsname{\color{black}}%
      \expandafter\def\csname LT0\endcsname{\color{black}}%
      \expandafter\def\csname LT1\endcsname{\color{black}}%
      \expandafter\def\csname LT2\endcsname{\color{black}}%
      \expandafter\def\csname LT3\endcsname{\color{black}}%
      \expandafter\def\csname LT4\endcsname{\color{black}}%
      \expandafter\def\csname LT5\endcsname{\color{black}}%
      \expandafter\def\csname LT6\endcsname{\color{black}}%
      \expandafter\def\csname LT7\endcsname{\color{black}}%
      \expandafter\def\csname LT8\endcsname{\color{black}}%
    \fi
  \fi
    \setlength{\unitlength}{0.0500bp}%
    \ifx\gptboxheight\undefined%
      \newlength{\gptboxheight}%
      \newlength{\gptboxwidth}%
      \newsavebox{\gptboxtext}%
    \fi%
    \setlength{\fboxrule}{0.5pt}%
    \setlength{\fboxsep}{1pt}%
\begin{picture}(5102.00,2550.00)%
    \gplgaddtomacro\gplbacktext{%
      \csname LTb\endcsname%
      \put(633,459){\makebox(0,0)[r]{\strut{}$15$}}%
      \put(633,750){\makebox(0,0)[r]{\strut{}$20$}}%
      \put(633,1042){\makebox(0,0)[r]{\strut{}$25$}}%
      \put(633,1333){\makebox(0,0)[r]{\strut{}$30$}}%
      \put(633,1624){\makebox(0,0)[r]{\strut{}$35$}}%
      \put(633,1915){\makebox(0,0)[r]{\strut{}$40$}}%
      \put(633,2207){\makebox(0,0)[r]{\strut{}$45$}}%
      \put(633,2498){\makebox(0,0)[r]{\strut{}$50$}}%
      \put(765,239){\makebox(0,0){\strut{}0.390}}%
      \put(1785,239){\makebox(0,0){\strut{}0.400}}%
      \put(2805,239){\makebox(0,0){\strut{}0.410}}%
      \put(3825,239){\makebox(0,0){\strut{}0.420}}%
      \put(4845,239){\makebox(0,0){\strut{}0.430}}%
    }%
    \gplgaddtomacro\gplfronttext{%
      \csname LTb\endcsname%
      \put(127,1478){\rotatebox{-270}{\makebox(0,0){\strut{}$\epsilon$ $(\degree)$}}}%
      \put(2805,-91){\makebox(0,0){\strut{}$\C$}}%
    }%
    \gplbacktext
    \put(0,0){\includegraphics{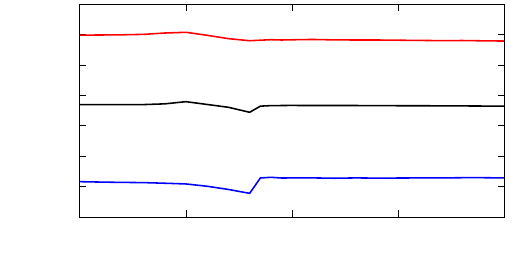}}%
    \gplfronttext
  \end{picture}%
\endgroup

%% file: figures/tex_figure12.tex
\begingroup
  \makeatletter
  \providecommand\color[2][]{%
    \GenericError{(gnuplot) \space\space\space\@spaces}{%
      Package color not loaded in conjunction with
      terminal option `colourtext'%
    }{See the gnuplot documentation for explanation.%
    }{Either use 'blacktext' in gnuplot or load the package
      color.sty in LaTeX.}%
    \renewcommand\color[2][]{}%
  }%
  \providecommand\includegraphics[2][]{%
    \GenericError{(gnuplot) \space\space\space\@spaces}{%
      Package graphicx or graphics not loaded%
    }{See the gnuplot documentation for explanation.%
    }{The gnuplot epslatex terminal needs graphicx.sty or graphics.sty.}%
    \renewcommand\includegraphics[2][]{}%
  }%
  \providecommand\rotatebox[2]{#2}%
  \@ifundefined{ifGPcolor}{%
    \newif\ifGPcolor
    \GPcolorfalse
  }{}%
  \@ifundefined{ifGPblacktext}{%
    \newif\ifGPblacktext
    \GPblacktextfalse
  }{}%
  \let\gplgaddtomacro\g@addto@macro
  \gdef\gplbacktext{}%
  \gdef\gplfronttext{}%
  \makeatother
  \ifGPblacktext
    \def\colorrgb#1{}%
    \def\colorgray#1{}%
  \else
    \ifGPcolor
      \def\colorrgb#1{\color[rgb]{#1}}%
      \def\colorgray#1{\color[gray]{#1}}%
      \expandafter\def\csname LTw\endcsname{\color{white}}%
      \expandafter\def\csname LTb\endcsname{\color{black}}%
      \expandafter\def\csname LTa\endcsname{\color{black}}%
      \expandafter\def\csname LT0\endcsname{\color[rgb]{1,0,0}}%
      \expandafter\def\csname LT1\endcsname{\color[rgb]{0,1,0}}%
      \expandafter\def\csname LT2\endcsname{\color[rgb]{0,0,1}}%
      \expandafter\def\csname LT3\endcsname{\color[rgb]{1,0,1}}%
      \expandafter\def\csname LT4\endcsname{\color[rgb]{0,1,1}}%
      \expandafter\def\csname LT5\endcsname{\color[rgb]{1,1,0}}%
      \expandafter\def\csname LT6\endcsname{\color[rgb]{0,0,0}}%
      \expandafter\def\csname LT7\endcsname{\color[rgb]{1,0.3,0}}%
      \expandafter\def\csname LT8\endcsname{\color[rgb]{0.5,0.5,0.5}}%
    \else
      \def\colorrgb#1{\color{black}}%
      \def\colorgray#1{\color[gray]{#1}}%
      \expandafter\def\csname LTw\endcsname{\color{white}}%
      \expandafter\def\csname LTb\endcsname{\color{black}}%
      \expandafter\def\csname LTa\endcsname{\color{black}}%
      \expandafter\def\csname LT0\endcsname{\color{black}}%
      \expandafter\def\csname LT1\endcsname{\color{black}}%
      \expandafter\def\csname LT2\endcsname{\color{black}}%
      \expandafter\def\csname LT3\endcsname{\color{black}}%
      \expandafter\def\csname LT4\endcsname{\color{black}}%
      \expandafter\def\csname LT5\endcsname{\color{black}}%
      \expandafter\def\csname LT6\endcsname{\color{black}}%
      \expandafter\def\csname LT7\endcsname{\color{black}}%
      \expandafter\def\csname LT8\endcsname{\color{black}}%
    \fi
  \fi
    \setlength{\unitlength}{0.0500bp}%
    \ifx\gptboxheight\undefined%
      \newlength{\gptboxheight}%
      \newlength{\gptboxwidth}%
      \newsavebox{\gptboxtext}%
    \fi%
    \setlength{\fboxrule}{0.5pt}%
    \setlength{\fboxsep}{1pt}%
\begin{picture}(5102.00,3570.00)%
    \gplgaddtomacro\gplbacktext{%
    }%
    \gplgaddtomacro\gplfronttext{%
      \csname LTb\endcsname%
      \put(148,1195){\rotatebox{-270}{\makebox(0,0){\strut{}$\delta i$ $(\degree)$}}}%
      \put(2958,-51){\makebox(0,0){\strut{}time $(\Myr)$}}%
      \csname LTb\endcsname%
      \put(786,499){\makebox(0,0)[r]{\strut{}   -2}}%
      \put(786,847){\makebox(0,0)[r]{\strut{}   -1}}%
      \put(786,1195){\makebox(0,0)[r]{\strut{}    0}}%
      \put(786,1543){\makebox(0,0)[r]{\strut{}    1}}%
      \put(786,1891){\makebox(0,0)[r]{\strut{}    2}}%
      \put(918,279){\makebox(0,0){\strut{}$-20$}}%
      \put(1938,279){\makebox(0,0){\strut{}$-15$}}%
      \put(2958,279){\makebox(0,0){\strut{}$-10$}}%
      \put(3978,279){\makebox(0,0){\strut{}$-5$}}%
      \put(4998,279){\makebox(0,0){\strut{}$0$}}%
      \put(4794,1780){\makebox(0,0)[l]{\strut{}b)}}%
    }%
    \gplgaddtomacro\gplbacktext{%
    }%
    \gplgaddtomacro\gplfronttext{%
      \csname LTb\endcsname%
      \put(148,2801){\rotatebox{-270}{\makebox(0,0){\strut{}$\delta e$}}}%
      \put(2958,1820){\makebox(0,0){\strut{}}}%
      \csname LTb\endcsname%
      \put(786,2106){\makebox(0,0)[r]{\strut{}-0.12}}%
      \put(786,2454){\makebox(0,0)[r]{\strut{}-0.06}}%
      \put(786,2802){\makebox(0,0)[r]{\strut{}0.00}}%
      \put(786,3149){\makebox(0,0)[r]{\strut{}0.06}}%
      \put(786,3497){\makebox(0,0)[r]{\strut{}0.12}}%
      \put(918,1886){\makebox(0,0){\strut{}}}%
      \put(1938,1886){\makebox(0,0){\strut{}}}%
      \put(2958,1886){\makebox(0,0){\strut{}}}%
      \put(3978,1886){\makebox(0,0){\strut{}}}%
      \put(4998,1886){\makebox(0,0){\strut{}}}%
      \put(4794,3386){\makebox(0,0)[l]{\strut{}a)}}%
    }%
    \gplbacktext
    \put(0,0){\includegraphics{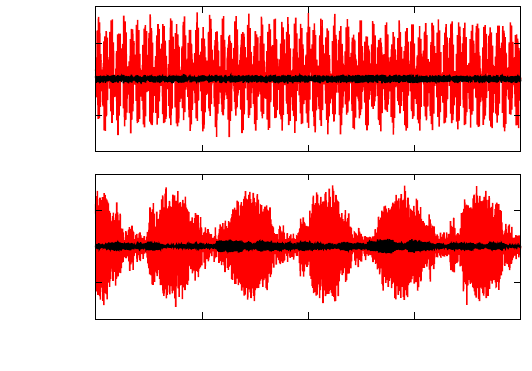}}%
    \gplfronttext
  \end{picture}%
\endgroup

%% file: figures/tex_figure13.tex
\begingroup
  \makeatletter
  \providecommand\color[2][]{%
    \GenericError{(gnuplot) \space\space\space\@spaces}{%
      Package color not loaded in conjunction with
      terminal option `colourtext'%
    }{See the gnuplot documentation for explanation.%
    }{Either use 'blacktext' in gnuplot or load the package
      color.sty in LaTeX.}%
    \renewcommand\color[2][]{}%
  }%
  \providecommand\includegraphics[2][]{%
    \GenericError{(gnuplot) \space\space\space\@spaces}{%
      Package graphicx or graphics not loaded%
    }{See the gnuplot documentation for explanation.%
    }{The gnuplot epslatex terminal needs graphicx.sty or graphics.sty.}%
    \renewcommand\includegraphics[2][]{}%
  }%
  \providecommand\rotatebox[2]{#2}%
  \@ifundefined{ifGPcolor}{%
    \newif\ifGPcolor
    \GPcolorfalse
  }{}%
  \@ifundefined{ifGPblacktext}{%
    \newif\ifGPblacktext
    \GPblacktextfalse
  }{}%
  \let\gplgaddtomacro\g@addto@macro
  \gdef\gplbacktext{}%
  \gdef\gplfronttext{}%
  \makeatother
  \ifGPblacktext
    \def\colorrgb#1{}%
    \def\colorgray#1{}%
  \else
    \ifGPcolor
      \def\colorrgb#1{\color[rgb]{#1}}%
      \def\colorgray#1{\color[gray]{#1}}%
      \expandafter\def\csname LTw\endcsname{\color{white}}%
      \expandafter\def\csname LTb\endcsname{\color{black}}%
      \expandafter\def\csname LTa\endcsname{\color{black}}%
      \expandafter\def\csname LT0\endcsname{\color[rgb]{1,0,0}}%
      \expandafter\def\csname LT1\endcsname{\color[rgb]{0,1,0}}%
      \expandafter\def\csname LT2\endcsname{\color[rgb]{0,0,1}}%
      \expandafter\def\csname LT3\endcsname{\color[rgb]{1,0,1}}%
      \expandafter\def\csname LT4\endcsname{\color[rgb]{0,1,1}}%
      \expandafter\def\csname LT5\endcsname{\color[rgb]{1,1,0}}%
      \expandafter\def\csname LT6\endcsname{\color[rgb]{0,0,0}}%
      \expandafter\def\csname LT7\endcsname{\color[rgb]{1,0.3,0}}%
      \expandafter\def\csname LT8\endcsname{\color[rgb]{0.5,0.5,0.5}}%
    \else
      \def\colorrgb#1{\color{black}}%
      \def\colorgray#1{\color[gray]{#1}}%
      \expandafter\def\csname LTw\endcsname{\color{white}}%
      \expandafter\def\csname LTb\endcsname{\color{black}}%
      \expandafter\def\csname LTa\endcsname{\color{black}}%
      \expandafter\def\csname LT0\endcsname{\color{black}}%
      \expandafter\def\csname LT1\endcsname{\color{black}}%
      \expandafter\def\csname LT2\endcsname{\color{black}}%
      \expandafter\def\csname LT3\endcsname{\color{black}}%
      \expandafter\def\csname LT4\endcsname{\color{black}}%
      \expandafter\def\csname LT5\endcsname{\color{black}}%
      \expandafter\def\csname LT6\endcsname{\color{black}}%
      \expandafter\def\csname LT7\endcsname{\color{black}}%
      \expandafter\def\csname LT8\endcsname{\color{black}}%
    \fi
  \fi
    \setlength{\unitlength}{0.0500bp}%
    \ifx\gptboxheight\undefined%
      \newlength{\gptboxheight}%
      \newlength{\gptboxwidth}%
      \newsavebox{\gptboxtext}%
    \fi%
    \setlength{\fboxrule}{0.5pt}%
    \setlength{\fboxsep}{1pt}%
\begin{picture}(5102.00,3570.00)%
    \gplgaddtomacro\gplbacktext{%
    }%
    \gplgaddtomacro\gplfronttext{%
      \csname LTb\endcsname%
      \put(148,1195){\rotatebox{-270}{\makebox(0,0){\strut{}$\delta i$ $(\degree)$}}}%
      \put(2958,-51){\makebox(0,0){\strut{}time $(\Myr)$}}%
      \csname LTb\endcsname%
      \put(786,499){\makebox(0,0)[r]{\strut{}   -3}}%
      \put(786,731){\makebox(0,0)[r]{\strut{}   -2}}%
      \put(786,963){\makebox(0,0)[r]{\strut{}   -1}}%
      \put(786,1195){\makebox(0,0)[r]{\strut{}    0}}%
      \put(786,1427){\makebox(0,0)[r]{\strut{}    1}}%
      \put(786,1659){\makebox(0,0)[r]{\strut{}    2}}%
      \put(786,1891){\makebox(0,0)[r]{\strut{}    3}}%
      \put(918,279){\makebox(0,0){\strut{}$-20$}}%
      \put(1938,279){\makebox(0,0){\strut{}$-15$}}%
      \put(2958,279){\makebox(0,0){\strut{}$-10$}}%
      \put(3978,279){\makebox(0,0){\strut{}$-5$}}%
      \put(4998,279){\makebox(0,0){\strut{}$0$}}%
      \put(4794,1780){\makebox(0,0)[l]{\strut{}b)}}%
    }%
    \gplgaddtomacro\gplbacktext{%
    }%
    \gplgaddtomacro\gplfronttext{%
      \csname LTb\endcsname%
      \put(148,2801){\rotatebox{-270}{\makebox(0,0){\strut{}$\delta e$}}}%
      \put(2958,1820){\makebox(0,0){\strut{}}}%
      \csname LTb\endcsname%
      \put(786,2106){\makebox(0,0)[r]{\strut{}-0.15}}%
      \put(786,2338){\makebox(0,0)[r]{\strut{}-0.10}}%
      \put(786,2570){\makebox(0,0)[r]{\strut{}-0.05}}%
      \put(786,2802){\makebox(0,0)[r]{\strut{}0.00}}%
      \put(786,3033){\makebox(0,0)[r]{\strut{}0.05}}%
      \put(786,3265){\makebox(0,0)[r]{\strut{}0.10}}%
      \put(786,3497){\makebox(0,0)[r]{\strut{}0.15}}%
      \put(918,1886){\makebox(0,0){\strut{}}}%
      \put(1938,1886){\makebox(0,0){\strut{}}}%
      \put(2958,1886){\makebox(0,0){\strut{}}}%
      \put(3978,1886){\makebox(0,0){\strut{}}}%
      \put(4998,1886){\makebox(0,0){\strut{}}}%
      \put(4794,3386){\makebox(0,0)[l]{\strut{}a)}}%
    }%
    \gplbacktext
    \put(0,0){\includegraphics{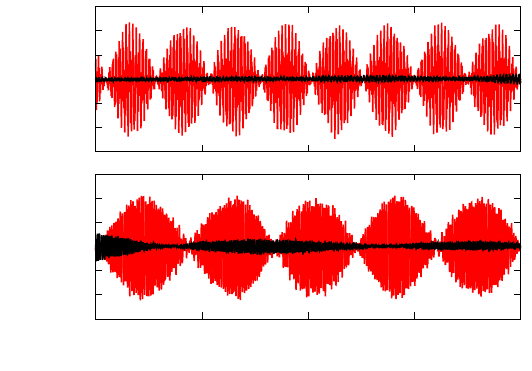}}%
    \gplfronttext
  \end{picture}%
\endgroup

%% file: figures/tex_figure14.tex
\begingroup
  \makeatletter
  \providecommand\color[2][]{%
    \GenericError{(gnuplot) \space\space\space\@spaces}{%
      Package color not loaded in conjunction with
      terminal option `colourtext'%
    }{See the gnuplot documentation for explanation.%
    }{Either use 'blacktext' in gnuplot or load the package
      color.sty in LaTeX.}%
    \renewcommand\color[2][]{}%
  }%
  \providecommand\includegraphics[2][]{%
    \GenericError{(gnuplot) \space\space\space\@spaces}{%
      Package graphicx or graphics not loaded%
    }{See the gnuplot documentation for explanation.%
    }{The gnuplot epslatex terminal needs graphicx.sty or graphics.sty.}%
    \renewcommand\includegraphics[2][]{}%
  }%
  \providecommand\rotatebox[2]{#2}%
  \@ifundefined{ifGPcolor}{%
    \newif\ifGPcolor
    \GPcolorfalse
  }{}%
  \@ifundefined{ifGPblacktext}{%
    \newif\ifGPblacktext
    \GPblacktextfalse
  }{}%
  \let\gplgaddtomacro\g@addto@macro
  \gdef\gplbacktext{}%
  \gdef\gplfronttext{}%
  \makeatother
  \ifGPblacktext
    \def\colorrgb#1{}%
    \def\colorgray#1{}%
  \else
    \ifGPcolor
      \def\colorrgb#1{\color[rgb]{#1}}%
      \def\colorgray#1{\color[gray]{#1}}%
      \expandafter\def\csname LTw\endcsname{\color{white}}%
      \expandafter\def\csname LTb\endcsname{\color{black}}%
      \expandafter\def\csname LTa\endcsname{\color{black}}%
      \expandafter\def\csname LT0\endcsname{\color[rgb]{1,0,0}}%
      \expandafter\def\csname LT1\endcsname{\color[rgb]{0,1,0}}%
      \expandafter\def\csname LT2\endcsname{\color[rgb]{0,0,1}}%
      \expandafter\def\csname LT3\endcsname{\color[rgb]{1,0,1}}%
      \expandafter\def\csname LT4\endcsname{\color[rgb]{0,1,1}}%
      \expandafter\def\csname LT5\endcsname{\color[rgb]{1,1,0}}%
      \expandafter\def\csname LT6\endcsname{\color[rgb]{0,0,0}}%
      \expandafter\def\csname LT7\endcsname{\color[rgb]{1,0.3,0}}%
      \expandafter\def\csname LT8\endcsname{\color[rgb]{0.5,0.5,0.5}}%
    \else
      \def\colorrgb#1{\color{black}}%
      \def\colorgray#1{\color[gray]{#1}}%
      \expandafter\def\csname LTw\endcsname{\color{white}}%
      \expandafter\def\csname LTb\endcsname{\color{black}}%
      \expandafter\def\csname LTa\endcsname{\color{black}}%
      \expandafter\def\csname LT0\endcsname{\color{black}}%
      \expandafter\def\csname LT1\endcsname{\color{black}}%
      \expandafter\def\csname LT2\endcsname{\color{black}}%
      \expandafter\def\csname LT3\endcsname{\color{black}}%
      \expandafter\def\csname LT4\endcsname{\color{black}}%
      \expandafter\def\csname LT5\endcsname{\color{black}}%
      \expandafter\def\csname LT6\endcsname{\color{black}}%
      \expandafter\def\csname LT7\endcsname{\color{black}}%
      \expandafter\def\csname LT8\endcsname{\color{black}}%
    \fi
  \fi
    \setlength{\unitlength}{0.0500bp}%
    \ifx\gptboxheight\undefined%
      \newlength{\gptboxheight}%
      \newlength{\gptboxwidth}%
      \newsavebox{\gptboxtext}%
    \fi%
    \setlength{\fboxrule}{0.5pt}%
    \setlength{\fboxsep}{1pt}%
\begin{picture}(5102.00,6802.00)%
    \gplgaddtomacro\gplbacktext{%
      \csname LTb\endcsname%
      \put(633,680){\makebox(0,0)[r]{\strut{}  50}}%
      \put(633,1224){\makebox(0,0)[r]{\strut{}  52}}%
      \put(633,1768){\makebox(0,0)[r]{\strut{}  54}}%
      \put(633,2312){\makebox(0,0)[r]{\strut{}  56}}%
      \put(765,324){\makebox(0,0){\strut{}$0$}}%
      \put(1471,324){\makebox(0,0){\strut{}$1$}}%
      \put(2176,324){\makebox(0,0){\strut{}$2$}}%
      \put(2882,324){\makebox(0,0){\strut{}$3$}}%
      \put(3587,324){\makebox(0,0){\strut{}$4$}}%
      \put(4293,324){\makebox(0,0){\strut{}$5$}}%
      \put(4998,324){\makebox(0,0){\strut{}$6$}}%
    }%
    \gplgaddtomacro\gplfronttext{%
      \csname LTb\endcsname%
      \put(127,1564){\rotatebox{-270}{\makebox(0,0){\strut{}$g_C$ $(\arc)$}}}%
      \put(2881,-6){\makebox(0,0){\strut{}$\mathcal{A}$ $(\arc)$}}%
      \csname LTb\endcsname%
      \put(4786,2482){\makebox(0,0)[l]{\strut{}c)}}%
    }%
    \gplgaddtomacro\gplbacktext{%
      \csname LTb\endcsname%
      \put(633,2924){\makebox(0,0)[r]{\strut{}   9}}%
      \put(633,3604){\makebox(0,0)[r]{\strut{}  10}}%
      \put(633,4284){\makebox(0,0)[r]{\strut{}  11}}%
      \put(765,2364){\makebox(0,0){\strut{}}}%
      \put(1471,2364){\makebox(0,0){\strut{}}}%
      \put(2176,2364){\makebox(0,0){\strut{}}}%
      \put(2882,2364){\makebox(0,0){\strut{}}}%
      \put(3587,2364){\makebox(0,0){\strut{}}}%
      \put(4293,2364){\makebox(0,0){\strut{}}}%
      \put(4998,2364){\makebox(0,0){\strut{}}}%
    }%
    \gplgaddtomacro\gplfronttext{%
      \csname LTb\endcsname%
      \put(127,3604){\rotatebox{-270}{\makebox(0,0){\strut{}$i$ $(\degree)$}}}%
      \put(2881,2298){\makebox(0,0){\strut{}}}%
      \csname LTb\endcsname%
      \put(4786,4522){\makebox(0,0)[l]{\strut{}b)}}%
    }%
    \gplgaddtomacro\gplbacktext{%
      \csname LTb\endcsname%
      \put(633,4689){\makebox(0,0)[r]{\strut{}0.00}}%
      \put(633,5326){\makebox(0,0)[r]{\strut{}0.10}}%
      \put(633,5963){\makebox(0,0)[r]{\strut{}0.20}}%
      \put(633,6600){\makebox(0,0)[r]{\strut{}0.30}}%
      \put(765,4405){\makebox(0,0){\strut{}}}%
      \put(1471,4405){\makebox(0,0){\strut{}}}%
      \put(2176,4405){\makebox(0,0){\strut{}}}%
      \put(2882,4405){\makebox(0,0){\strut{}}}%
      \put(3587,4405){\makebox(0,0){\strut{}}}%
      \put(4293,4405){\makebox(0,0){\strut{}}}%
      \put(4998,4405){\makebox(0,0){\strut{}}}%
    }%
    \gplgaddtomacro\gplfronttext{%
      \csname LTb\endcsname%
      \put(127,5644){\rotatebox{-270}{\makebox(0,0){\strut{}$e$}}}%
      \put(2881,4339){\makebox(0,0){\strut{}}}%
      \csname LTb\endcsname%
      \put(4786,6562){\makebox(0,0)[l]{\strut{}a)}}%
    }%
    \gplbacktext
    \put(0,0){\includegraphics{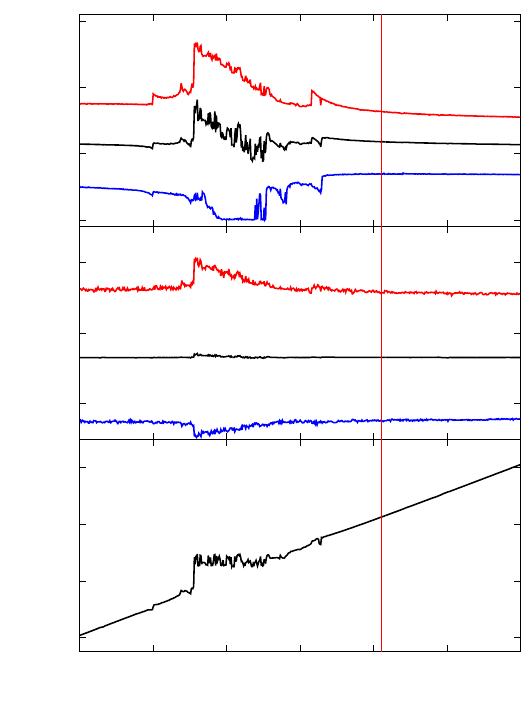}}%
    \gplfronttext
  \end{picture}%
\endgroup

%% file: figures/tex_figure15.tex
\begingroup
  \makeatletter
  \providecommand\color[2][]{%
    \GenericError{(gnuplot) \space\space\space\@spaces}{%
      Package color not loaded in conjunction with
      terminal option `colourtext'%
    }{See the gnuplot documentation for explanation.%
    }{Either use 'blacktext' in gnuplot or load the package
      color.sty in LaTeX.}%
    \renewcommand\color[2][]{}%
  }%
  \providecommand\includegraphics[2][]{%
    \GenericError{(gnuplot) \space\space\space\@spaces}{%
      Package graphicx or graphics not loaded%
    }{See the gnuplot documentation for explanation.%
    }{The gnuplot epslatex terminal needs graphicx.sty or graphics.sty.}%
    \renewcommand\includegraphics[2][]{}%
  }%
  \providecommand\rotatebox[2]{#2}%
  \@ifundefined{ifGPcolor}{%
    \newif\ifGPcolor
    \GPcolorfalse
  }{}%
  \@ifundefined{ifGPblacktext}{%
    \newif\ifGPblacktext
    \GPblacktextfalse
  }{}%
  \let\gplgaddtomacro\g@addto@macro
  \gdef\gplbacktext{}%
  \gdef\gplfronttext{}%
  \makeatother
  \ifGPblacktext
    \def\colorrgb#1{}%
    \def\colorgray#1{}%
  \else
    \ifGPcolor
      \def\colorrgb#1{\color[rgb]{#1}}%
      \def\colorgray#1{\color[gray]{#1}}%
      \expandafter\def\csname LTw\endcsname{\color{white}}%
      \expandafter\def\csname LTb\endcsname{\color{black}}%
      \expandafter\def\csname LTa\endcsname{\color{black}}%
      \expandafter\def\csname LT0\endcsname{\color[rgb]{1,0,0}}%
      \expandafter\def\csname LT1\endcsname{\color[rgb]{0,1,0}}%
      \expandafter\def\csname LT2\endcsname{\color[rgb]{0,0,1}}%
      \expandafter\def\csname LT3\endcsname{\color[rgb]{1,0,1}}%
      \expandafter\def\csname LT4\endcsname{\color[rgb]{0,1,1}}%
      \expandafter\def\csname LT5\endcsname{\color[rgb]{1,1,0}}%
      \expandafter\def\csname LT6\endcsname{\color[rgb]{0,0,0}}%
      \expandafter\def\csname LT7\endcsname{\color[rgb]{1,0.3,0}}%
      \expandafter\def\csname LT8\endcsname{\color[rgb]{0.5,0.5,0.5}}%
    \else
      \def\colorrgb#1{\color{black}}%
      \def\colorgray#1{\color[gray]{#1}}%
      \expandafter\def\csname LTw\endcsname{\color{white}}%
      \expandafter\def\csname LTb\endcsname{\color{black}}%
      \expandafter\def\csname LTa\endcsname{\color{black}}%
      \expandafter\def\csname LT0\endcsname{\color{black}}%
      \expandafter\def\csname LT1\endcsname{\color{black}}%
      \expandafter\def\csname LT2\endcsname{\color{black}}%
      \expandafter\def\csname LT3\endcsname{\color{black}}%
      \expandafter\def\csname LT4\endcsname{\color{black}}%
      \expandafter\def\csname LT5\endcsname{\color{black}}%
      \expandafter\def\csname LT6\endcsname{\color{black}}%
      \expandafter\def\csname LT7\endcsname{\color{black}}%
      \expandafter\def\csname LT8\endcsname{\color{black}}%
    \fi
  \fi
    \setlength{\unitlength}{0.0500bp}%
    \ifx\gptboxheight\undefined%
      \newlength{\gptboxheight}%
      \newlength{\gptboxwidth}%
      \newsavebox{\gptboxtext}%
    \fi%
    \setlength{\fboxrule}{0.5pt}%
    \setlength{\fboxsep}{1pt}%
\begin{picture}(5102.00,6802.00)%
    \gplgaddtomacro\gplbacktext{%
      \csname LTb\endcsname%
      \put(633,816){\makebox(0,0)[r]{\strut{} -62}}%
      \put(633,1360){\makebox(0,0)[r]{\strut{} -60}}%
      \put(633,1904){\makebox(0,0)[r]{\strut{} -58}}%
      \put(633,2448){\makebox(0,0)[r]{\strut{} -56}}%
      \put(765,324){\makebox(0,0){\strut{}$-3$}}%
      \put(1471,324){\makebox(0,0){\strut{}$-2$}}%
      \put(2176,324){\makebox(0,0){\strut{}$-1$}}%
      \put(2882,324){\makebox(0,0){\strut{}$0$}}%
      \put(3587,324){\makebox(0,0){\strut{}$1$}}%
      \put(4293,324){\makebox(0,0){\strut{}$2$}}%
      \put(4998,324){\makebox(0,0){\strut{}$3$}}%
    }%
    \gplgaddtomacro\gplfronttext{%
      \csname LTb\endcsname%
      \put(127,1564){\rotatebox{-270}{\makebox(0,0){\strut{}$s_C$ $(\arc)$}}}%
      \put(2881,-6){\makebox(0,0){\strut{}$\mathcal{B}$ $(\arc)$}}%
      \csname LTb\endcsname%
      \put(4786,2482){\makebox(0,0)[l]{\strut{}c)}}%
    }%
    \gplgaddtomacro\gplbacktext{%
      \csname LTb\endcsname%
      \put(633,2924){\makebox(0,0)[r]{\strut{}   9}}%
      \put(633,3604){\makebox(0,0)[r]{\strut{}  10}}%
      \put(633,4284){\makebox(0,0)[r]{\strut{}  11}}%
      \put(765,2364){\makebox(0,0){\strut{}}}%
      \put(1471,2364){\makebox(0,0){\strut{}}}%
      \put(2176,2364){\makebox(0,0){\strut{}}}%
      \put(2882,2364){\makebox(0,0){\strut{}}}%
      \put(3587,2364){\makebox(0,0){\strut{}}}%
      \put(4293,2364){\makebox(0,0){\strut{}}}%
      \put(4998,2364){\makebox(0,0){\strut{}}}%
    }%
    \gplgaddtomacro\gplfronttext{%
      \csname LTb\endcsname%
      \put(127,3604){\rotatebox{-270}{\makebox(0,0){\strut{}$i$ $(\degree)$}}}%
      \put(2881,2298){\makebox(0,0){\strut{}}}%
      \csname LTb\endcsname%
      \put(4786,4522){\makebox(0,0)[l]{\strut{}b)}}%
    }%
    \gplgaddtomacro\gplbacktext{%
      \csname LTb\endcsname%
      \put(633,4718){\makebox(0,0)[r]{\strut{}0.00}}%
      \put(633,5645){\makebox(0,0)[r]{\strut{}0.10}}%
      \put(633,6571){\makebox(0,0)[r]{\strut{}0.20}}%
      \put(765,4405){\makebox(0,0){\strut{}}}%
      \put(1471,4405){\makebox(0,0){\strut{}}}%
      \put(2176,4405){\makebox(0,0){\strut{}}}%
      \put(2882,4405){\makebox(0,0){\strut{}}}%
      \put(3587,4405){\makebox(0,0){\strut{}}}%
      \put(4293,4405){\makebox(0,0){\strut{}}}%
      \put(4998,4405){\makebox(0,0){\strut{}}}%
    }%
    \gplgaddtomacro\gplfronttext{%
      \csname LTb\endcsname%
      \put(127,5644){\rotatebox{-270}{\makebox(0,0){\strut{}$e$}}}%
      \put(2881,4339){\makebox(0,0){\strut{}}}%
      \csname LTb\endcsname%
      \put(4786,6562){\makebox(0,0)[l]{\strut{}a)}}%
    }%
    \gplbacktext
    \put(0,0){\includegraphics{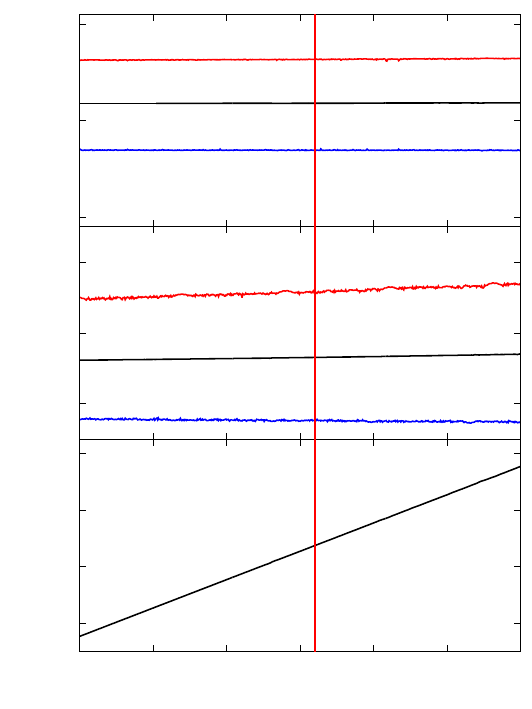}}%
    \gplfronttext
  \end{picture}%
\endgroup

%% file: figures/tex_figure16.tex
\begingroup
  \makeatletter
  \providecommand\color[2][]{%
    \GenericError{(gnuplot) \space\space\space\@spaces}{%
      Package color not loaded in conjunction with
      terminal option `colourtext'%
    }{See the gnuplot documentation for explanation.%
    }{Either use 'blacktext' in gnuplot or load the package
      color.sty in LaTeX.}%
    \renewcommand\color[2][]{}%
  }%
  \providecommand\includegraphics[2][]{%
    \GenericError{(gnuplot) \space\space\space\@spaces}{%
      Package graphicx or graphics not loaded%
    }{See the gnuplot documentation for explanation.%
    }{The gnuplot epslatex terminal needs graphicx.sty or graphics.sty.}%
    \renewcommand\includegraphics[2][]{}%
  }%
  \providecommand\rotatebox[2]{#2}%
  \@ifundefined{ifGPcolor}{%
    \newif\ifGPcolor
    \GPcolorfalse
  }{}%
  \@ifundefined{ifGPblacktext}{%
    \newif\ifGPblacktext
    \GPblacktextfalse
  }{}%
  \let\gplgaddtomacro\g@addto@macro
  \gdef\gplbacktext{}%
  \gdef\gplfronttext{}%
  \makeatother
  \ifGPblacktext
    \def\colorrgb#1{}%
    \def\colorgray#1{}%
  \else
    \ifGPcolor
      \def\colorrgb#1{\color[rgb]{#1}}%
      \def\colorgray#1{\color[gray]{#1}}%
      \expandafter\def\csname LTw\endcsname{\color{white}}%
      \expandafter\def\csname LTb\endcsname{\color{black}}%
      \expandafter\def\csname LTa\endcsname{\color{black}}%
      \expandafter\def\csname LT0\endcsname{\color[rgb]{1,0,0}}%
      \expandafter\def\csname LT1\endcsname{\color[rgb]{0,1,0}}%
      \expandafter\def\csname LT2\endcsname{\color[rgb]{0,0,1}}%
      \expandafter\def\csname LT3\endcsname{\color[rgb]{1,0,1}}%
      \expandafter\def\csname LT4\endcsname{\color[rgb]{0,1,1}}%
      \expandafter\def\csname LT5\endcsname{\color[rgb]{1,1,0}}%
      \expandafter\def\csname LT6\endcsname{\color[rgb]{0,0,0}}%
      \expandafter\def\csname LT7\endcsname{\color[rgb]{1,0.3,0}}%
      \expandafter\def\csname LT8\endcsname{\color[rgb]{0.5,0.5,0.5}}%
    \else
      \def\colorrgb#1{\color{black}}%
      \def\colorgray#1{\color[gray]{#1}}%
      \expandafter\def\csname LTw\endcsname{\color{white}}%
      \expandafter\def\csname LTb\endcsname{\color{black}}%
      \expandafter\def\csname LTa\endcsname{\color{black}}%
      \expandafter\def\csname LT0\endcsname{\color{black}}%
      \expandafter\def\csname LT1\endcsname{\color{black}}%
      \expandafter\def\csname LT2\endcsname{\color{black}}%
      \expandafter\def\csname LT3\endcsname{\color{black}}%
      \expandafter\def\csname LT4\endcsname{\color{black}}%
      \expandafter\def\csname LT5\endcsname{\color{black}}%
      \expandafter\def\csname LT6\endcsname{\color{black}}%
      \expandafter\def\csname LT7\endcsname{\color{black}}%
      \expandafter\def\csname LT8\endcsname{\color{black}}%
    \fi
  \fi
    \setlength{\unitlength}{0.0500bp}%
    \ifx\gptboxheight\undefined%
      \newlength{\gptboxheight}%
      \newlength{\gptboxwidth}%
      \newsavebox{\gptboxtext}%
    \fi%
    \setlength{\fboxrule}{0.5pt}%
    \setlength{\fboxsep}{1pt}%
\begin{picture}(5102.00,6802.00)%
    \gplgaddtomacro\gplbacktext{%
      \csname LTb\endcsname%
      \put(633,816){\makebox(0,0)[r]{\strut{}  34}}%
      \put(633,1360){\makebox(0,0)[r]{\strut{}  36}}%
      \put(633,1904){\makebox(0,0)[r]{\strut{}  38}}%
      \put(633,2448){\makebox(0,0)[r]{\strut{}  40}}%
      \put(765,324){\makebox(0,0){\strut{}$-3$}}%
      \put(1471,324){\makebox(0,0){\strut{}$-2$}}%
      \put(2176,324){\makebox(0,0){\strut{}$-1$}}%
      \put(2882,324){\makebox(0,0){\strut{}$0$}}%
      \put(3587,324){\makebox(0,0){\strut{}$1$}}%
      \put(4293,324){\makebox(0,0){\strut{}$2$}}%
      \put(4998,324){\makebox(0,0){\strut{}$3$}}%
    }%
    \gplgaddtomacro\gplfronttext{%
      \csname LTb\endcsname%
      \put(127,1564){\rotatebox{-270}{\makebox(0,0){\strut{}$g_V$ $(\arc)$}}}%
      \put(2881,-6){\makebox(0,0){\strut{}$\mathcal{A}$ $(\arc)$}}%
      \csname LTb\endcsname%
      \put(4786,2482){\makebox(0,0)[l]{\strut{}c)}}%
    }%
    \gplgaddtomacro\gplbacktext{%
      \csname LTb\endcsname%
      \put(633,2839){\makebox(0,0)[r]{\strut{}   5}}%
      \put(633,3349){\makebox(0,0)[r]{\strut{}   6}}%
      \put(633,3859){\makebox(0,0)[r]{\strut{}   7}}%
      \put(633,4369){\makebox(0,0)[r]{\strut{}   8}}%
      \put(765,2364){\makebox(0,0){\strut{}}}%
      \put(1471,2364){\makebox(0,0){\strut{}}}%
      \put(2176,2364){\makebox(0,0){\strut{}}}%
      \put(2882,2364){\makebox(0,0){\strut{}}}%
      \put(3587,2364){\makebox(0,0){\strut{}}}%
      \put(4293,2364){\makebox(0,0){\strut{}}}%
      \put(4998,2364){\makebox(0,0){\strut{}}}%
    }%
    \gplgaddtomacro\gplfronttext{%
      \csname LTb\endcsname%
      \put(127,3604){\rotatebox{-270}{\makebox(0,0){\strut{}$i$ $(\degree)$}}}%
      \put(2881,2298){\makebox(0,0){\strut{}}}%
      \csname LTb\endcsname%
      \put(4786,4522){\makebox(0,0)[l]{\strut{}b)}}%
    }%
    \gplgaddtomacro\gplbacktext{%
      \csname LTb\endcsname%
      \put(633,4718){\makebox(0,0)[r]{\strut{}0.00}}%
      \put(633,5645){\makebox(0,0)[r]{\strut{}0.10}}%
      \put(633,6571){\makebox(0,0)[r]{\strut{}0.20}}%
      \put(765,4405){\makebox(0,0){\strut{}}}%
      \put(1471,4405){\makebox(0,0){\strut{}}}%
      \put(2176,4405){\makebox(0,0){\strut{}}}%
      \put(2882,4405){\makebox(0,0){\strut{}}}%
      \put(3587,4405){\makebox(0,0){\strut{}}}%
      \put(4293,4405){\makebox(0,0){\strut{}}}%
      \put(4998,4405){\makebox(0,0){\strut{}}}%
    }%
    \gplgaddtomacro\gplfronttext{%
      \csname LTb\endcsname%
      \put(127,5644){\rotatebox{-270}{\makebox(0,0){\strut{}$e$}}}%
      \put(2881,4339){\makebox(0,0){\strut{}}}%
      \csname LTb\endcsname%
      \put(4786,6562){\makebox(0,0)[l]{\strut{}a)}}%
    }%
    \gplbacktext
    \put(0,0){\includegraphics{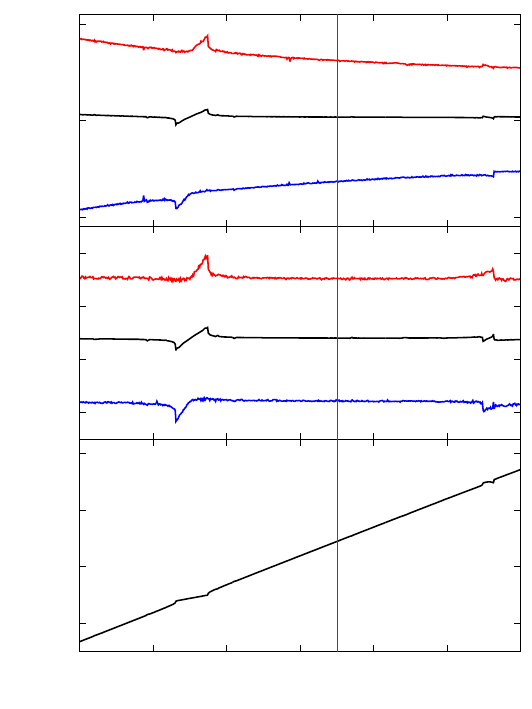}}%
    \gplfronttext
  \end{picture}%
\endgroup

%% file: figures/tex_figure17.tex
\begingroup
  \makeatletter
  \providecommand\color[2][]{%
    \GenericError{(gnuplot) \space\space\space\@spaces}{%
      Package color not loaded in conjunction with
      terminal option `colourtext'%
    }{See the gnuplot documentation for explanation.%
    }{Either use 'blacktext' in gnuplot or load the package
      color.sty in LaTeX.}%
    \renewcommand\color[2][]{}%
  }%
  \providecommand\includegraphics[2][]{%
    \GenericError{(gnuplot) \space\space\space\@spaces}{%
      Package graphicx or graphics not loaded%
    }{See the gnuplot documentation for explanation.%
    }{The gnuplot epslatex terminal needs graphicx.sty or graphics.sty.}%
    \renewcommand\includegraphics[2][]{}%
  }%
  \providecommand\rotatebox[2]{#2}%
  \@ifundefined{ifGPcolor}{%
    \newif\ifGPcolor
    \GPcolorfalse
  }{}%
  \@ifundefined{ifGPblacktext}{%
    \newif\ifGPblacktext
    \GPblacktextfalse
  }{}%
  \let\gplgaddtomacro\g@addto@macro
  \gdef\gplbacktext{}%
  \gdef\gplfronttext{}%
  \makeatother
  \ifGPblacktext
    \def\colorrgb#1{}%
    \def\colorgray#1{}%
  \else
    \ifGPcolor
      \def\colorrgb#1{\color[rgb]{#1}}%
      \def\colorgray#1{\color[gray]{#1}}%
      \expandafter\def\csname LTw\endcsname{\color{white}}%
      \expandafter\def\csname LTb\endcsname{\color{black}}%
      \expandafter\def\csname LTa\endcsname{\color{black}}%
      \expandafter\def\csname LT0\endcsname{\color[rgb]{1,0,0}}%
      \expandafter\def\csname LT1\endcsname{\color[rgb]{0,1,0}}%
      \expandafter\def\csname LT2\endcsname{\color[rgb]{0,0,1}}%
      \expandafter\def\csname LT3\endcsname{\color[rgb]{1,0,1}}%
      \expandafter\def\csname LT4\endcsname{\color[rgb]{0,1,1}}%
      \expandafter\def\csname LT5\endcsname{\color[rgb]{1,1,0}}%
      \expandafter\def\csname LT6\endcsname{\color[rgb]{0,0,0}}%
      \expandafter\def\csname LT7\endcsname{\color[rgb]{1,0.3,0}}%
      \expandafter\def\csname LT8\endcsname{\color[rgb]{0.5,0.5,0.5}}%
    \else
      \def\colorrgb#1{\color{black}}%
      \def\colorgray#1{\color[gray]{#1}}%
      \expandafter\def\csname LTw\endcsname{\color{white}}%
      \expandafter\def\csname LTb\endcsname{\color{black}}%
      \expandafter\def\csname LTa\endcsname{\color{black}}%
      \expandafter\def\csname LT0\endcsname{\color{black}}%
      \expandafter\def\csname LT1\endcsname{\color{black}}%
      \expandafter\def\csname LT2\endcsname{\color{black}}%
      \expandafter\def\csname LT3\endcsname{\color{black}}%
      \expandafter\def\csname LT4\endcsname{\color{black}}%
      \expandafter\def\csname LT5\endcsname{\color{black}}%
      \expandafter\def\csname LT6\endcsname{\color{black}}%
      \expandafter\def\csname LT7\endcsname{\color{black}}%
      \expandafter\def\csname LT8\endcsname{\color{black}}%
    \fi
  \fi
    \setlength{\unitlength}{0.0500bp}%
    \ifx\gptboxheight\undefined%
      \newlength{\gptboxheight}%
      \newlength{\gptboxwidth}%
      \newsavebox{\gptboxtext}%
    \fi%
    \setlength{\fboxrule}{0.5pt}%
    \setlength{\fboxsep}{1pt}%
\begin{picture}(5102.00,6802.00)%
    \gplgaddtomacro\gplbacktext{%
      \csname LTb\endcsname%
      \put(633,690){\makebox(0,0)[r]{\strut{} -42}}%
      \put(633,1273){\makebox(0,0)[r]{\strut{} -40}}%
      \put(633,1855){\makebox(0,0)[r]{\strut{} -38}}%
      \put(633,2438){\makebox(0,0)[r]{\strut{} -36}}%
      \put(765,324){\makebox(0,0){\strut{}$-3$}}%
      \put(1471,324){\makebox(0,0){\strut{}$-2$}}%
      \put(2176,324){\makebox(0,0){\strut{}$-1$}}%
      \put(2882,324){\makebox(0,0){\strut{}$0$}}%
      \put(3587,324){\makebox(0,0){\strut{}$1$}}%
      \put(4293,324){\makebox(0,0){\strut{}$2$}}%
      \put(4998,324){\makebox(0,0){\strut{}$3$}}%
    }%
    \gplgaddtomacro\gplfronttext{%
      \csname LTb\endcsname%
      \put(127,1564){\rotatebox{-270}{\makebox(0,0){\strut{}$s_V$ $(\arc)$}}}%
      \put(2881,-6){\makebox(0,0){\strut{}$\mathcal{B}$ $(\arc)$}}%
      \csname LTb\endcsname%
      \put(4786,2482){\makebox(0,0)[l]{\strut{}c)}}%
    }%
    \gplgaddtomacro\gplbacktext{%
      \csname LTb\endcsname%
      \put(633,2824){\makebox(0,0)[r]{\strut{}   5}}%
      \put(633,3304){\makebox(0,0)[r]{\strut{}   6}}%
      \put(633,3784){\makebox(0,0)[r]{\strut{}   7}}%
      \put(633,4264){\makebox(0,0)[r]{\strut{}   8}}%
      \put(765,2364){\makebox(0,0){\strut{}}}%
      \put(1471,2364){\makebox(0,0){\strut{}}}%
      \put(2176,2364){\makebox(0,0){\strut{}}}%
      \put(2882,2364){\makebox(0,0){\strut{}}}%
      \put(3587,2364){\makebox(0,0){\strut{}}}%
      \put(4293,2364){\makebox(0,0){\strut{}}}%
      \put(4998,2364){\makebox(0,0){\strut{}}}%
    }%
    \gplgaddtomacro\gplfronttext{%
      \csname LTb\endcsname%
      \put(127,3604){\rotatebox{-270}{\makebox(0,0){\strut{}$i$ $(\degree)$}}}%
      \put(2881,2298){\makebox(0,0){\strut{}}}%
      \csname LTb\endcsname%
      \put(4786,4522){\makebox(0,0)[l]{\strut{}b)}}%
    }%
    \gplgaddtomacro\gplbacktext{%
      \csname LTb\endcsname%
      \put(633,4718){\makebox(0,0)[r]{\strut{}0.00}}%
      \put(633,5645){\makebox(0,0)[r]{\strut{}0.10}}%
      \put(633,6571){\makebox(0,0)[r]{\strut{}0.20}}%
      \put(765,4405){\makebox(0,0){\strut{}}}%
      \put(1471,4405){\makebox(0,0){\strut{}}}%
      \put(2176,4405){\makebox(0,0){\strut{}}}%
      \put(2882,4405){\makebox(0,0){\strut{}}}%
      \put(3587,4405){\makebox(0,0){\strut{}}}%
      \put(4293,4405){\makebox(0,0){\strut{}}}%
      \put(4998,4405){\makebox(0,0){\strut{}}}%
    }%
    \gplgaddtomacro\gplfronttext{%
      \csname LTb\endcsname%
      \put(127,5644){\rotatebox{-270}{\makebox(0,0){\strut{}$e$}}}%
      \put(2881,4339){\makebox(0,0){\strut{}}}%
      \csname LTb\endcsname%
      \put(4786,6562){\makebox(0,0)[l]{\strut{}a)}}%
    }%
    \gplbacktext
    \put(0,0){\includegraphics{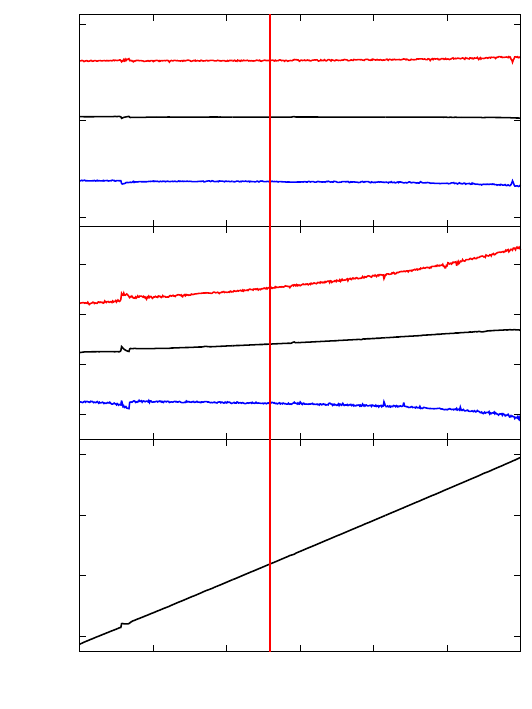}}%
    \gplfronttext
  \end{picture}%
\endgroup

%% file: figures/tex_figure18.tex
\begingroup
  \makeatletter
  \providecommand\color[2][]{%
    \GenericError{(gnuplot) \space\space\space\@spaces}{%
      Package color not loaded in conjunction with
      terminal option `colourtext'%
    }{See the gnuplot documentation for explanation.%
    }{Either use 'blacktext' in gnuplot or load the package
      color.sty in LaTeX.}%
    \renewcommand\color[2][]{}%
  }%
  \providecommand\includegraphics[2][]{%
    \GenericError{(gnuplot) \space\space\space\@spaces}{%
      Package graphicx or graphics not loaded%
    }{See the gnuplot documentation for explanation.%
    }{The gnuplot epslatex terminal needs graphicx.sty or graphics.sty.}%
    \renewcommand\includegraphics[2][]{}%
  }%
  \providecommand\rotatebox[2]{#2}%
  \@ifundefined{ifGPcolor}{%
    \newif\ifGPcolor
    \GPcolorfalse
  }{}%
  \@ifundefined{ifGPblacktext}{%
    \newif\ifGPblacktext
    \GPblacktextfalse
  }{}%
  \let\gplgaddtomacro\g@addto@macro
  \gdef\gplbacktext{}%
  \gdef\gplfronttext{}%
  \makeatother
  \ifGPblacktext
    \def\colorrgb#1{}%
    \def\colorgray#1{}%
  \else
    \ifGPcolor
      \def\colorrgb#1{\color[rgb]{#1}}%
      \def\colorgray#1{\color[gray]{#1}}%
      \expandafter\def\csname LTw\endcsname{\color{white}}%
      \expandafter\def\csname LTb\endcsname{\color{black}}%
      \expandafter\def\csname LTa\endcsname{\color{black}}%
      \expandafter\def\csname LT0\endcsname{\color[rgb]{1,0,0}}%
      \expandafter\def\csname LT1\endcsname{\color[rgb]{0,1,0}}%
      \expandafter\def\csname LT2\endcsname{\color[rgb]{0,0,1}}%
      \expandafter\def\csname LT3\endcsname{\color[rgb]{1,0,1}}%
      \expandafter\def\csname LT4\endcsname{\color[rgb]{0,1,1}}%
      \expandafter\def\csname LT5\endcsname{\color[rgb]{1,1,0}}%
      \expandafter\def\csname LT6\endcsname{\color[rgb]{0,0,0}}%
      \expandafter\def\csname LT7\endcsname{\color[rgb]{1,0.3,0}}%
      \expandafter\def\csname LT8\endcsname{\color[rgb]{0.5,0.5,0.5}}%
    \else
      \def\colorrgb#1{\color{black}}%
      \def\colorgray#1{\color[gray]{#1}}%
      \expandafter\def\csname LTw\endcsname{\color{white}}%
      \expandafter\def\csname LTb\endcsname{\color{black}}%
      \expandafter\def\csname LTa\endcsname{\color{black}}%
      \expandafter\def\csname LT0\endcsname{\color{black}}%
      \expandafter\def\csname LT1\endcsname{\color{black}}%
      \expandafter\def\csname LT2\endcsname{\color{black}}%
      \expandafter\def\csname LT3\endcsname{\color{black}}%
      \expandafter\def\csname LT4\endcsname{\color{black}}%
      \expandafter\def\csname LT5\endcsname{\color{black}}%
      \expandafter\def\csname LT6\endcsname{\color{black}}%
      \expandafter\def\csname LT7\endcsname{\color{black}}%
      \expandafter\def\csname LT8\endcsname{\color{black}}%
    \fi
  \fi
    \setlength{\unitlength}{0.0500bp}%
    \ifx\gptboxheight\undefined%
      \newlength{\gptboxheight}%
      \newlength{\gptboxwidth}%
      \newsavebox{\gptboxtext}%
    \fi%
    \setlength{\fboxrule}{0.5pt}%
    \setlength{\fboxsep}{1pt}%
\begin{picture}(5102.00,5102.00)%
    \gplgaddtomacro\gplbacktext{%
      \csname LTb\endcsname%
      \put(633,635){\makebox(0,0)[r]{\strut{}  -4}}%
      \put(633,884){\makebox(0,0)[r]{\strut{}  -3}}%
      \put(633,1133){\makebox(0,0)[r]{\strut{}  -2}}%
      \put(633,1383){\makebox(0,0)[r]{\strut{}  -1}}%
      \put(633,1632){\makebox(0,0)[r]{\strut{}   0}}%
      \put(633,1881){\makebox(0,0)[r]{\strut{}   1}}%
      \put(633,2131){\makebox(0,0)[r]{\strut{}   2}}%
      \put(633,2380){\makebox(0,0)[r]{\strut{}   3}}%
      \put(633,2629){\makebox(0,0)[r]{\strut{}   4}}%
      \put(765,290){\makebox(0,0){\strut{}$-20$}}%
      \put(1823,290){\makebox(0,0){\strut{}$-15$}}%
      \put(2882,290){\makebox(0,0){\strut{}$-10$}}%
      \put(3940,290){\makebox(0,0){\strut{}$-5$}}%
      \put(4998,290){\makebox(0,0){\strut{}$0$}}%
    }%
    \gplgaddtomacro\gplfronttext{%
      \csname LTb\endcsname%
      \put(127,1632){\rotatebox{-270}{\makebox(0,0){\strut{}$\delta \epsilon$ $(\degree)$}}}%
      \put(2881,-40){\makebox(0,0){\strut{}time $(\Myr)$}}%
      \csname LTb\endcsname%
      \put(4786,2642){\makebox(0,0)[l]{\strut{}b)}}%
    }%
    \gplgaddtomacro\gplbacktext{%
      \csname LTb\endcsname%
      \put(633,2915){\makebox(0,0)[r]{\strut{}-1.5}}%
      \put(633,3236){\makebox(0,0)[r]{\strut{}-1.0}}%
      \put(633,3556){\makebox(0,0)[r]{\strut{}-0.5}}%
      \put(633,3877){\makebox(0,0)[r]{\strut{}0.0}}%
      \put(633,4197){\makebox(0,0)[r]{\strut{}0.5}}%
      \put(633,4517){\makebox(0,0)[r]{\strut{}1.0}}%
      \put(633,4838){\makebox(0,0)[r]{\strut{}1.5}}%
      \put(765,2535){\makebox(0,0){\strut{}}}%
      \put(1823,2535){\makebox(0,0){\strut{}}}%
      \put(2882,2535){\makebox(0,0){\strut{}}}%
      \put(3940,2535){\makebox(0,0){\strut{}}}%
      \put(4998,2535){\makebox(0,0){\strut{}}}%
    }%
    \gplgaddtomacro\gplfronttext{%
      \csname LTb\endcsname%
      \put(127,3876){\rotatebox{-270}{\makebox(0,0){\strut{}$\delta \epsilon$ $(\degree)$}}}%
      \put(2881,2469){\makebox(0,0){\strut{}}}%
      \csname LTb\endcsname%
      \put(4786,4886){\makebox(0,0)[l]{\strut{}a)}}%
    }%
    \gplbacktext
    \put(0,0){\includegraphics{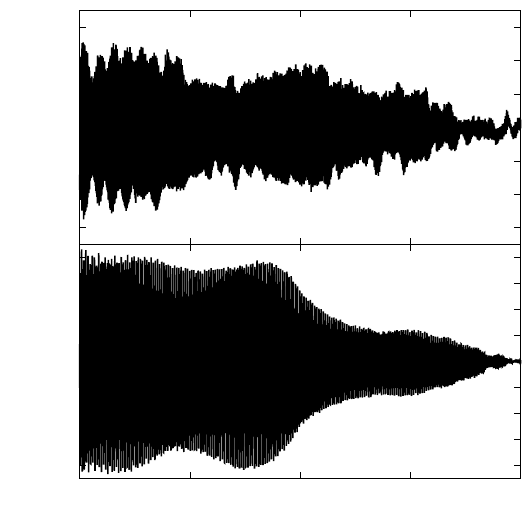}}%
    \gplfronttext
  \end{picture}%
\endgroup

%% file: figures/tex_figure19.tex
\begingroup
  \makeatletter
  \providecommand\color[2][]{%
    \GenericError{(gnuplot) \space\space\space\@spaces}{%
      Package color not loaded in conjunction with
      terminal option `colourtext'%
    }{See the gnuplot documentation for explanation.%
    }{Either use 'blacktext' in gnuplot or load the package
      color.sty in LaTeX.}%
    \renewcommand\color[2][]{}%
  }%
  \providecommand\includegraphics[2][]{%
    \GenericError{(gnuplot) \space\space\space\@spaces}{%
      Package graphicx or graphics not loaded%
    }{See the gnuplot documentation for explanation.%
    }{The gnuplot epslatex terminal needs graphicx.sty or graphics.sty.}%
    \renewcommand\includegraphics[2][]{}%
  }%
  \providecommand\rotatebox[2]{#2}%
  \@ifundefined{ifGPcolor}{%
    \newif\ifGPcolor
    \GPcolorfalse
  }{}%
  \@ifundefined{ifGPblacktext}{%
    \newif\ifGPblacktext
    \GPblacktextfalse
  }{}%
  \let\gplgaddtomacro\g@addto@macro
  \gdef\gplbacktext{}%
  \gdef\gplfronttext{}%
  \makeatother
  \ifGPblacktext
    \def\colorrgb#1{}%
    \def\colorgray#1{}%
  \else
    \ifGPcolor
      \def\colorrgb#1{\color[rgb]{#1}}%
      \def\colorgray#1{\color[gray]{#1}}%
      \expandafter\def\csname LTw\endcsname{\color{white}}%
      \expandafter\def\csname LTb\endcsname{\color{black}}%
      \expandafter\def\csname LTa\endcsname{\color{black}}%
      \expandafter\def\csname LT0\endcsname{\color[rgb]{1,0,0}}%
      \expandafter\def\csname LT1\endcsname{\color[rgb]{0,1,0}}%
      \expandafter\def\csname LT2\endcsname{\color[rgb]{0,0,1}}%
      \expandafter\def\csname LT3\endcsname{\color[rgb]{1,0,1}}%
      \expandafter\def\csname LT4\endcsname{\color[rgb]{0,1,1}}%
      \expandafter\def\csname LT5\endcsname{\color[rgb]{1,1,0}}%
      \expandafter\def\csname LT6\endcsname{\color[rgb]{0,0,0}}%
      \expandafter\def\csname LT7\endcsname{\color[rgb]{1,0.3,0}}%
      \expandafter\def\csname LT8\endcsname{\color[rgb]{0.5,0.5,0.5}}%
    \else
      \def\colorrgb#1{\color{black}}%
      \def\colorgray#1{\color[gray]{#1}}%
      \expandafter\def\csname LTw\endcsname{\color{white}}%
      \expandafter\def\csname LTb\endcsname{\color{black}}%
      \expandafter\def\csname LTa\endcsname{\color{black}}%
      \expandafter\def\csname LT0\endcsname{\color{black}}%
      \expandafter\def\csname LT1\endcsname{\color{black}}%
      \expandafter\def\csname LT2\endcsname{\color{black}}%
      \expandafter\def\csname LT3\endcsname{\color{black}}%
      \expandafter\def\csname LT4\endcsname{\color{black}}%
      \expandafter\def\csname LT5\endcsname{\color{black}}%
      \expandafter\def\csname LT6\endcsname{\color{black}}%
      \expandafter\def\csname LT7\endcsname{\color{black}}%
      \expandafter\def\csname LT8\endcsname{\color{black}}%
    \fi
  \fi
    \setlength{\unitlength}{0.0500bp}%
    \ifx\gptboxheight\undefined%
      \newlength{\gptboxheight}%
      \newlength{\gptboxwidth}%
      \newsavebox{\gptboxtext}%
    \fi%
    \setlength{\fboxrule}{0.5pt}%
    \setlength{\fboxsep}{1pt}%
\begin{picture}(5102.00,5102.00)%
    \gplgaddtomacro\gplbacktext{%
      \csname LTb\endcsname%
      \put(480,650){\makebox(0,0)[r]{\strut{}-7}}%
      \put(480,931){\makebox(0,0)[r]{\strut{}-6}}%
      \put(480,1211){\makebox(0,0)[r]{\strut{}-5}}%
      \put(480,1492){\makebox(0,0)[r]{\strut{}-4}}%
      \put(480,1772){\makebox(0,0)[r]{\strut{}-3}}%
      \put(480,2053){\makebox(0,0)[r]{\strut{}-2}}%
      \put(480,2333){\makebox(0,0)[r]{\strut{}-1}}%
      \put(480,2614){\makebox(0,0)[r]{\strut{}0}}%
      \put(612,290){\makebox(0,0){\strut{}$0$}}%
      \put(1479,290){\makebox(0,0){\strut{}$20$}}%
      \put(2346,290){\makebox(0,0){\strut{}$40$}}%
      \put(3213,290){\makebox(0,0){\strut{}$60$}}%
      \put(4080,290){\makebox(0,0){\strut{}$80$}}%
      \put(4947,290){\makebox(0,0){\strut{}$100$}}%
    }%
    \gplgaddtomacro\gplfronttext{%
      \csname LTb\endcsname%
      \put(106,1632){\rotatebox{-270}{\makebox(0,0){\strut{}$\log_{10}(\sigma)$}}}%
      \put(2779,-40){\makebox(0,0){\strut{}$\epsilon$ $(\degree)$}}%
      \csname LTb\endcsname%
      \put(4730,2642){\makebox(0,0)[l]{\strut{}b)}}%
    }%
    \gplgaddtomacro\gplbacktext{%
      \csname LTb\endcsname%
      \put(480,2895){\makebox(0,0)[r]{\strut{}-7}}%
      \put(480,3176){\makebox(0,0)[r]{\strut{}-6}}%
      \put(480,3456){\makebox(0,0)[r]{\strut{}-5}}%
      \put(480,3736){\makebox(0,0)[r]{\strut{}-4}}%
      \put(480,4017){\makebox(0,0)[r]{\strut{}-3}}%
      \put(480,4297){\makebox(0,0)[r]{\strut{}-2}}%
      \put(480,4577){\makebox(0,0)[r]{\strut{}-1}}%
      \put(480,4858){\makebox(0,0)[r]{\strut{}0}}%
      \put(612,2535){\makebox(0,0){\strut{}}}%
      \put(1479,2535){\makebox(0,0){\strut{}}}%
      \put(2346,2535){\makebox(0,0){\strut{}}}%
      \put(3213,2535){\makebox(0,0){\strut{}}}%
      \put(4080,2535){\makebox(0,0){\strut{}}}%
      \put(4947,2535){\makebox(0,0){\strut{}}}%
    }%
    \gplgaddtomacro\gplfronttext{%
      \csname LTb\endcsname%
      \put(106,3876){\rotatebox{-270}{\makebox(0,0){\strut{}$\log_{10}(\sigma)$}}}%
      \put(2779,2469){\makebox(0,0){\strut{}}}%
      \csname LTb\endcsname%
      \put(4730,4886){\makebox(0,0)[l]{\strut{}a)}}%
    }%
    \gplbacktext
    \put(0,0){\includegraphics{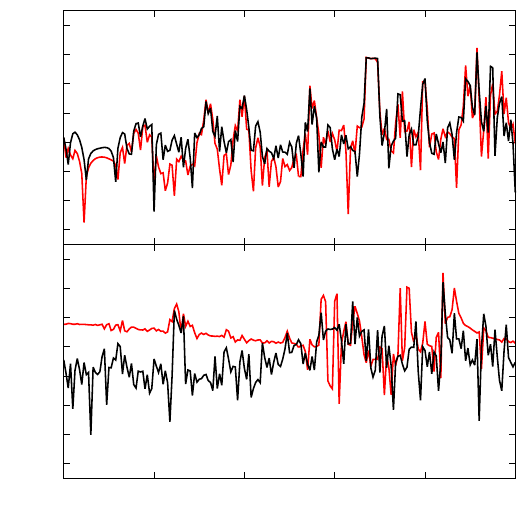}}%
    \gplfronttext
  \end{picture}%
\endgroup

%% file: figures/tex_figure20.tex
\begingroup
  \makeatletter
  \providecommand\color[2][]{%
    \GenericError{(gnuplot) \space\space\space\@spaces}{%
      Package color not loaded in conjunction with
      terminal option `colourtext'%
    }{See the gnuplot documentation for explanation.%
    }{Either use 'blacktext' in gnuplot or load the package
      color.sty in LaTeX.}%
    \renewcommand\color[2][]{}%
  }%
  \providecommand\includegraphics[2][]{%
    \GenericError{(gnuplot) \space\space\space\@spaces}{%
      Package graphicx or graphics not loaded%
    }{See the gnuplot documentation for explanation.%
    }{The gnuplot epslatex terminal needs graphicx.sty or graphics.sty.}%
    \renewcommand\includegraphics[2][]{}%
  }%
  \providecommand\rotatebox[2]{#2}%
  \@ifundefined{ifGPcolor}{%
    \newif\ifGPcolor
    \GPcolorfalse
  }{}%
  \@ifundefined{ifGPblacktext}{%
    \newif\ifGPblacktext
    \GPblacktextfalse
  }{}%
  \let\gplgaddtomacro\g@addto@macro
  \gdef\gplbacktext{}%
  \gdef\gplfronttext{}%
  \makeatother
  \ifGPblacktext
    \def\colorrgb#1{}%
    \def\colorgray#1{}%
  \else
    \ifGPcolor
      \def\colorrgb#1{\color[rgb]{#1}}%
      \def\colorgray#1{\color[gray]{#1}}%
      \expandafter\def\csname LTw\endcsname{\color{white}}%
      \expandafter\def\csname LTb\endcsname{\color{black}}%
      \expandafter\def\csname LTa\endcsname{\color{black}}%
      \expandafter\def\csname LT0\endcsname{\color[rgb]{1,0,0}}%
      \expandafter\def\csname LT1\endcsname{\color[rgb]{0,1,0}}%
      \expandafter\def\csname LT2\endcsname{\color[rgb]{0,0,1}}%
      \expandafter\def\csname LT3\endcsname{\color[rgb]{1,0,1}}%
      \expandafter\def\csname LT4\endcsname{\color[rgb]{0,1,1}}%
      \expandafter\def\csname LT5\endcsname{\color[rgb]{1,1,0}}%
      \expandafter\def\csname LT6\endcsname{\color[rgb]{0,0,0}}%
      \expandafter\def\csname LT7\endcsname{\color[rgb]{1,0.3,0}}%
      \expandafter\def\csname LT8\endcsname{\color[rgb]{0.5,0.5,0.5}}%
    \else
      \def\colorrgb#1{\color{black}}%
      \def\colorgray#1{\color[gray]{#1}}%
      \expandafter\def\csname LTw\endcsname{\color{white}}%
      \expandafter\def\csname LTb\endcsname{\color{black}}%
      \expandafter\def\csname LTa\endcsname{\color{black}}%
      \expandafter\def\csname LT0\endcsname{\color{black}}%
      \expandafter\def\csname LT1\endcsname{\color{black}}%
      \expandafter\def\csname LT2\endcsname{\color{black}}%
      \expandafter\def\csname LT3\endcsname{\color{black}}%
      \expandafter\def\csname LT4\endcsname{\color{black}}%
      \expandafter\def\csname LT5\endcsname{\color{black}}%
      \expandafter\def\csname LT6\endcsname{\color{black}}%
      \expandafter\def\csname LT7\endcsname{\color{black}}%
      \expandafter\def\csname LT8\endcsname{\color{black}}%
    \fi
  \fi
    \setlength{\unitlength}{0.0500bp}%
    \ifx\gptboxheight\undefined%
      \newlength{\gptboxheight}%
      \newlength{\gptboxwidth}%
      \newsavebox{\gptboxtext}%
    \fi%
    \setlength{\fboxrule}{0.5pt}%
    \setlength{\fboxsep}{1pt}%
\begin{picture}(5102.00,6802.00)%
    \gplgaddtomacro\gplbacktext{%
    }%
    \gplgaddtomacro\gplfronttext{%
      \csname LTb\endcsname%
      \put(127,5644){\rotatebox{-270}{\makebox(0,0){\strut{}$\epsilon$ $(\degree)$}}}%
      \put(2881,4339){\makebox(0,0){\strut{}}}%
      \csname LTb\endcsname%
      \put(633,4718){\makebox(0,0)[r]{\strut{} 0}}%
      \put(633,5181){\makebox(0,0)[r]{\strut{}10}}%
      \put(633,5645){\makebox(0,0)[r]{\strut{}20}}%
      \put(633,6108){\makebox(0,0)[r]{\strut{}30}}%
      \put(633,6571){\makebox(0,0)[r]{\strut{}40}}%
      \put(765,4405){\makebox(0,0){\strut{}}}%
      \put(1471,4405){\makebox(0,0){\strut{}}}%
      \put(2176,4405){\makebox(0,0){\strut{}}}%
      \put(2882,4405){\makebox(0,0){\strut{}}}%
      \put(3587,4405){\makebox(0,0){\strut{}}}%
      \put(4293,4405){\makebox(0,0){\strut{}}}%
      \put(4998,4405){\makebox(0,0){\strut{}}}%
      \csname LTb\endcsname%
      \put(4786,6562){\makebox(0,0)[l]{\strut{}a)}}%
      \put(2811,2380){\makebox(0,0)[l]{\strut{}A}}%
      \put(2726,680){\makebox(0,0)[l]{\strut{}B}}%
    }%
    \gplgaddtomacro\gplbacktext{%
    }%
    \gplgaddtomacro\gplfronttext{%
      \csname LTb\endcsname%
      \put(127,3604){\rotatebox{-270}{\makebox(0,0){\strut{}$f_C$ $(\arc)$}}}%
      \put(2881,2298){\makebox(0,0){\strut{}}}%
      \csname LTb\endcsname%
      \put(633,2730){\makebox(0,0)[r]{\strut{}-12}}%
      \put(633,3021){\makebox(0,0)[r]{\strut{}-10}}%
      \put(633,3313){\makebox(0,0)[r]{\strut{}-8}}%
      \put(633,3604){\makebox(0,0)[r]{\strut{}-6}}%
      \put(633,3895){\makebox(0,0)[r]{\strut{}-4}}%
      \put(633,4187){\makebox(0,0)[r]{\strut{}-2}}%
      \put(633,4478){\makebox(0,0)[r]{\strut{} 0}}%
      \put(765,2364){\makebox(0,0){\strut{}}}%
      \put(1471,2364){\makebox(0,0){\strut{}}}%
      \put(2176,2364){\makebox(0,0){\strut{}}}%
      \put(2882,2364){\makebox(0,0){\strut{}}}%
      \put(3587,2364){\makebox(0,0){\strut{}}}%
      \put(4293,2364){\makebox(0,0){\strut{}}}%
      \put(4998,2364){\makebox(0,0){\strut{}}}%
      \csname LTb\endcsname%
      \put(4786,4522){\makebox(0,0)[l]{\strut{}b)}}%
      \put(2811,2380){\makebox(0,0)[l]{\strut{}A}}%
      \put(2726,680){\makebox(0,0)[l]{\strut{}B}}%
    }%
    \gplgaddtomacro\gplbacktext{%
    }%
    \gplgaddtomacro\gplfronttext{%
      \csname LTb\endcsname%
      \put(127,1564){\rotatebox{-270}{\makebox(0,0){\strut{}$\log_{10}(\sigma)$}}}%
      \put(2881,-6){\makebox(0,0){\strut{}$\alpha$ $(\arc)$}}%
      \csname LTb\endcsname%
      \put(633,617){\makebox(0,0)[r]{\strut{}-7}}%
      \put(633,908){\makebox(0,0)[r]{\strut{}-6}}%
      \put(633,1200){\makebox(0,0)[r]{\strut{}-5}}%
      \put(633,1491){\makebox(0,0)[r]{\strut{}-4}}%
      \put(633,1783){\makebox(0,0)[r]{\strut{}-3}}%
      \put(633,2074){\makebox(0,0)[r]{\strut{}-2}}%
      \put(633,2365){\makebox(0,0)[r]{\strut{}-1}}%
      \put(765,324){\makebox(0,0){\strut{}0}}%
      \put(1471,324){\makebox(0,0){\strut{}2}}%
      \put(2176,324){\makebox(0,0){\strut{}4}}%
      \put(2882,324){\makebox(0,0){\strut{}6}}%
      \put(3587,324){\makebox(0,0){\strut{}8}}%
      \put(4293,324){\makebox(0,0){\strut{}10}}%
      \put(4998,324){\makebox(0,0){\strut{}12}}%
      \csname LTb\endcsname%
      \put(4786,2482){\makebox(0,0)[l]{\strut{}c)}}%
      \put(2811,2380){\makebox(0,0)[l]{\strut{}A}}%
      \put(2726,680){\makebox(0,0)[l]{\strut{}B}}%
    }%
    \gplbacktext
    \put(0,0){\includegraphics{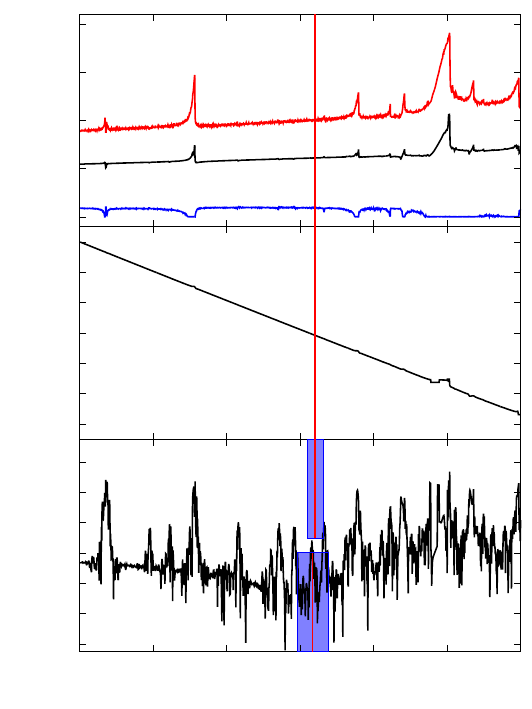}}%
    \gplfronttext
  \end{picture}%
\endgroup

%% file: figures/tex_figure21.tex
\begingroup
  \makeatletter
  \providecommand\color[2][]{%
    \GenericError{(gnuplot) \space\space\space\@spaces}{%
      Package color not loaded in conjunction with
      terminal option `colourtext'%
    }{See the gnuplot documentation for explanation.%
    }{Either use 'blacktext' in gnuplot or load the package
      color.sty in LaTeX.}%
    \renewcommand\color[2][]{}%
  }%
  \providecommand\includegraphics[2][]{%
    \GenericError{(gnuplot) \space\space\space\@spaces}{%
      Package graphicx or graphics not loaded%
    }{See the gnuplot documentation for explanation.%
    }{The gnuplot epslatex terminal needs graphicx.sty or graphics.sty.}%
    \renewcommand\includegraphics[2][]{}%
  }%
  \providecommand\rotatebox[2]{#2}%
  \@ifundefined{ifGPcolor}{%
    \newif\ifGPcolor
    \GPcolorfalse
  }{}%
  \@ifundefined{ifGPblacktext}{%
    \newif\ifGPblacktext
    \GPblacktextfalse
  }{}%
  \let\gplgaddtomacro\g@addto@macro
  \gdef\gplbacktext{}%
  \gdef\gplfronttext{}%
  \makeatother
  \ifGPblacktext
    \def\colorrgb#1{}%
    \def\colorgray#1{}%
  \else
    \ifGPcolor
      \def\colorrgb#1{\color[rgb]{#1}}%
      \def\colorgray#1{\color[gray]{#1}}%
      \expandafter\def\csname LTw\endcsname{\color{white}}%
      \expandafter\def\csname LTb\endcsname{\color{black}}%
      \expandafter\def\csname LTa\endcsname{\color{black}}%
      \expandafter\def\csname LT0\endcsname{\color[rgb]{1,0,0}}%
      \expandafter\def\csname LT1\endcsname{\color[rgb]{0,1,0}}%
      \expandafter\def\csname LT2\endcsname{\color[rgb]{0,0,1}}%
      \expandafter\def\csname LT3\endcsname{\color[rgb]{1,0,1}}%
      \expandafter\def\csname LT4\endcsname{\color[rgb]{0,1,1}}%
      \expandafter\def\csname LT5\endcsname{\color[rgb]{1,1,0}}%
      \expandafter\def\csname LT6\endcsname{\color[rgb]{0,0,0}}%
      \expandafter\def\csname LT7\endcsname{\color[rgb]{1,0.3,0}}%
      \expandafter\def\csname LT8\endcsname{\color[rgb]{0.5,0.5,0.5}}%
    \else
      \def\colorrgb#1{\color{black}}%
      \def\colorgray#1{\color[gray]{#1}}%
      \expandafter\def\csname LTw\endcsname{\color{white}}%
      \expandafter\def\csname LTb\endcsname{\color{black}}%
      \expandafter\def\csname LTa\endcsname{\color{black}}%
      \expandafter\def\csname LT0\endcsname{\color{black}}%
      \expandafter\def\csname LT1\endcsname{\color{black}}%
      \expandafter\def\csname LT2\endcsname{\color{black}}%
      \expandafter\def\csname LT3\endcsname{\color{black}}%
      \expandafter\def\csname LT4\endcsname{\color{black}}%
      \expandafter\def\csname LT5\endcsname{\color{black}}%
      \expandafter\def\csname LT6\endcsname{\color{black}}%
      \expandafter\def\csname LT7\endcsname{\color{black}}%
      \expandafter\def\csname LT8\endcsname{\color{black}}%
    \fi
  \fi
    \setlength{\unitlength}{0.0500bp}%
    \ifx\gptboxheight\undefined%
      \newlength{\gptboxheight}%
      \newlength{\gptboxwidth}%
      \newsavebox{\gptboxtext}%
    \fi%
    \setlength{\fboxrule}{0.5pt}%
    \setlength{\fboxsep}{1pt}%
\begin{picture}(5102.00,6802.00)%
    \gplgaddtomacro\gplbacktext{%
    }%
    \gplgaddtomacro\gplfronttext{%
      \csname LTb\endcsname%
      \put(127,5644){\rotatebox{-270}{\makebox(0,0){\strut{}$\epsilon$ $(\degree)$}}}%
      \put(2881,4339){\makebox(0,0){\strut{}}}%
      \csname LTb\endcsname%
      \put(633,4693){\makebox(0,0)[r]{\strut{} 0}}%
      \put(633,5033){\makebox(0,0)[r]{\strut{}10}}%
      \put(633,5373){\makebox(0,0)[r]{\strut{}20}}%
      \put(633,5712){\makebox(0,0)[r]{\strut{}30}}%
      \put(633,6052){\makebox(0,0)[r]{\strut{}40}}%
      \put(633,6392){\makebox(0,0)[r]{\strut{}50}}%
      \put(765,4405){\makebox(0,0){\strut{}}}%
      \put(1471,4405){\makebox(0,0){\strut{}}}%
      \put(2176,4405){\makebox(0,0){\strut{}}}%
      \put(2882,4405){\makebox(0,0){\strut{}}}%
      \put(3587,4405){\makebox(0,0){\strut{}}}%
      \put(4293,4405){\makebox(0,0){\strut{}}}%
      \put(4998,4405){\makebox(0,0){\strut{}}}%
      \csname LTb\endcsname%
      \put(4786,6562){\makebox(0,0)[l]{\strut{}a)}}%
      \put(2341,2380){\makebox(0,0)[l]{\strut{}A}}%
      \put(991,2380){\makebox(0,0)[l]{\strut{}B}}%
    }%
    \gplgaddtomacro\gplbacktext{%
    }%
    \gplgaddtomacro\gplfronttext{%
      \csname LTb\endcsname%
      \put(127,3604){\rotatebox{-270}{\makebox(0,0){\strut{}$f_V$ $(\arc)$}}}%
      \put(2881,2298){\makebox(0,0){\strut{}}}%
      \csname LTb\endcsname%
      \put(633,2754){\makebox(0,0)[r]{\strut{}-18}}%
      \put(633,3094){\makebox(0,0)[r]{\strut{}-16}}%
      \put(633,3434){\makebox(0,0)[r]{\strut{}-14}}%
      \put(633,3774){\makebox(0,0)[r]{\strut{}-12}}%
      \put(633,4114){\makebox(0,0)[r]{\strut{}-10}}%
      \put(633,4454){\makebox(0,0)[r]{\strut{}-8}}%
      \put(765,2364){\makebox(0,0){\strut{}}}%
      \put(1471,2364){\makebox(0,0){\strut{}}}%
      \put(2176,2364){\makebox(0,0){\strut{}}}%
      \put(2882,2364){\makebox(0,0){\strut{}}}%
      \put(3587,2364){\makebox(0,0){\strut{}}}%
      \put(4293,2364){\makebox(0,0){\strut{}}}%
      \put(4998,2364){\makebox(0,0){\strut{}}}%
      \csname LTb\endcsname%
      \put(4786,4522){\makebox(0,0)[l]{\strut{}b)}}%
      \put(2341,2380){\makebox(0,0)[l]{\strut{}A}}%
      \put(991,2380){\makebox(0,0)[l]{\strut{}B}}%
    }%
    \gplgaddtomacro\gplbacktext{%
    }%
    \gplgaddtomacro\gplfronttext{%
      \csname LTb\endcsname%
      \put(127,1564){\rotatebox{-270}{\makebox(0,0){\strut{}$\log_{10}(\sigma)$}}}%
      \put(2881,-6){\makebox(0,0){\strut{}$\alpha$ $(\arc)$}}%
      \csname LTb\endcsname%
      \put(633,651){\makebox(0,0)[r]{\strut{}-5}}%
      \put(633,1081){\makebox(0,0)[r]{\strut{}-4}}%
      \put(633,1510){\makebox(0,0)[r]{\strut{}-3}}%
      \put(633,1940){\makebox(0,0)[r]{\strut{}-2}}%
      \put(633,2369){\makebox(0,0)[r]{\strut{}-1}}%
      \put(765,324){\makebox(0,0){\strut{}10}}%
      \put(1471,324){\makebox(0,0){\strut{}12}}%
      \put(2176,324){\makebox(0,0){\strut{}14}}%
      \put(2882,324){\makebox(0,0){\strut{}16}}%
      \put(3587,324){\makebox(0,0){\strut{}18}}%
      \put(4293,324){\makebox(0,0){\strut{}20}}%
      \put(4998,324){\makebox(0,0){\strut{}22}}%
      \csname LTb\endcsname%
      \put(4786,2482){\makebox(0,0)[l]{\strut{}c)}}%
      \put(2341,2380){\makebox(0,0)[l]{\strut{}A}}%
      \put(991,2380){\makebox(0,0)[l]{\strut{}B}}%
    }%
    \gplbacktext
    \put(0,0){\includegraphics{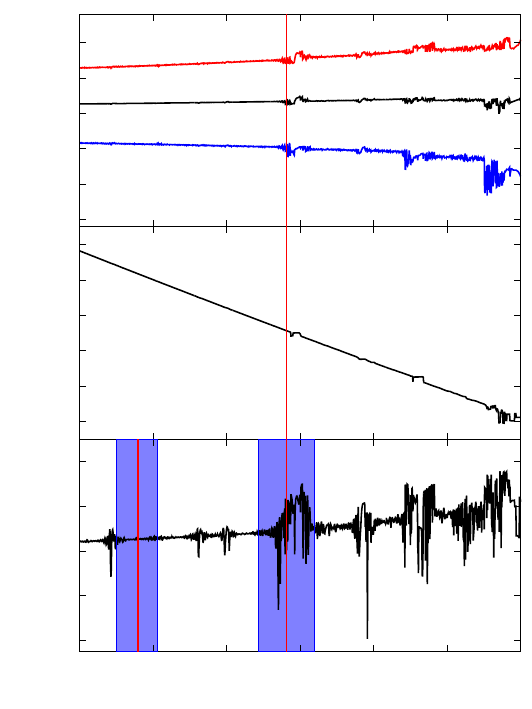}}%
    \gplfronttext
  \end{picture}%
\endgroup

%% file: figures/tex_figure22a.tex
\begingroup
  \makeatletter
  \providecommand\color[2][]{%
    \GenericError{(gnuplot) \space\space\space\@spaces}{%
      Package color not loaded in conjunction with
      terminal option `colourtext'%
    }{See the gnuplot documentation for explanation.%
    }{Either use 'blacktext' in gnuplot or load the package
      color.sty in LaTeX.}%
    \renewcommand\color[2][]{}%
  }%
  \providecommand\includegraphics[2][]{%
    \GenericError{(gnuplot) \space\space\space\@spaces}{%
      Package graphicx or graphics not loaded%
    }{See the gnuplot documentation for explanation.%
    }{The gnuplot epslatex terminal needs graphicx.sty or graphics.sty.}%
    \renewcommand\includegraphics[2][]{}%
  }%
  \providecommand\rotatebox[2]{#2}%
  \@ifundefined{ifGPcolor}{%
    \newif\ifGPcolor
    \GPcolorfalse
  }{}%
  \@ifundefined{ifGPblacktext}{%
    \newif\ifGPblacktext
    \GPblacktextfalse
  }{}%
  \let\gplgaddtomacro\g@addto@macro
  \gdef\gplbacktext{}%
  \gdef\gplfronttext{}%
  \makeatother
  \ifGPblacktext
    \def\colorrgb#1{}%
    \def\colorgray#1{}%
  \else
    \ifGPcolor
      \def\colorrgb#1{\color[rgb]{#1}}%
      \def\colorgray#1{\color[gray]{#1}}%
      \expandafter\def\csname LTw\endcsname{\color{white}}%
      \expandafter\def\csname LTb\endcsname{\color{black}}%
      \expandafter\def\csname LTa\endcsname{\color{black}}%
      \expandafter\def\csname LT0\endcsname{\color[rgb]{1,0,0}}%
      \expandafter\def\csname LT1\endcsname{\color[rgb]{0,1,0}}%
      \expandafter\def\csname LT2\endcsname{\color[rgb]{0,0,1}}%
      \expandafter\def\csname LT3\endcsname{\color[rgb]{1,0,1}}%
      \expandafter\def\csname LT4\endcsname{\color[rgb]{0,1,1}}%
      \expandafter\def\csname LT5\endcsname{\color[rgb]{1,1,0}}%
      \expandafter\def\csname LT6\endcsname{\color[rgb]{0,0,0}}%
      \expandafter\def\csname LT7\endcsname{\color[rgb]{1,0.3,0}}%
      \expandafter\def\csname LT8\endcsname{\color[rgb]{0.5,0.5,0.5}}%
    \else
      \def\colorrgb#1{\color{black}}%
      \def\colorgray#1{\color[gray]{#1}}%
      \expandafter\def\csname LTw\endcsname{\color{white}}%
      \expandafter\def\csname LTb\endcsname{\color{black}}%
      \expandafter\def\csname LTa\endcsname{\color{black}}%
      \expandafter\def\csname LT0\endcsname{\color{black}}%
      \expandafter\def\csname LT1\endcsname{\color{black}}%
      \expandafter\def\csname LT2\endcsname{\color{black}}%
      \expandafter\def\csname LT3\endcsname{\color{black}}%
      \expandafter\def\csname LT4\endcsname{\color{black}}%
      \expandafter\def\csname LT5\endcsname{\color{black}}%
      \expandafter\def\csname LT6\endcsname{\color{black}}%
      \expandafter\def\csname LT7\endcsname{\color{black}}%
      \expandafter\def\csname LT8\endcsname{\color{black}}%
    \fi
  \fi
    \setlength{\unitlength}{0.0500bp}%
    \ifx\gptboxheight\undefined%
      \newlength{\gptboxheight}%
      \newlength{\gptboxwidth}%
      \newsavebox{\gptboxtext}%
    \fi%
    \setlength{\fboxrule}{0.5pt}%
    \setlength{\fboxsep}{1pt}%
\begin{picture}(5102.00,12754.00)%
    \gplgaddtomacro\gplbacktext{%
    }%
    \gplgaddtomacro\gplfronttext{%
      \csname LTb\endcsname%
      \put(1192,2423){\rotatebox{-270}{\makebox(0,0){\strut{}$\alpha$ $(\arc)$}}}%
      \put(3289,-40){\makebox(0,0){\strut{}$\epsilon_0$ $(\degree)$}}%
      \csname LTb\endcsname%
      \put(1500,510){\makebox(0,0)[r]{\strut{} 0}}%
      \put(1500,1145){\makebox(0,0)[r]{\strut{}10}}%
      \put(1500,1780){\makebox(0,0)[r]{\strut{}20}}%
      \put(1500,2415){\makebox(0,0)[r]{\strut{}30}}%
      \put(1500,3050){\makebox(0,0)[r]{\strut{}40}}%
      \put(1500,3685){\makebox(0,0)[r]{\strut{}50}}%
      \put(1500,4320){\makebox(0,0)[r]{\strut{}60}}%
      \put(1640,290){\makebox(0,0){\strut{}$0$}}%
      \put(1970,290){\makebox(0,0){\strut{}$10$}}%
      \put(2300,290){\makebox(0,0){\strut{}$20$}}%
      \put(2630,290){\makebox(0,0){\strut{}$30$}}%
      \put(2960,290){\makebox(0,0){\strut{}$40$}}%
      \put(3290,290){\makebox(0,0){\strut{}$50$}}%
      \put(3619,290){\makebox(0,0){\strut{}$60$}}%
      \put(3949,290){\makebox(0,0){\strut{}$70$}}%
      \put(4279,290){\makebox(0,0){\strut{}$80$}}%
      \put(4609,290){\makebox(0,0){\strut{}$90$}}%
      \put(4939,290){\makebox(0,0){\strut{}$100$}}%
      \csname LTb\endcsname%
      \put(51,12498){\makebox(0,0)[l]{\strut{}a)}}%
      \put(51,8417){\makebox(0,0)[l]{\strut{}b)}}%
      \put(51,4336){\makebox(0,0)[l]{\strut{}c)}}%
    }%
    \gplgaddtomacro\gplbacktext{%
    }%
    \gplgaddtomacro\gplfronttext{%
      \csname LTb\endcsname%
      \put(142,2423){\rotatebox{-270}{\makebox(0,0){\strut{}$f_C$ $(\arc)$}}}%
      \put(841,224){\makebox(0,0){\strut{}}}%
      \csname LTb\endcsname%
      \put(582,510){\makebox(0,0)[r]{\strut{}-37}}%
      \put(582,1147){\makebox(0,0)[r]{\strut{}-30}}%
      \put(582,2107){\makebox(0,0)[r]{\strut{}-20}}%
      \put(582,3068){\makebox(0,0)[r]{\strut{}-10}}%
      \put(582,4028){\makebox(0,0)[r]{\strut{}  0}}%
      \put(582,4336){\makebox(0,0)[r]{\strut{}  3}}%
      \put(714,290){\makebox(0,0){\strut{}}}%
      \put(969,290){\makebox(0,0){\strut{}}}%
      \put(51,12498){\makebox(0,0)[l]{\strut{}a)}}%
      \put(51,8417){\makebox(0,0)[l]{\strut{}b)}}%
      \put(51,4336){\makebox(0,0)[l]{\strut{}c)}}%
    }%
    \gplgaddtomacro\gplbacktext{%
    }%
    \gplgaddtomacro\gplfronttext{%
      \csname LTb\endcsname%
      \put(1192,6503){\rotatebox{-270}{\makebox(0,0){\strut{}$\alpha$ $(\arc)$}}}%
      \put(3289,4305){\makebox(0,0){\strut{}}}%
      \csname LTb\endcsname%
      \put(1500,4591){\makebox(0,0)[r]{\strut{} 0}}%
      \put(1500,5226){\makebox(0,0)[r]{\strut{}10}}%
      \put(1500,5861){\makebox(0,0)[r]{\strut{}20}}%
      \put(1500,6496){\makebox(0,0)[r]{\strut{}30}}%
      \put(1500,7130){\makebox(0,0)[r]{\strut{}40}}%
      \put(1500,7765){\makebox(0,0)[r]{\strut{}50}}%
      \put(1500,8400){\makebox(0,0)[r]{\strut{}60}}%
      \put(1640,4371){\makebox(0,0){\strut{}}}%
      \put(1970,4371){\makebox(0,0){\strut{}}}%
      \put(2300,4371){\makebox(0,0){\strut{}}}%
      \put(2630,4371){\makebox(0,0){\strut{}}}%
      \put(2960,4371){\makebox(0,0){\strut{}}}%
      \put(3290,4371){\makebox(0,0){\strut{}}}%
      \put(3619,4371){\makebox(0,0){\strut{}}}%
      \put(3949,4371){\makebox(0,0){\strut{}}}%
      \put(4279,4371){\makebox(0,0){\strut{}}}%
      \put(4609,4371){\makebox(0,0){\strut{}}}%
      \put(4939,4371){\makebox(0,0){\strut{}}}%
      \csname LTb\endcsname%
      \put(51,12498){\makebox(0,0)[l]{\strut{}a)}}%
      \put(51,8417){\makebox(0,0)[l]{\strut{}b)}}%
      \put(51,4336){\makebox(0,0)[l]{\strut{}c)}}%
    }%
    \gplgaddtomacro\gplbacktext{%
    }%
    \gplgaddtomacro\gplfronttext{%
      \csname LTb\endcsname%
      \put(142,6503){\rotatebox{-270}{\makebox(0,0){\strut{}$\epsilon$ $(\degree)$}}}%
      \put(841,4305){\makebox(0,0){\strut{}}}%
      \csname LTb\endcsname%
      \put(582,4591){\makebox(0,0)[r]{\strut{} 15}}%
      \put(582,4916){\makebox(0,0)[r]{\strut{} 20}}%
      \put(582,5566){\makebox(0,0)[r]{\strut{} 30}}%
      \put(582,6216){\makebox(0,0)[r]{\strut{} 40}}%
      \put(582,6866){\makebox(0,0)[r]{\strut{} 50}}%
      \put(582,7516){\makebox(0,0)[r]{\strut{} 60}}%
      \put(582,8166){\makebox(0,0)[r]{\strut{} 70}}%
      \put(582,8416){\makebox(0,0)[r]{\strut{} 74}}%
      \put(714,4371){\makebox(0,0){\strut{}}}%
      \put(969,4371){\makebox(0,0){\strut{}}}%
      \put(51,12498){\makebox(0,0)[l]{\strut{}a)}}%
      \put(51,8417){\makebox(0,0)[l]{\strut{}b)}}%
      \put(51,4336){\makebox(0,0)[l]{\strut{}c)}}%
    }%
    \gplgaddtomacro\gplbacktext{%
    }%
    \gplgaddtomacro\gplfronttext{%
      \csname LTb\endcsname%
      \put(1192,10584){\rotatebox{-270}{\makebox(0,0){\strut{}$\alpha$ $(\arc)$}}}%
      \put(3289,8386){\makebox(0,0){\strut{}}}%
      \csname LTb\endcsname%
      \put(1500,8672){\makebox(0,0)[r]{\strut{} 0}}%
      \put(1500,9307){\makebox(0,0)[r]{\strut{}10}}%
      \put(1500,9942){\makebox(0,0)[r]{\strut{}20}}%
      \put(1500,10577){\makebox(0,0)[r]{\strut{}30}}%
      \put(1500,11211){\makebox(0,0)[r]{\strut{}40}}%
      \put(1500,11846){\makebox(0,0)[r]{\strut{}50}}%
      \put(1500,12481){\makebox(0,0)[r]{\strut{}60}}%
      \put(1640,8452){\makebox(0,0){\strut{}}}%
      \put(1970,8452){\makebox(0,0){\strut{}}}%
      \put(2300,8452){\makebox(0,0){\strut{}}}%
      \put(2630,8452){\makebox(0,0){\strut{}}}%
      \put(2960,8452){\makebox(0,0){\strut{}}}%
      \put(3290,8452){\makebox(0,0){\strut{}}}%
      \put(3619,8452){\makebox(0,0){\strut{}}}%
      \put(3949,8452){\makebox(0,0){\strut{}}}%
      \put(4279,8452){\makebox(0,0){\strut{}}}%
      \put(4609,8452){\makebox(0,0){\strut{}}}%
      \put(4939,8452){\makebox(0,0){\strut{}}}%
      \csname LTb\endcsname%
      \put(51,12498){\makebox(0,0)[l]{\strut{}a)}}%
      \put(51,8417){\makebox(0,0)[l]{\strut{}b)}}%
      \put(51,4336){\makebox(0,0)[l]{\strut{}c)}}%
    }%
    \gplgaddtomacro\gplbacktext{%
    }%
    \gplgaddtomacro\gplfronttext{%
      \csname LTb\endcsname%
      \put(142,10584){\rotatebox{-270}{\makebox(0,0){\strut{}$\log_{10}(\sigma)$}}}%
      \put(841,8386){\makebox(0,0){\strut{}}}%
      \csname LTb\endcsname%
      \put(582,8672){\makebox(0,0)[r]{\strut{} -6}}%
      \put(582,9437){\makebox(0,0)[r]{\strut{} -5}}%
      \put(582,10202){\makebox(0,0)[r]{\strut{} -4}}%
      \put(582,10967){\makebox(0,0)[r]{\strut{} -3}}%
      \put(582,11732){\makebox(0,0)[r]{\strut{} -2}}%
      \put(582,12497){\makebox(0,0)[r]{\strut{} -1}}%
      \put(714,8452){\makebox(0,0){\strut{}}}%
      \put(969,8452){\makebox(0,0){\strut{}}}%
      \put(51,12498){\makebox(0,0)[l]{\strut{}a)}}%
      \put(51,8417){\makebox(0,0)[l]{\strut{}b)}}%
      \put(51,4336){\makebox(0,0)[l]{\strut{}c)}}%
    }%
    \gplbacktext
    \put(0,0){\includegraphics{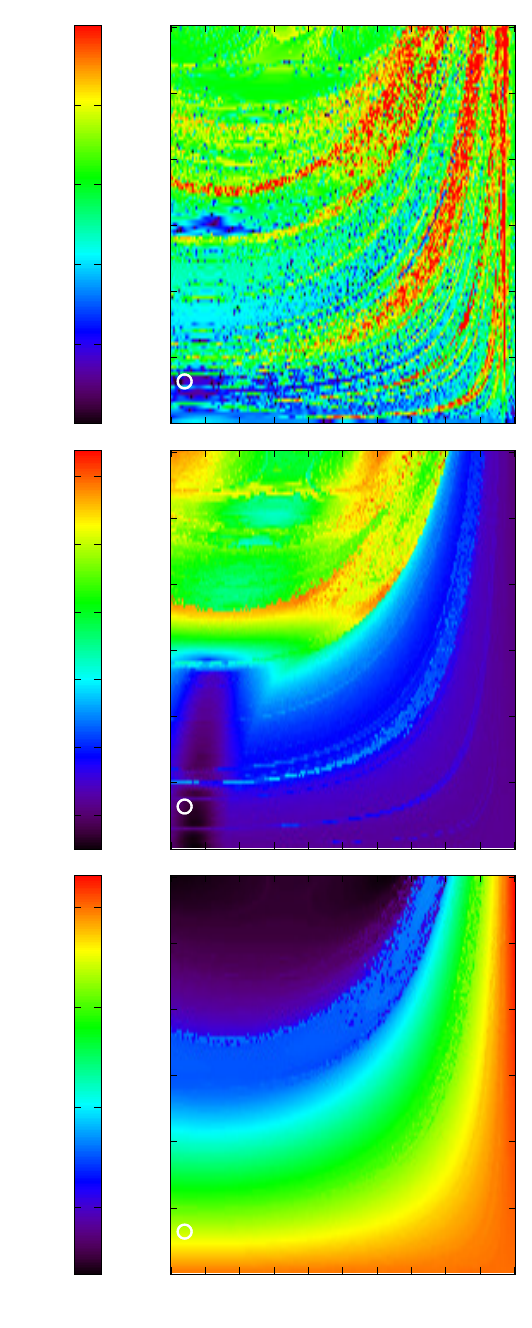}}%
    \gplfronttext
  \end{picture}%
\endgroup

%% file: figures/tex_figure22b.tex
\begingroup
  \makeatletter
  \providecommand\color[2][]{%
    \GenericError{(gnuplot) \space\space\space\@spaces}{%
      Package color not loaded in conjunction with
      terminal option `colourtext'%
    }{See the gnuplot documentation for explanation.%
    }{Either use 'blacktext' in gnuplot or load the package
      color.sty in LaTeX.}%
    \renewcommand\color[2][]{}%
  }%
  \providecommand\includegraphics[2][]{%
    \GenericError{(gnuplot) \space\space\space\@spaces}{%
      Package graphicx or graphics not loaded%
    }{See the gnuplot documentation for explanation.%
    }{The gnuplot epslatex terminal needs graphicx.sty or graphics.sty.}%
    \renewcommand\includegraphics[2][]{}%
  }%
  \providecommand\rotatebox[2]{#2}%
  \@ifundefined{ifGPcolor}{%
    \newif\ifGPcolor
    \GPcolorfalse
  }{}%
  \@ifundefined{ifGPblacktext}{%
    \newif\ifGPblacktext
    \GPblacktextfalse
  }{}%
  \let\gplgaddtomacro\g@addto@macro
  \gdef\gplbacktext{}%
  \gdef\gplfronttext{}%
  \makeatother
  \ifGPblacktext
    \def\colorrgb#1{}%
    \def\colorgray#1{}%
  \else
    \ifGPcolor
      \def\colorrgb#1{\color[rgb]{#1}}%
      \def\colorgray#1{\color[gray]{#1}}%
      \expandafter\def\csname LTw\endcsname{\color{white}}%
      \expandafter\def\csname LTb\endcsname{\color{black}}%
      \expandafter\def\csname LTa\endcsname{\color{black}}%
      \expandafter\def\csname LT0\endcsname{\color[rgb]{1,0,0}}%
      \expandafter\def\csname LT1\endcsname{\color[rgb]{0,1,0}}%
      \expandafter\def\csname LT2\endcsname{\color[rgb]{0,0,1}}%
      \expandafter\def\csname LT3\endcsname{\color[rgb]{1,0,1}}%
      \expandafter\def\csname LT4\endcsname{\color[rgb]{0,1,1}}%
      \expandafter\def\csname LT5\endcsname{\color[rgb]{1,1,0}}%
      \expandafter\def\csname LT6\endcsname{\color[rgb]{0,0,0}}%
      \expandafter\def\csname LT7\endcsname{\color[rgb]{1,0.3,0}}%
      \expandafter\def\csname LT8\endcsname{\color[rgb]{0.5,0.5,0.5}}%
    \else
      \def\colorrgb#1{\color{black}}%
      \def\colorgray#1{\color[gray]{#1}}%
      \expandafter\def\csname LTw\endcsname{\color{white}}%
      \expandafter\def\csname LTb\endcsname{\color{black}}%
      \expandafter\def\csname LTa\endcsname{\color{black}}%
      \expandafter\def\csname LT0\endcsname{\color{black}}%
      \expandafter\def\csname LT1\endcsname{\color{black}}%
      \expandafter\def\csname LT2\endcsname{\color{black}}%
      \expandafter\def\csname LT3\endcsname{\color{black}}%
      \expandafter\def\csname LT4\endcsname{\color{black}}%
      \expandafter\def\csname LT5\endcsname{\color{black}}%
      \expandafter\def\csname LT6\endcsname{\color{black}}%
      \expandafter\def\csname LT7\endcsname{\color{black}}%
      \expandafter\def\csname LT8\endcsname{\color{black}}%
    \fi
  \fi
    \setlength{\unitlength}{0.0500bp}%
    \ifx\gptboxheight\undefined%
      \newlength{\gptboxheight}%
      \newlength{\gptboxwidth}%
      \newsavebox{\gptboxtext}%
    \fi%
    \setlength{\fboxrule}{0.5pt}%
    \setlength{\fboxsep}{1pt}%
\begin{picture}(5102.00,12754.00)%
    \gplgaddtomacro\gplbacktext{%
    }%
    \gplgaddtomacro\gplfronttext{%
      \csname LTb\endcsname%
      \put(1192,2423){\rotatebox{-270}{\makebox(0,0){\strut{}$\alpha$ $(\arc)$}}}%
      \put(3289,-40){\makebox(0,0){\strut{}$\epsilon_0$ $(\degree)$}}%
      \csname LTb\endcsname%
      \put(1500,510){\makebox(0,0)[r]{\strut{} 0}}%
      \put(1500,1145){\makebox(0,0)[r]{\strut{}10}}%
      \put(1500,1780){\makebox(0,0)[r]{\strut{}20}}%
      \put(1500,2415){\makebox(0,0)[r]{\strut{}30}}%
      \put(1500,3050){\makebox(0,0)[r]{\strut{}40}}%
      \put(1500,3685){\makebox(0,0)[r]{\strut{}50}}%
      \put(1500,4320){\makebox(0,0)[r]{\strut{}60}}%
      \put(1640,290){\makebox(0,0){\strut{}$0$}}%
      \put(1970,290){\makebox(0,0){\strut{}$10$}}%
      \put(2300,290){\makebox(0,0){\strut{}$20$}}%
      \put(2630,290){\makebox(0,0){\strut{}$30$}}%
      \put(2960,290){\makebox(0,0){\strut{}$40$}}%
      \put(3290,290){\makebox(0,0){\strut{}$50$}}%
      \put(3619,290){\makebox(0,0){\strut{}$60$}}%
      \put(3949,290){\makebox(0,0){\strut{}$70$}}%
      \put(4279,290){\makebox(0,0){\strut{}$80$}}%
      \put(4609,290){\makebox(0,0){\strut{}$90$}}%
      \put(4939,290){\makebox(0,0){\strut{}$100$}}%
      \csname LTb\endcsname%
      \put(51,12498){\makebox(0,0)[l]{\strut{}d)}}%
      \put(51,8417){\makebox(0,0)[l]{\strut{}e)}}%
      \put(51,4336){\makebox(0,0)[l]{\strut{}f)}}%
    }%
    \gplgaddtomacro\gplbacktext{%
    }%
    \gplgaddtomacro\gplfronttext{%
      \csname LTb\endcsname%
      \put(142,2423){\rotatebox{-270}{\makebox(0,0){\strut{}$f_V$ $(\arc)$}}}%
      \put(841,224){\makebox(0,0){\strut{}}}%
      \csname LTb\endcsname%
      \put(582,510){\makebox(0,0)[r]{\strut{}-57}}%
      \put(582,914){\makebox(0,0)[r]{\strut{}-50}}%
      \put(582,1458){\makebox(0,0)[r]{\strut{}-40}}%
      \put(582,2002){\makebox(0,0)[r]{\strut{}-30}}%
      \put(582,2545){\makebox(0,0)[r]{\strut{}-20}}%
      \put(582,3089){\makebox(0,0)[r]{\strut{}-10}}%
      \put(582,3633){\makebox(0,0)[r]{\strut{}  0}}%
      \put(582,4177){\makebox(0,0)[r]{\strut{} 10}}%
      \put(582,4336){\makebox(0,0)[r]{\strut{} 13}}%
      \put(714,290){\makebox(0,0){\strut{}}}%
      \put(969,290){\makebox(0,0){\strut{}}}%
      \put(51,12498){\makebox(0,0)[l]{\strut{}d)}}%
      \put(51,8417){\makebox(0,0)[l]{\strut{}e)}}%
      \put(51,4336){\makebox(0,0)[l]{\strut{}f)}}%
    }%
    \gplgaddtomacro\gplbacktext{%
    }%
    \gplgaddtomacro\gplfronttext{%
      \csname LTb\endcsname%
      \put(1192,6503){\rotatebox{-270}{\makebox(0,0){\strut{}$\alpha$ $(\arc)$}}}%
      \put(3289,4305){\makebox(0,0){\strut{}}}%
      \csname LTb\endcsname%
      \put(1500,4591){\makebox(0,0)[r]{\strut{} 0}}%
      \put(1500,5226){\makebox(0,0)[r]{\strut{}10}}%
      \put(1500,5861){\makebox(0,0)[r]{\strut{}20}}%
      \put(1500,6496){\makebox(0,0)[r]{\strut{}30}}%
      \put(1500,7130){\makebox(0,0)[r]{\strut{}40}}%
      \put(1500,7765){\makebox(0,0)[r]{\strut{}50}}%
      \put(1500,8400){\makebox(0,0)[r]{\strut{}60}}%
      \put(1640,4371){\makebox(0,0){\strut{}}}%
      \put(1970,4371){\makebox(0,0){\strut{}}}%
      \put(2300,4371){\makebox(0,0){\strut{}}}%
      \put(2630,4371){\makebox(0,0){\strut{}}}%
      \put(2960,4371){\makebox(0,0){\strut{}}}%
      \put(3290,4371){\makebox(0,0){\strut{}}}%
      \put(3619,4371){\makebox(0,0){\strut{}}}%
      \put(3949,4371){\makebox(0,0){\strut{}}}%
      \put(4279,4371){\makebox(0,0){\strut{}}}%
      \put(4609,4371){\makebox(0,0){\strut{}}}%
      \put(4939,4371){\makebox(0,0){\strut{}}}%
      \csname LTb\endcsname%
      \put(51,12498){\makebox(0,0)[l]{\strut{}d)}}%
      \put(51,8417){\makebox(0,0)[l]{\strut{}e)}}%
      \put(51,4336){\makebox(0,0)[l]{\strut{}f)}}%
    }%
    \gplgaddtomacro\gplbacktext{%
    }%
    \gplgaddtomacro\gplfronttext{%
      \csname LTb\endcsname%
      \put(142,6503){\rotatebox{-270}{\makebox(0,0){\strut{}$\epsilon$ $(\degree)$}}}%
      \put(841,4305){\makebox(0,0){\strut{}}}%
      \csname LTb\endcsname%
      \put(582,4591){\makebox(0,0)[r]{\strut{} 12}}%
      \put(582,4998){\makebox(0,0)[r]{\strut{} 20}}%
      \put(582,5502){\makebox(0,0)[r]{\strut{} 30}}%
      \put(582,6005){\makebox(0,0)[r]{\strut{} 40}}%
      \put(582,6509){\makebox(0,0)[r]{\strut{} 50}}%
      \put(582,7012){\makebox(0,0)[r]{\strut{} 60}}%
      \put(582,7516){\makebox(0,0)[r]{\strut{} 70}}%
      \put(582,8019){\makebox(0,0)[r]{\strut{} 80}}%
      \put(582,8416){\makebox(0,0)[r]{\strut{} 88}}%
      \put(714,4371){\makebox(0,0){\strut{}}}%
      \put(969,4371){\makebox(0,0){\strut{}}}%
      \put(51,12498){\makebox(0,0)[l]{\strut{}d)}}%
      \put(51,8417){\makebox(0,0)[l]{\strut{}e)}}%
      \put(51,4336){\makebox(0,0)[l]{\strut{}f)}}%
    }%
    \gplgaddtomacro\gplbacktext{%
    }%
    \gplgaddtomacro\gplfronttext{%
      \csname LTb\endcsname%
      \put(1192,10584){\rotatebox{-270}{\makebox(0,0){\strut{}$\alpha$ $(\arc)$}}}%
      \put(3289,8386){\makebox(0,0){\strut{}}}%
      \csname LTb\endcsname%
      \put(1500,8672){\makebox(0,0)[r]{\strut{} 0}}%
      \put(1500,9307){\makebox(0,0)[r]{\strut{}10}}%
      \put(1500,9942){\makebox(0,0)[r]{\strut{}20}}%
      \put(1500,10577){\makebox(0,0)[r]{\strut{}30}}%
      \put(1500,11211){\makebox(0,0)[r]{\strut{}40}}%
      \put(1500,11846){\makebox(0,0)[r]{\strut{}50}}%
      \put(1500,12481){\makebox(0,0)[r]{\strut{}60}}%
      \put(1640,8452){\makebox(0,0){\strut{}}}%
      \put(1970,8452){\makebox(0,0){\strut{}}}%
      \put(2300,8452){\makebox(0,0){\strut{}}}%
      \put(2630,8452){\makebox(0,0){\strut{}}}%
      \put(2960,8452){\makebox(0,0){\strut{}}}%
      \put(3290,8452){\makebox(0,0){\strut{}}}%
      \put(3619,8452){\makebox(0,0){\strut{}}}%
      \put(3949,8452){\makebox(0,0){\strut{}}}%
      \put(4279,8452){\makebox(0,0){\strut{}}}%
      \put(4609,8452){\makebox(0,0){\strut{}}}%
      \put(4939,8452){\makebox(0,0){\strut{}}}%
      \csname LTb\endcsname%
      \put(51,12498){\makebox(0,0)[l]{\strut{}d)}}%
      \put(51,8417){\makebox(0,0)[l]{\strut{}e)}}%
      \put(51,4336){\makebox(0,0)[l]{\strut{}f)}}%
    }%
    \gplgaddtomacro\gplbacktext{%
    }%
    \gplgaddtomacro\gplfronttext{%
      \csname LTb\endcsname%
      \put(142,10584){\rotatebox{-270}{\makebox(0,0){\strut{}$\log_{10}(\sigma)$}}}%
      \put(841,8386){\makebox(0,0){\strut{}}}%
      \csname LTb\endcsname%
      \put(582,8672){\makebox(0,0)[r]{\strut{} -6}}%
      \put(582,9437){\makebox(0,0)[r]{\strut{} -5}}%
      \put(582,10202){\makebox(0,0)[r]{\strut{} -4}}%
      \put(582,10967){\makebox(0,0)[r]{\strut{} -3}}%
      \put(582,11732){\makebox(0,0)[r]{\strut{} -2}}%
      \put(582,12497){\makebox(0,0)[r]{\strut{} -1}}%
      \put(714,8452){\makebox(0,0){\strut{}}}%
      \put(969,8452){\makebox(0,0){\strut{}}}%
      \put(51,12498){\makebox(0,0)[l]{\strut{}d)}}%
      \put(51,8417){\makebox(0,0)[l]{\strut{}e)}}%
      \put(51,4336){\makebox(0,0)[l]{\strut{}f)}}%
    }%
    \gplbacktext
    \put(0,0){\includegraphics{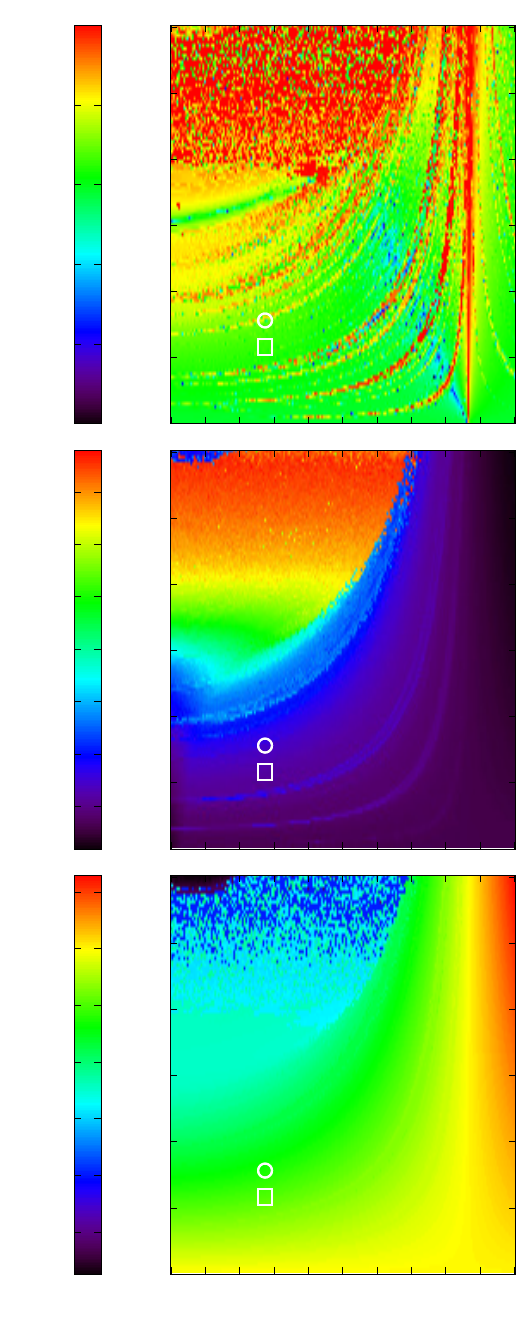}}%
    \gplfronttext
  \end{picture}%
\endgroup